%% file: ComaLFpaper.tex
\documentclass[twocolumn,trackchanges,twocolappendix]{aastex701}
\shorttitle{Galaxy Luminosity Function of the Coma Cluster}
\shortauthors{Zöller et al.}
\graphicspath{{./}{figures/}}
\usepackage{amsmath}
\usepackage{todonotes}

\begin{document}

\title{Galaxy Luminosity Function of the Coma Cluster from Deep $u'$--$g'$--$r'$ Wendelstein Imaging Data}

\author[0000-0002-0938-5686]{Raphael Zöller}
\thanks{rzoeller@mpe.mpg.de}
\email{rzoeller@mpe.mpg.de}
\affiliation{University Observatory, Faculty of Physics, Ludwig-Maximilians-Universität München, Scheinerstr. 1, 81679 Munich, Germany}
\affiliation{Max Planck Institute for Extraterrestrial Physics, Giessenbachstrasse, D-85748 Garching, Germany}
\author[0000-0002-9618-2552]{Matthias Kluge}
\email{mkluge@mpe.mpg.de}
\affiliation{Max Planck Institute for Extraterrestrial Physics, Giessenbachstrasse, D-85748 Garching, Germany}
\author[0000-0001-7179-0626]{Ralf Bender}
\email{bender@mpe.mpg.de}
\affiliation{University Observatory, Faculty of Physics, Ludwig-Maximilians-Universität München, Scheinerstr. 1, 81679 Munich, Germany}
\affiliation{Max Planck Institute for Extraterrestrial Physics, Giessenbachstrasse, D-85748 Garching, Germany}
\author[0009-0006-9461-002X]{Jan-Niklas Pippert}
\email{jpippert@mpe.mpg.de}
\affiliation{Max Planck Institute for Extraterrestrial Physics, Giessenbachstrasse, D-85748 Garching, Germany}
\author[0009-0000-3688-4379]{Benjamin Seidel}
\email{bseidel@usm.lmu.de}
\affiliation{University Observatory, Faculty of Physics, Ludwig-Maximilians-Universität München, Scheinerstr. 1, 81679 Munich, Germany}
\author[0000-0002-2152-6277]{Claus Gössl}
\email{claus.goessl@physik.lmu.de}
\affiliation{University Observatory, Faculty of Physics, Ludwig-Maximilians-Universität München, Scheinerstr. 1, 81679 Munich, Germany}
\author[0000-0003-1008-225X]{Ulrich Hopp}
\email{hopp@usm.lmu.de}
\affiliation{University Observatory, Faculty of Physics, Ludwig-Maximilians-Universität München, Scheinerstr. 1, 81679 Munich, Germany}
\author[0009-0006-3527-0424]{Hanna Kellermann}
\email{hanna.kellermann@physik.lmu.de}
\affiliation{University Observatory, Faculty of Physics, Ludwig-Maximilians-Universität München, Scheinerstr. 1, 81679 Munich, Germany}
\author{Christoph Ries}
\email{cries@usm.lmu.de}
\affiliation{University Observatory, Faculty of Physics, Ludwig-Maximilians-Universität München, Scheinerstr. 1, 81679 Munich, Germany}
\author[0000-0002-5466-3892]{Arno Riffeser}
\email{riffeser@physik.lmu.de}
\affiliation{University Observatory, Faculty of Physics, Ludwig-Maximilians-Universität München, Scheinerstr. 1, 81679 Munich, Germany}
\author{Michael Schmidt}
\email{mschmidt@usm.lmu.de}
\affiliation{University Observatory, Faculty of Physics, Ludwig-Maximilians-Universität München, Scheinerstr. 1, 81679 Munich, Germany}
\author[0009-0006-1571-0306]{Luis Thomas}
\email{lthomas@mpe.mpg.de}
\affiliation{University Observatory, Faculty of Physics, Ludwig-Maximilians-Universität München, Scheinerstr. 1, 81679 Munich, Germany}
\affiliation{Max Planck Institute for Extraterrestrial Physics, Giessenbachstrasse, D-85748 Garching, Germany}


\begin{abstract}
We derive the $g'$ band galaxy luminosity function (GLF) of quenched resolved galaxies in the Coma cluster from a deep-imaging survey with $\approx1.5\,\mathrm{deg^2}$ around the cluster center. The dataset comprises deep $u'$-, $g'$-, and $r'$-band data obtained with the Wendelstein Wide Field Imager on the 2.1\,m Fraunhofer Wendelstein Telescope reaching median $3\sigma$ surface brightness limits in $10\arcsec\times10\arcsec$ boxes of  $\mathrm{ (30.0\,u',\,\,29.6\,g',\,\,28.7\,r')\,mag\,arcsec^{-2}}$. We measure structural parameters across a large dynamic range in galaxy brightness ($-24.5\,g'\,\mathrm{mag} \lessapprox M\lessapprox-11.3\,g'\,\mathrm{mag}$), from the brightest cluster galaxy to low-luminosity dwarfs, including compact dwarf galaxies and ultra-diffuse galaxies.
We automatically identify more than 6000 cluster member candidates based on their membership on the quiescent sequence in the $u'-g'$ versus $g'-r'$ color--color diagram.
The structural parameters of bright galaxies are obtained via isophotal modeling, and fully automated parametric image fitting for faint ones. Injection-recovery tests and two identically analyzed reference fields provide statistical corrections for completeness and contamination, yielding a representative GLF that reliably probes the faint end and may serve as a benchmark for future studies. We report a best-fit single Schechter $g'$ band GLF with $M^*=-21.60^{+0.25}_{-0.27}\,g'\,\mathrm{mag}$, $\log_{10}(\phi^*\,[\mathrm{deg}^{-2}\,\mathrm{mag}^{-1}])=1.378^{+0.076}_{-0.081}$, and a comparatively steep faint-end slope $\alpha=-1.445^{+0.016}_{-0.017}$. 
A directly matched comparison with the Coma counterpart in SLOW, a constrained cosmological simulation using CDM, shows broad agreement between the GLFs down to the simulation limit of $M = -15.5\,g'\,\mathrm{mag}$, despite a deficit of bright galaxies.
\end{abstract}



\section{Introduction} \label{sec:intro}

The galaxy luminosity function (GLF) and the galaxy stellar mass function (GSMF) quantify the number density of galaxies per magnitude or mass interval, respectively, and are usually described by a \cite{Schechter1976} function. They are important tools for understanding galaxy populations and galaxy evolution, as well as constraining cosmological models and dark matter models, by comparing observed GLFs \citep[e.g.,][]{Binggeli1985,Efstathiou1988,Bernstein1995,Trentham1998,Andreon2002,Mobasher2003,
Blanton2005,Faber2007,Yamanoi2012,Ferrarese2016,Wright2017,CuillandreLF2025} with the predicted GLFs and GSMFs from simulations \citep[e.g.,][]{Kauffmann1993,Pillepich2018, Behroozi2019,Dolag2025}. The tension between the predicted steep faint-end slope from first-generation cosmological simulations and observed flat slopes caused discussions about the "missing satellite problem" \citep[e.g.,][]{Mateo1998,Klypin1999,Moore1999}. For observations, the major limitations are the corrections for completeness and contamination, particularly for imaging-based GLFs relying on statistical foreground/background subtraction, with cosmic variance further contributing to the contamination uncertainty \citep[e.g., ][]{Driver2010}. Morphology-selected imaging samples can reduce contamination, but require sufficient spatial resolution and may introduce observer-dependent classification biases and a less well-defined selection function \citep[see e.g., ][]{Marleau2025,CuillandreLF2025}, while spectroscopic samples provide cleaner membership information but are limited by depth, target selection, and spatial coverage \citep[e.g., ][]{Mobasher2003}.

For simulations, the treatment of baryonic physics is crucial as stellar winds and supernova feedback suppress the star formation in low mass galaxies, whereas AGN feedback suppresses the star formation in massive galaxies and hence affect the stellar mass -- halo mass function, and the stellar mass function, leading to fewer low-mass galaxies and, indirectly, to fewer luminous galaxies, with the luminosity dependence also set by stellar populations and mass-to-light ratios \citep[e.g.,][]{Moster2010,Moster2013,Moster2018,Behroozi2019}.
Furthermore, the slope of the GLF and GSMF are powerful tools to probe dark matter models, i.e., cold dark matter predicts a high abundance of dwarf galaxies, whereas warm dark matter suppresses the formation of small halos \citep{Menci2012,Bullock2017}. 

Recent deep imaging surveys have greatly increased the number of identified low-surface brightness (LSB) galaxies such as ultra-diffuse galaxies (UDGs) \citep[e.g.,][]{vanDokkum2015,Yagi2016, Venhola2019, Marleau2021,Zoeller2024,Marleau2025}. In this study, we measure the GLF in the Coma cluster, including those most diffuse galaxies but also compact (though resolved) dwarf galaxies.

In this study, we focus on the GLF, but our Coma cluster galaxy sample also provides an excellent dataset to study the distribution of galaxies in various parameter spaces, which will be discussed in a complementary work in \cite{Zoellerinprep}. Both our goals, determining the GLF and structural parameter correlation densities, require a correction for the selection function, i.e., correcting for contamination and completeness. Especially for dwarf galaxies, detecting and separating them from interloping galaxies to classify them as cluster members is a challenge. Many dwarf galaxy studies rely on visual inspection to identify them, either fully relying on visual inspection of images, or using a semi-automatic procedure, first automatically preselecting candidates, followed by a visual inspection \citep[e.g.,][]{Binggeli1985,Mueller2017,Wittmann2017,Venhola2017,Habas2020,Marleau2025,Marleau2026}. Despite this giving relatively clean and complete dwarf galaxy samples, determining a selection function for visual inspection is problematic. In addition, upscaling to upcoming deep surveys such as Euclid \citep{EuclidLaureijs2011,EuclidScaramella2022,EuclidMellier2025} and LSST \citep{LSST2019} will require full automation. 

For a massive galaxy cluster like the Coma cluster with $M_{200}=1.88\times10^{15}\,h^{-1}\,\mathrm{M_\odot}$ within $r_{200}=1.99\,h^{-1}\,\mathrm{Mpc}$ \citep{Kubo2007}, we expect that the galaxy population in the central region of the cluster is dominated by quenched 
elliptical galaxies, S0 galaxies, dwarf spheroidal galaxies, and dwarf elliptical galaxies \citep[e.g.,][]{Dressler1980,Lewis2002,Gomez2003} that follow the red sequence \citep[e.g.,][]{Visvanathan1977,Bower1992}. \cite{Zoeller2024} presented an automatic approach to detect cluster member galaxies using \verb+SExtractor+ \citep{Sextractor} for the object detection and separation of them from background objects based on quenched sequence membership in the $u'-g'$ vs. $g'-r'$ color-color diagram \citep{Williams,McIntosh} and red sequence membership \citep{Visvanathan1977}. However, this previous study was based on images with significantly varying depth over the field and aimed for a clean sample rather than a complete sample using manual improvement of automatically created masks and manual fit rejection. In the present study, we build on this approach while revising the cluster-member selection and analysis procedure, as discussed in the following.

For this work, we obtained additional observations of the Coma cluster to achieve a larger field of view (FoV), and, more importantly, a homogeneous depth. Furthermore, we adjust the procedure presented in \cite{Zoeller2024} for full automation of the dwarf galaxy cluster member candidate selection and their structural parameter measurements, also allowing more interloping objects and then statistically correcting for contamination using the background-contamination function (BCF), defined from the identical analysis of two reference fields.
Furthermore, we correct for completeness using an injection-recovery test.

Throughout this paper, we follow \citet{Scolnic2025} and use $z_{helio}=0.0232$, $z_{CMB}=0.02445\pm0.00024$, $m-M=(34.97\pm0.05)\,\mathrm{mag}$, and $D_L=(98.5\pm2.2)\,\mathrm{Mpc}$, yielding $D_A=(93.9\pm2.1)\,\mathrm{Mpc}$ and a physical scale of $\mathrm{(0.455\pm0.010)\,kpc\,arcsec^{-1}}$ for the Coma cluster.

\section{Data} \label{sec:data}
Our observations have been carried out with the Wendelstein Wide Field Imager  \citep[WWFI;][]{WWFI} at the 2.1\,m-Fraunhofer Wendelstein Telescope \citep{Hopp2014}. The WWFI covers a field of view of $27.6\arcmin\times28.9\arcmin$. It consists of four CCDs in a $2\times2$ mosaic and each CCD has $4096\times4109$ pixels with a pixel scale of $\mathrm{0.2\,arcsec\,px^{-1}}$.

\subsection{Sample and Observing Strategy}
Our objective is to acquire deep $u'$-, $g'$-, and $r'$-band images of the Coma cluster with uniform depth in order to minimize sample selection biases. The $u'-g'-r'$ color information is used to select the member galaxies (see Section \ref{sec:selection}). For this, we start with the data taken for the Coma cluster studies of the brightest cluster galaxy (BCG), intracluster light (ICL), and dwarf galaxies by \cite{kluge,Kluge2021} and \cite{Zoeller2024}. The dithering strategy for this archival data was optimized to measure the ICL around BCGs, and, therefore, delivers a large spatial coverage, allowing us to study the whole galaxy population in the Coma cluster. However, the depth is inhomogeneous.
One full dither pattern consists of 52 dither steps, as described in \cite{kluge} and \cite{Zoeller2024}.
The observations are centered between the two brightest galaxies in the Coma cluster, NGC\,4874 and NGC\,4889. 
The pattern is built from 13 four-exposure blocks. Within each block, the cluster center is placed once on each of the four CCDs. From one block to the next, the pointing is shifted by 2\arcmin, in either the R.A.\ or decl.\ direction, sampling different off-center configurations. Thus, one full dither pattern contains $13\times4=52$ science exposures.
For this central pointing, the ICL of the Coma cluster spans nearly the entire field of view under the large dither pattern, making it impossible to derive accurate night-sky flats from the science exposures alone. Therefore, dedicated sky pointings are necessary for the central pointing. These sky pointings were obtained between the science dither steps and used to construct night-sky flats.

This dither pattern was designed to optimize the background subtraction, but it does not provide a uniform depth, especially near the image corners. In order to obtain a more homogeneous depth and to increase the FoV, we took additional data with the same dither strategy at eight pointings around the central Coma pointing. For those pointing, additional sky pointing are not necessary. The central coordinates of the different pointings are given in Table \ref{tab:Coords}.
\begin{deluxetable}{c|cc}[t]
    \tabletypesize{\small}
    \tablecaption{Coordinates of Pointing Centers}
    \label{tab:Coords}
    \tablehead{
        \colhead{Pointing} & \colhead{R.A.} & \colhead{Decl.}    
    }
    \startdata
    Coma Central &12:59:48 &27:58:48\\
    Coma Sky & 12:56:38 & 28:08:27\\
    Coma P0&13:01:48 & 27:31:48\\
    Coma P1&13:01:48 & 27:58:48\\
    Coma P2&13:01:48 & 28:25:48\\
    Coma P3&12:59:48 & 28:25:48\\
    Coma P4&12:57:48 & 28:26:49\\
    Coma P5&12:57:48 & 27:58:48\\
    Coma P6&12:57:48 & 27:31:48\\
    Coma P7&12:59:48 & 27:31:48\\
    Ref1 (SDSSJ1433+6007) & 14:32:29 & 60:12:26\\
    Ref2 (SDSSJ0909+4449) & 9:10:11 & 44:45:58\\
    \enddata
    \tablecomments{Coordinates are given as right ascension and declination in the ICRS reference.}
\end{deluxetable}  

Furthermore, we observed two reference fields to investigate the sample contamination by interloping galaxies. For this, we chose a pointing around the lensed quasar SDSSJ1433+6007 (from now on referred to as Ref1) as we already had archival deep $g'$-band data available for the time-delay cosmography studies of this quasar \citep{Queirolo} with additional $u'$ and $r'$ band data from \cite{Zoeller2024}. The other reference pointing is located around the lensed quasar SDSSJ0909+4449 (from now on referred to as Ref2). For our reference pointings, we stick to the dither pattern applied by \citet{Queirolo}. Here, we dither only 8\arcsec~ per dither step without centering the pointing on the different CCDs. This gives us a relatively uniform depth over a smaller FoV.

All observations were scheduled under photometric conditions during dark time, with a zenith sky brightness fainter than $21.3\,V\mathrm{\,mag\,arcsec^{-2}}$. Exposure times were chosen to ensure that sky photon noise dominated over readout noise. For the $g'$ and the $r'$ bands, we used 60\,s exposures in the fast readout mode. For the $u'$ band, we opted for 600\,s exposures in the slow readout mode, which reduces readout noise to about one-quarter with respect to the fast mode, albeit with a fourfold increase in readout time. For the $g'$-band observations of Ref1, \cite{Queirolo} used longer individual exposure times of 240\,s. We emphasize that the different individual exposure times of the target and reference pointings have no considerable impact on the depth, as in both cases, the depth is limited by the photon noise of the sky. For each pointing, we took one full dither pattern (each 52 images) in the $u'$ band, four full dither patterns in the $g'$ band, and two full dither patterns in the $r'$ band.

\subsection{Data Reduction} \label{sec:datareduction}
The data was processed using the WWFI data reduction pipeline \citep{kluge,klugeDiss}, which incorporates \verb+fitstools+ \citep{fitstools}, SExtractor \citep{Sextractor}, SCAMP \citep{scamp}, and SWarp \citep{swarp}.
The pipeline performs various tasks, such as bias subtraction, flat-fielding (using twilight flats), automatic masking of charge persistence, bad pixels, and cosmic rays. At the operating temperature of $\mathrm{-115^{\circ}C}$, dark current is negligible for the WWFI \citep{WWFI}. Photometric zero-points for the $g'$ and $r'$ bands are determined by comparing the flux of point sources in 5\arcsec\, diameter apertures ($\mathrm{ZP_{5}}$) to the 
PSF magnitudes of the corresponding stacked objects in the PS1 DR2 catalog\footnote{The catalog is accessible at \url{https://catalogs.mast.stsci.edu/panstarrs/} }.

Our $u'$-band data is calibrated to the Sloan Digital Sky Survey \citep[SDSS;][]{SDSS_overview_2000,SDSS_DR12_2015} photometric system because Pan-STARRS does not provide coverage in this band. Individual $u'$-band exposures are first calibrated relative to each other with SCAMP using 5\arcsec\, diameter apertures and of order 400 stars. A direct stellar zero-point calibration of the stacked image relative to SDSS is not possible because suitable stars in the SDSS images are saturated in our WWFI exposures owing to the longer WWFI exposure times, better seeing, and low detector gain. Therefore, the stacked $u'$ image is calibrated relative to SDSS DR12 $u$-band imaging by minimizing residuals for bright extended sources, after resampling both images to a pixel scale of 2\arcsec\,pixel$^{-1}$ in order to suppress seeing-related differences. The SDSS-to-AB correction is not applied; thus, the $u'$ band calibration remains tied to the SDSS photometric system. No color-term correction is applied to the $u'$ calibration. This does not affect our results, since the color selection is based on WWFI colors only.

\defcitealias{GAIAeDR3}{Gaia Collaboration et al.\ 2021} 
Afterward, we subtract extended point-spread function (PSF) models and ghosts from bright stars to improve the flatness of the background. For this, we subtract all stars contained in the GAIA EDR3 catalog ~\mbox{(\citetalias{GAIAeDR3})} brighter than $14\,G\,\mathrm{mag}$. As the ghosts have some substructure at the level of a few percent that is not covered by the model, we additionally decided to mask all regions where the model exceeds 0.5 ADU ($\approx27\,\mathrm{mag\,arcsec^{-2}}$) in the $g'$ and $r'$ bands, and 10 ADU ($\approx24.5\,\mathrm{mag\,arcsec^{-2}}$) in the $u'$ band.  Due to the large dither pattern and the resulting spatial shifts of the ghosts, as well as the number of individual exposures, the resulting masks do not remain in the final stack. As a consequence, the background, especially around the brightest stars, becomes more homogeneous. Therefore, we no longer need to mask large areas around the brightest stars.

Then, we create night-sky flats for each night, scale them to the individual exposures, and subtract them. For details, we refer to \cite{klugeDiss}

\begin{figure*}[t]
    \begin{center}
        \includegraphics[width=\textwidth]{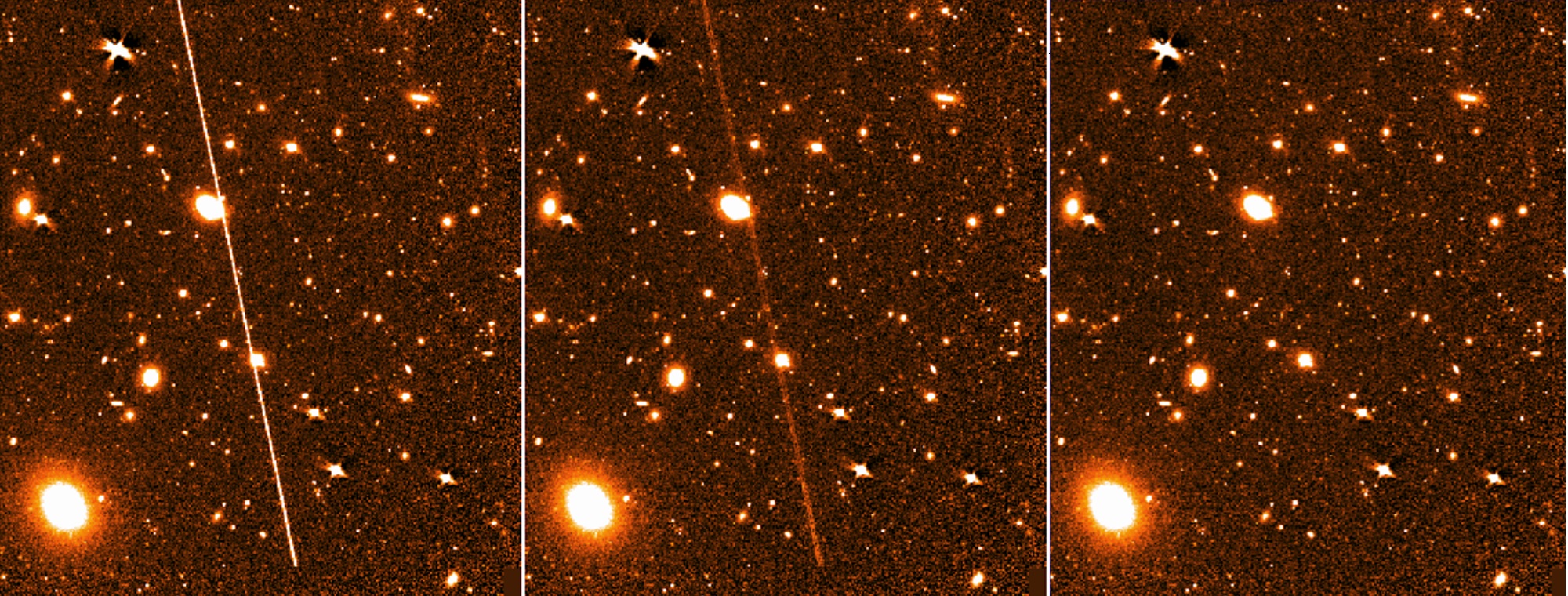} 
        \includegraphics[width=\textwidth]{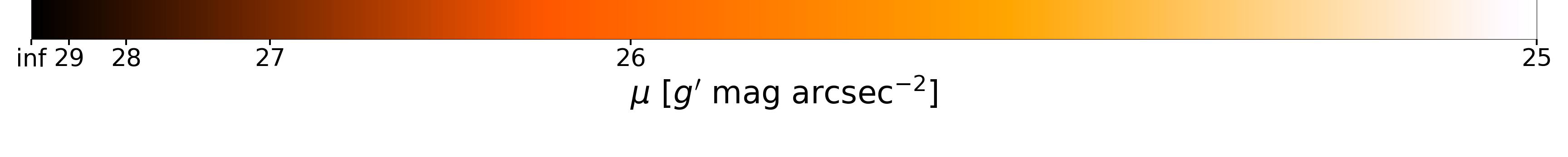}  
    \caption{$10\times10$ times binned cutout of the Coma $g'$ band stack created without sigma clipping (left), with $3-\sigma$-clipping applied (middle), and with our masking method applied (right).} 
    \label{fig:satmasks}
    \end{center}
\end{figure*}
In order to remove artifacts like satellite trails, reflections of stars at the edges of the CCDs, or non-masked charge persistence, we use the robust clipped-mean stacking method implemented in SWarp \citep{swarp,Gruen2014} with $\sigma=3$. However, this method only removes outliers that are $3\sigma$ above the background, and, hence, the outer PSF wings of satellite trails are not masked, and also faint parts of stray-light streaks from stars just outside the CCD footprint are not removed. Due to this issue, we developed an automatic method to remove those low-surface-brightness parts of artifacts. Manual masking of individual exposures, as it was done for \cite{kluge} and \cite{Zoeller2024} is not preferred anymore due to the large amount ($\approx 4000$) of individual images. As a basis for our masking method, we use the clip log file provided by SWarp and create from this log file a clipped-pixel mask (binary; 1=unmasked, 0=masked) for each individual resampled image. Then, we run a $3\times3$ pixel median filter over those images to remove isolated clipped pixels and one-pixel-wide structures before identifying extended artifacts for mask expansion. Afterward, we remove those with fewer than a thousand connected pixels, as satellite trails are considerably larger. %
The remaining masks contain mainly satellite trails, stray-light streaks from stars just outside the CCD footprint, and the centers of the brightest stars and their diffraction spikes. Then, we expand those remaining masks by convolving them with a circular tophat kernel with a diameter of 51 pixels, resulting in a more conservative mask covering also the low surface brightness features of the artifacts. Finally, the expanded mask is combined with the original clip-mask, i.e., the binary clipped-pixel mask before median filtering and size selection. Thus, connected clipped regions larger than 1000 pixels are expanded with the 51-pixel tophat kernel, while smaller clipped regions are retained in their original form. These combined masks are then applied to the resampled images, their weights ($\mathrm{1/BGRMS^2}$) and error images, and then combined using the weighted average combination in SWarp. In Figure \ref{fig:satmasks} we show a $10\times10$ times binned cutout of the Coma $g'$ band stack without sigma clipping (left), with $3-\sigma$-clipping applied (middle), and with our masking method applied (right). This method reliably removes the faint outer parts of bright artifacts. However, this method does not work for faint satellite trails when they do not exceed the clipping threshold. Only a small number of such remaining satellite trails are manually masked in the final stacks after visual inspection of $10\times10$ binned images.

Of all of the data taken, we reject some exposures due to low sky transparency, bad seeing, stray light, light pollution, or significantly varying night-sky patterns.
Specifically, images are rejected if the relative transparency is below 90\% in $g'$ and $r'$ or below 80\% in $u'$, or if the FWHM exceeds 1.3\arcsec, in $g'$ and $r'$ or 1.5\arcsec, in $u'$. Additional images are rejected after visual inspection of the all-sky camera and the sky background if they are affected by cirrus, stray light, light pollution, or strongly varying night-sky patterns.
Additional observations were taken to account for that and to decrease the variation of the depth due to varying transparency and photon noise of the sky. However, we note that perfect homogeneity is not reached due to the limit of available observation time with the required observing conditions. The effect on our results is taken into account by an injection and recovery test (see Section \ref{sec:injrec}).

Finally, all remaining resampled images for each filter are combined using the weighted-average combination in SWarp, with the masks applied through the corresponding weight maps and without additional sigma clipping, and zero-points are determined for the stacks using a larger 10\arcsec\, aperture, as seeing variation can slightly affect the relative matching using $\mathrm{ZP_5}$. We assume seeing independence beyond $\mathrm{ZP_{10}}$, as the seeing-dependent Moffat term \citep{Trujillo2001} of the PSF surface brightness profile is subdominant beyond $r>3.84"$ compared to the other terms that describe the extended PSF wings \citep{klugeDiss}. We report a typical std of the $\mathrm{ZP_{10}}$ for the individual stars of $\approx 0.02-0.03\,\mathrm{mag}$ in the final stack, which also includes the uncertainties in the reference catalogs and variations of the stars. Additionally, new zero-points for the $g'$ band are computed, taking into account the flux lost beyond the 10\arcsec\, aperture following \citet{klugeDiss}:
\begin{equation}
    \mathrm{ZP_{inf}=ZP_{10}+0.1155}\,g'\,\mathrm{mag}.
\end{equation}
Here, we only correct the $g'$ band for this effect, as this correction factor is only available for this filter. Since the $g'$ band serves as our reference measurement band and we use the $u'$- and $r'$-bands solely to derive aperture colors, this correction is not necessary for the latter. Unless otherwise specified, all $g'$-band total magnitudes and surface brightnesses are calibrated using $\mathrm{ZP_{inf}}$. All colors and aperture magnitudes are calibrated using $\mathrm{ZP_{10}}$. Magnitudes calibrated with $\mathrm{ZP_{10}}$ are consistent with those from Pan-STARRS \citep{kluge}.

Additionally, a new astrometric solution is computed for the final stacks using the GAIA EDR3 catalog \citep{GAIAeDR3}.
\begin{deluxetable}{c|cccc}[t]
    \tabletypesize{\small}
    \tablecaption{Exposure Times, Depths, Seeing, and Effective Area}
    \label{tab:exposuretimes}
    \tablehead{
        \colhead{}& \colhead{} & \colhead{Coma} & \colhead{Ref1} & \colhead{Ref2}   
    }
    \startdata
        exp. time  & $u'$ & 4940 & 570 & 570 \\
        (min) & $g'$ & 1812 & 216 & 195 \\
        & $r'$ & 926 & 113 &  99 \\
        \hline median  & $u'$ & 30.0  & 30.0 & 30.0 \\
        $3\sigma$ depth& $g'$ & 29.6 & 30.1 & 29.7 \\
        $(\mathrm{mag\,arcsec^{-2}})$& $r'$ & 28.7 & 28.9  &28.8\\
        \hline median  & $u'$ & 1.15  & 0.93 &  1.10\\
        FWHM& $g'$ & 0.96 & 1.17 &  1.08\\
        $(\mathrm{arcsec})$& $r'$ & 0.95 & 1.03  &1.07\\
        \hline effective area  &  & 1.41  & 0.12 &  0.15\\
        $(\mathrm{deg^2})$&  &  &  &  \\
    \enddata
    \tablecomments{Total exposure time of the images included in the final stacks, median $3\sigma$ depth on a $10\arcsec\times10\arcsec$ scale, and median FWHM of our Coma cluster and the two reference field observations for the individual filter bands. The total area covered by the Coma cluster footprint is $1.49\,\mathrm{deg^2}$}.
\end{deluxetable}   

\begin{figure*}[ht]
    \begin{center}
        \includegraphics[width=\textwidth]{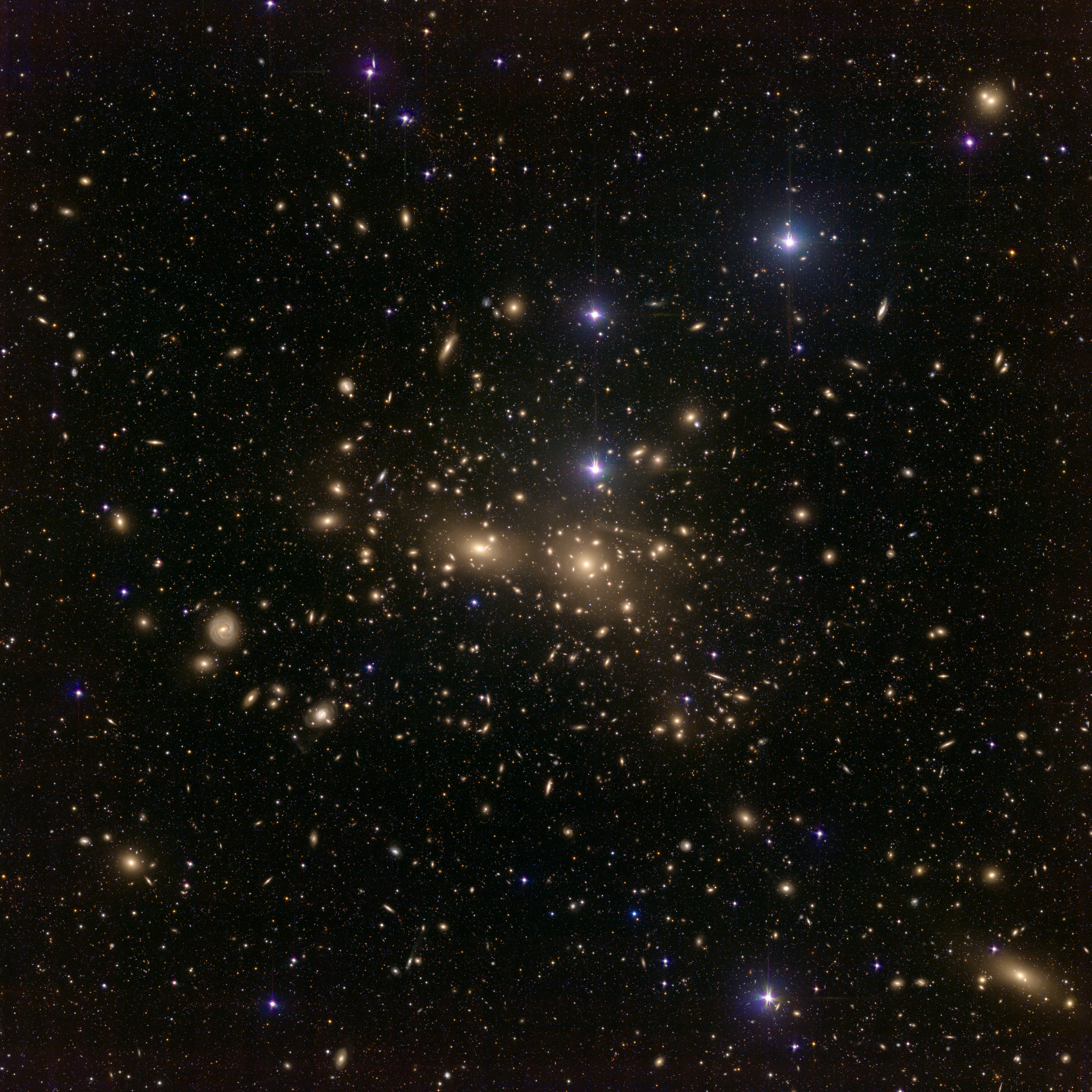}
    \caption{$6\times6$ binned $u'-g'-r'$ color image of the Coma cluster survey region ($\mathrm{1.22\,deg \times 1.22\,deg}$). Note that the bright star models were reinjected into the actual science stacks with one-third of their actual flux for visualization only.}
    \label{fig:Comargb}
    \end{center}
\end{figure*}

For the Coma cluster images, we finally restrict the area to $\mathrm{1.22\,deg \times 1.22\,deg}$, in order to yield a fairly homogeneous depth over the field of view, excluding the low S/N regions in the outskirts. The resulting survey footprint is visualized as a color image in Figure \ref{fig:Comargb}.

\begin{figure*}[ht]
    \begin{center}
        \includegraphics[width=0.9\textwidth]{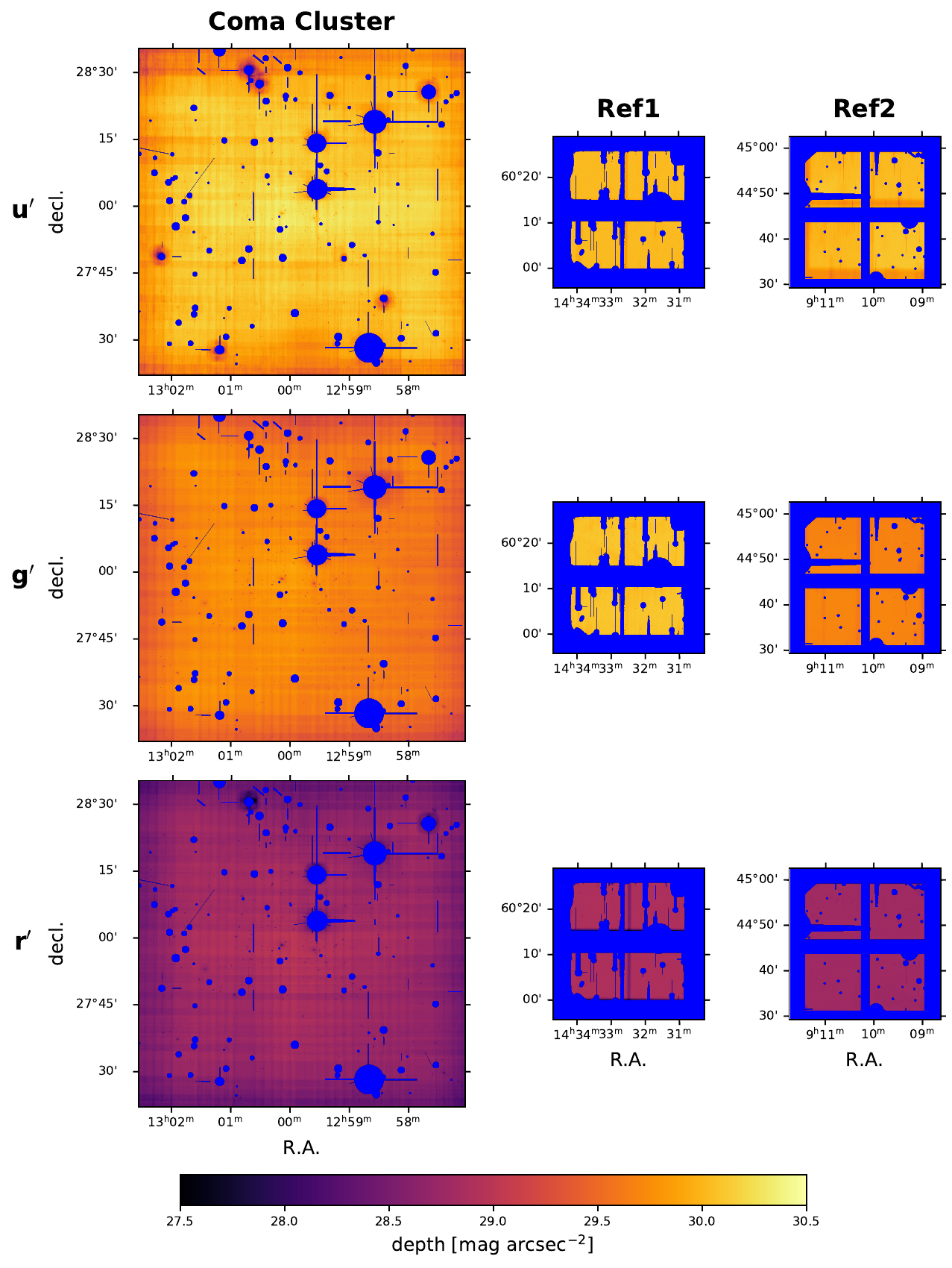}
    \caption{$3\sigma$ depth on a $10\arcsec \times 10 \arcsec$ scale of our Coma cluster data (left), as well as of our two reference fields Ref1 (middle), and Ref2 (right) in the $u'$ (top), $g'$ (middle), and $r'$-band (bottom).}
    \label{fig:depths}
    \end{center}
\end{figure*}

The total exposure times and the median $3\sigma$ surface brightness depths, measured on a $10\arcsec \times 10 \arcsec$ scale, are listed in Table \ref{tab:exposuretimes} for the Coma cluster stack and the reference fields across all filters. 
This depth corresponds to the surface brightness at which the integrated flux over a $10\arcsec\times10\arcsec$ area has $S/N=3$.
The depths are calculated following \citet{Roman2020}:
\begin{equation}
    \mu_{\mathrm{lim}}(3\sigma;10\arcsec\times10\arcsec)=-2.5\mathrm{log\left(\frac{3\sigma}{pxs\times10}\right)+ZP},
\end{equation}
where "pxs" is the pixel size in arcseconds, and $\sigma$ is obtained from the corresponding error image rather than from the background rms of the coadded science image. The error images contain the statistical uncertainties propagated from the individual exposures through the reduction and are resampled to the science-image grid and combined in quadrature using the corresponding stacking weights, including masked regions (see also \citealt{klugeDiss}). The quoted depths therefore trace the propagated statistical uncertainties and do not include additional uncertainties arising from residual large-scale background variations. Since these uncertainties are propagated from the individual exposures through the resampling and coaddition, rather than estimated from the pixel-to-pixel rms of the final coadded images, the depth estimate does not depend on the resampling-modified single-pixel scatter. In Appendix ~\ref{app:correlatednoise}, we explicitly test the propagated error model by comparing it with the measured standard deviation of N×N box means in the final coadded image.


For the $g'$ band we use $\mathrm{ZP_{inf}}$ and for the $u'$ and $r'$ band we use $\mathrm{ZP_{10}}$. The depth maps for the Coma cluster and reference fields in the $u'$, $g'$, and $r'$ bands are visualized in Figure \ref{fig:depths}, illustrating the achieved homogeneity. For all fields, the final survey footprint is restricted to the common unmasked area in all filters to ensure consistent multi-band photometry.

\section{Data Processing Pipeline} \label{sec:dataprocessingpipeline}
\begin{figure}
    \begin{center}
        \includegraphics[width=0.47\textwidth]{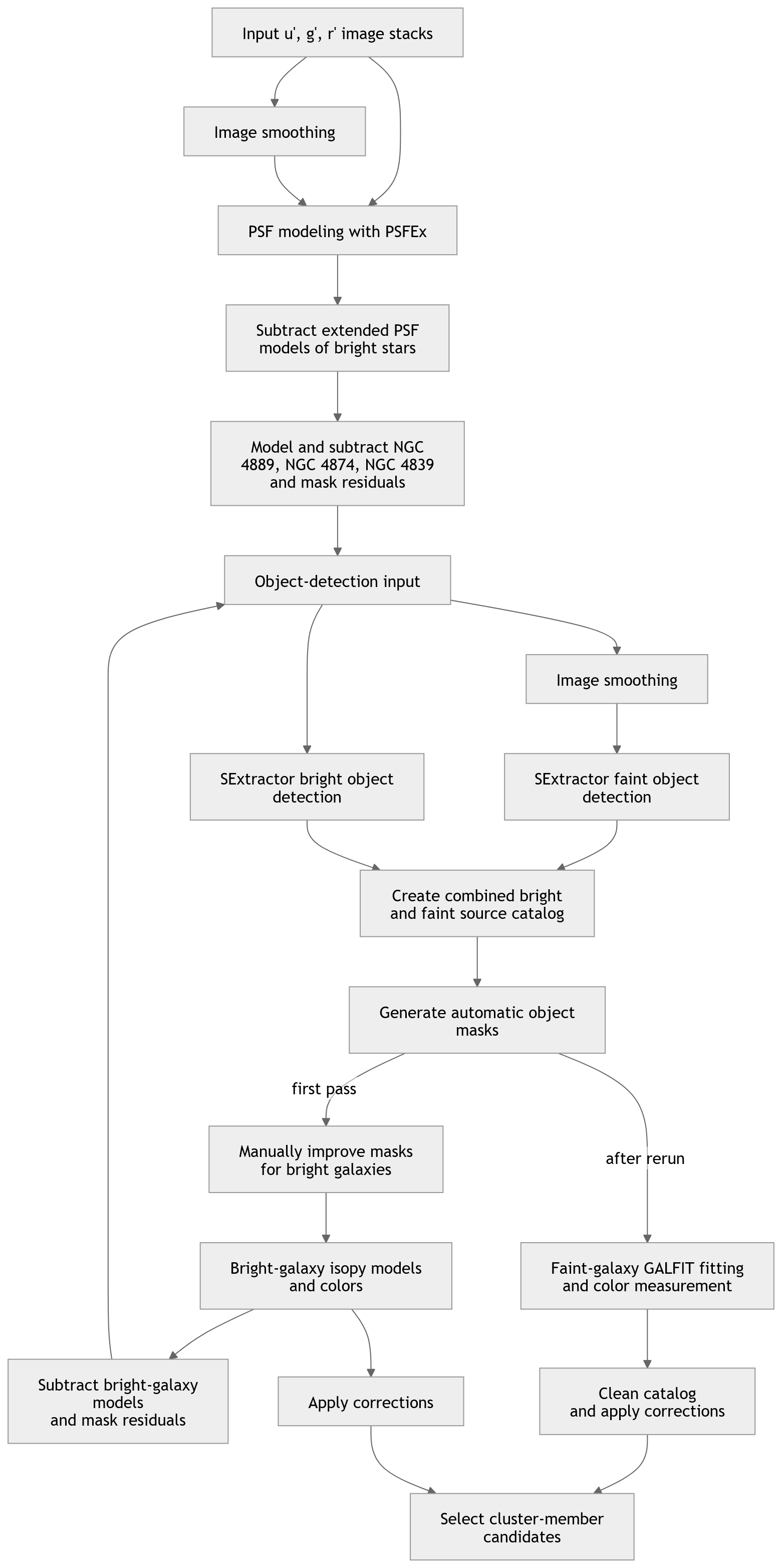}
    \caption{Simplified flow chart of the data-processing pipeline used to detect, model, and select Coma cluster galaxy candidates, including PSF modeling, bright-source subtraction, source detection, masking, structural modeling, catalog cleaning, and color-based member selection.}
    \label{fig:flowchart}
    \end{center}
\end{figure}
In order to automatically identify and separate Coma cluster member candidates from diffuse background galaxies and measure their structural parameters, we use the pipeline presented in \citet{Zoeller2024} with a few improvements. In the following, we will give a brief overview of the pipeline, which is summarized in simplified form in Figure \ref{fig:flowchart}; for details, we refer to \cite{Zoeller2024}. The basic pipeline includes accurate measurements of the PSF over the whole field of view using \verb+PSFEx+ \citep{PSFEx}.

\begin{figure*}[ht]
    \begin{center}
        \includegraphics[width=\textwidth]{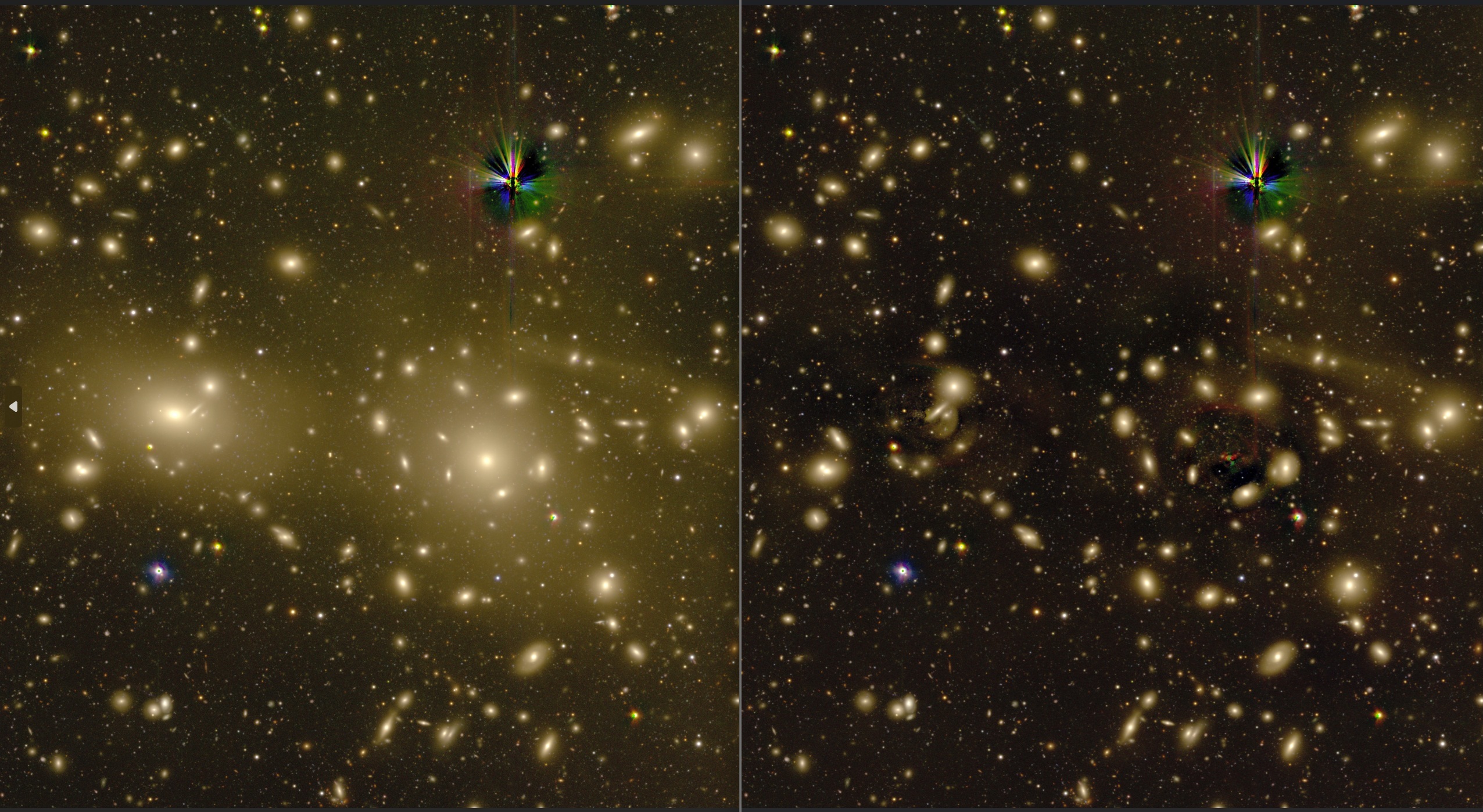} 
    \caption{$u'-g'-r'$ image $\mathrm{(0.28\,deg \times 0.31\,deg)}$ of the central region of the Coma cluster before subtracting NGC\,4874 (including the ICL) and NGC\,4889 (left) and after (right).}
    \label{fig:BCGsub}
    \end{center}
\end{figure*}

Furthermore, we improve the flattening of the background by subtracting bright stars ($m \lesssim 16\,g'$\,mag) using an extended PSF model, as well as models of the brightest three cluster galaxies NGC\,4874 (including the ICL), NGC\,4889, and NGC\,4839.

To derive models for these galaxies, we use the isophote fitting tool \verb+isopy+ \citep{klugeisophotespy,Kluge2023rhea}, which is based on the python package \verb+photutils+ \citep{photutils}. Here, the ellipticity, position angle, and center of the isophotes can vary.  
Beyond the largest radius for which the ellipse parameters can be robustly fitted, the ellipticity and position angle are held fixed, while the isophotes are sampled at successively larger semi-major-axis radii. The $a_4/a$ parameter is still fitted for all sampled isophotes, but beyond the fixing radius it is not considered reliable, since it can reflect deviations from the adopted fixed ellipse geometry rather than intrinsic boxiness or diskiness.
Model images are created by assigning the median measured flux value along these ellipses.
For NGC\,4889 and NGC\,4874, we use masks from \cite{Zoeller2024}, and for NGC\,4839, we create new masks using the same masking method to obtain masks for small and medium-sized objects and manually improve the masks.
NGC\,4889 and NGC\,4874 are modeled iteratively, similar to the approach in \cite{Zoeller2024} and \cite{KlugeERO}. For that, we first apply the masks from \citet{Zoeller2024}, manually mask the fainter NGC\,4889, and then create a first model of NGC\,4874. This model is subtracted from the original stack, and the residuals are masked. Using the resulting image, the model of NGC\,4889 is created and subtracted from the original stack, and the residuals are masked. Afterward, we create a second iteration model of NGC\,4874. Then, the procedure is repeated to create a second version of the NGC\,4889 model and a third iteration of the NGC\,4874 model, which are also the final models. For NGC\,4839, no iteration is needed. The models of NGC\,4889, NGC\,4874, and NGC\,4839 are subtracted from the star-subtracted image stack. 
Figure \ref{fig:BCGsub} shows color images of the center of the Coma cluster before (left) and after (right) the subtraction. The image demonstrates how the removal of NGC\,4889 and NGC\,4874 reduces contamination, enabling more accurate detection and measurement of the underlying galaxies. Residuals are manually masked.

\subsection{Object Detection}
\label{sec:detection}
Source catalogs are created using \verb+SExtractor+ \citep{Sextractor}, detecting objects in the $g'$ band images. The catalogs contain first estimates of the structural parameters and positions of the objects. They are in the following used to preselect cluster member candidates and as initial parameters for a more precise measurement using \verb+GALFIT+ \citep{galfit}. We create two object catalogs, one for large and bright sources and one for relatively small and faint sources, and combine them afterward; only the combined catalog is used in the following. The \verb+SExtractor+ parameters for faint objects were optimized to reliably detect UDGs in the Coma cluster while avoiding obvious false detections. In this run, we employ smoothed images (convolved with a Gaussian kernel with $\sigma=2\,\mathrm{px=0.4\,arcsec}$) to reduce false detections of noise peaks while simultaneously increasing the number of detected UDGs. The parameters for large objects were tuned for large elliptical galaxies. For an in-depth discussion of the creation of object catalogs, we refer to \cite{Zoeller2024}. Compared to \cite{Zoeller2024}, the Coma data has a more homogeneous depth, which allows us to slightly tweak the detection parameters of the \verb+SExtractor+ call for small objects. \verb+DETECT_THRESH+ was increased from $27.2\,g'\,\mathrm{mag \,arcsec^{-2}}$ to $27.6\,g'\,\mathrm{mag \,arcsec^{-2}}$, and the \verb+BACK_SIZE+ was increased from 32\,px to 64\,px.

\subsection{Masking}
\label{sec:masking}

To create individual object masks, we follow the masking procedure presented in \cite{Zoeller2024}. Here, a set of multiple masks covering different scales of objects is combined while always unmasking the central object. These individual masks were created using two methods. The first method is to strongly smooth the images and afterward mask connected pixels above a certain local threshold if their area exceeds the detection area. The second method involves running \verb+SExtractor+ with various detection parameters and background subtractions, and then utilizing the segmentation maps as masks. For details and the parameters used, we refer to \cite{Zoeller2024}.
\begin{figure*}[ht]
    \centering
    \includegraphics[width=\textwidth]{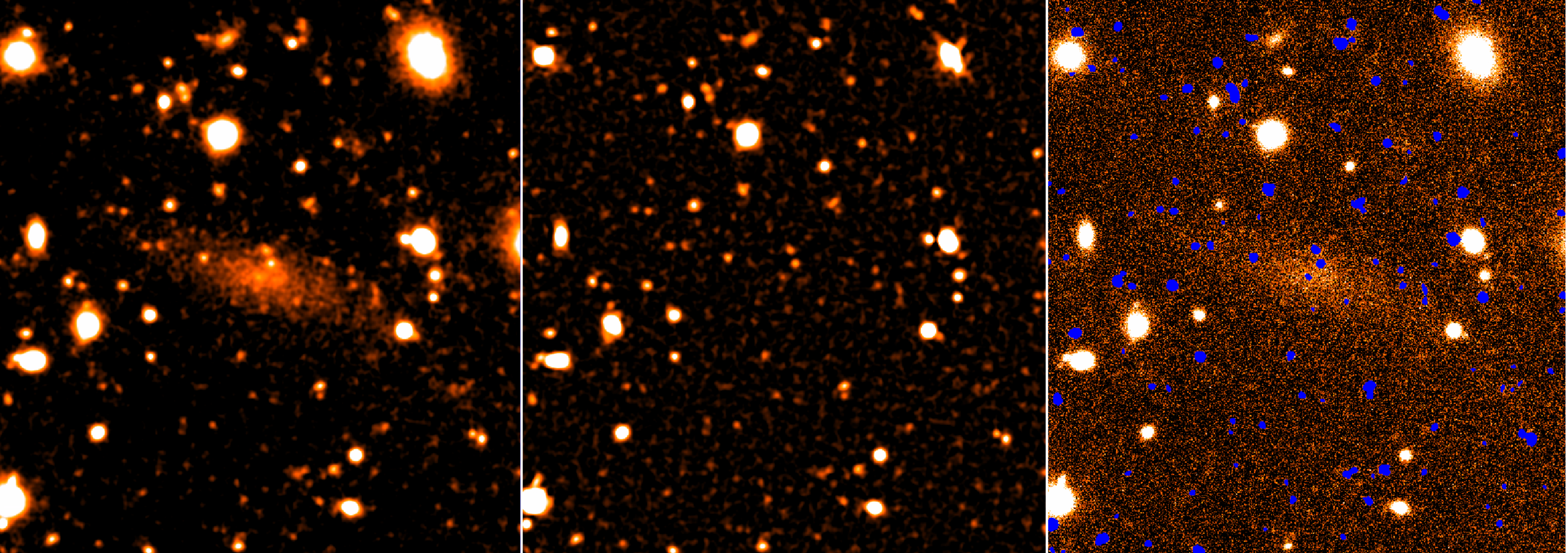}
    \includegraphics[width=\textwidth]{scale0to4g.png}
    \caption{Smoothed $g'$ band image cutout $(\mathrm{2.1\arcmin\times2.1\arcmin})$ before (left panel) and after (middle panel) the aggressive background subtraction. The created masks (blue) for faint point sources are shown in the right panel, applied to the non-smoothed image.}
    \label{fig:tinymasks}
\end{figure*}

In addition to those masks, we create a mask focused on faint point sources. This was necessary, as non-masked faint point sources were the main reason for manual masking in \cite{Zoeller2024}. Automatically masking them is especially problematic when they overlap with extended sources. Even if they are detected as individual objects, their segmentation maps can contain too much of the stellar body of a dwarf galaxy, and hence result in unnecessarily large masks. Therefore, we create another \verb+SExtractor+ segmentation map for the $g'$ band with an aggressive background subtraction using  \verb+BACK_SIZE=8+, \verb+BACK_FILTERSIZE=3+, \verb+DETECT_MINAREA=16+, and a detection threshold of $27.2\,g'\,\mathrm{mag \,arcsec^{-2}}$ . This \verb+SExtractor+ run uses the smoothed images to mitigate false detections and to create more conservative masks. This background subtraction efficiently removes the outskirts of galaxies or even a whole diffuse galaxy such as DF15 \citep{vanDokkum2015} as shown in Figure \ref{fig:tinymasks} (middle panel), allowing the detection and masking of faint point sources. We remove all masks containing more than 200 pixels, as this background subtraction also leads to artifacts around the centers of bright galaxies and does not provide proper masks for them, and because we only want to create a mask for faint point sources. The resulting mask applied to a non-smoothed cutout around DF15 is shown in blue on the right panel of Figure \ref{fig:tinymasks}. These masks are expanded by a convolution with a tophat kernel of a diameter of 11 pixels during the creation of the individual object masks, after unmasking the central object in order to obtain more conservative masks. For bright galaxies, this saves significant time for manual masking, but the main benefit is for the fits of faint objects, as opposed to \cite{Zoeller2024} we rely on full automation of the masking and fitting due to the enormous amount of fitted objects (see Section \ref{sec:udgmeasurements}).

\subsection{Modeling and Subtraction of Bright Galaxies} \label{sec:EllS0}
We aim to measure the luminosity function and structural parameter number densities of a large dynamical range of galaxy brightnesses in the Coma cluster, ranging from the BCG to the dwarf galaxy regime. However, these galaxies show very different morphologies. Most dwarf galaxies in galaxy clusters have a simple morphology that can be well modeled with a single \citet{sersic} model, whereas brighter galaxies can be more complicated. S0 galaxies consist of a bulge and a disk component, and elliptical galaxies frequently show isophote shifts and twists, have a varying ellipticity, differ from perfect elliptical isophotes (i.e., disky or boxy isophotes), or require multiple Sérsic components.
Hence, we chose to use two different ways to model the galaxies. For dwarf galaxies, we use single Sérsic GALFIT \citep{galfit} fits, and for brighter galaxies such as ellipticals, S0s, and spirals, we choose to use the more versatile isophote fitting approach using \verb+isopy+ \citep{klugeisophotespy,Kluge2023rhea}. This approach also leads to fewer residuals than just fitting a simple single or double Sérsic profile. As automatically detecting dwarf galaxies that overlap with brighter galaxies is challenging, we also use these accurate isophote models to subtract the bright galaxies from the images to improve the detectability and the structural parameter measurements of such galaxies.

All objects with $m\leq18\,g'$\,mag and a star-galaxy classifier (S/G) lower than 0.4 are considered bright galaxies (403 objects) and modeled with \verb+isopy+. For all objects, the masks are manually checked and improved. 
\begin{figure*}[ht]
    \centering
    \includegraphics[width=\textwidth]{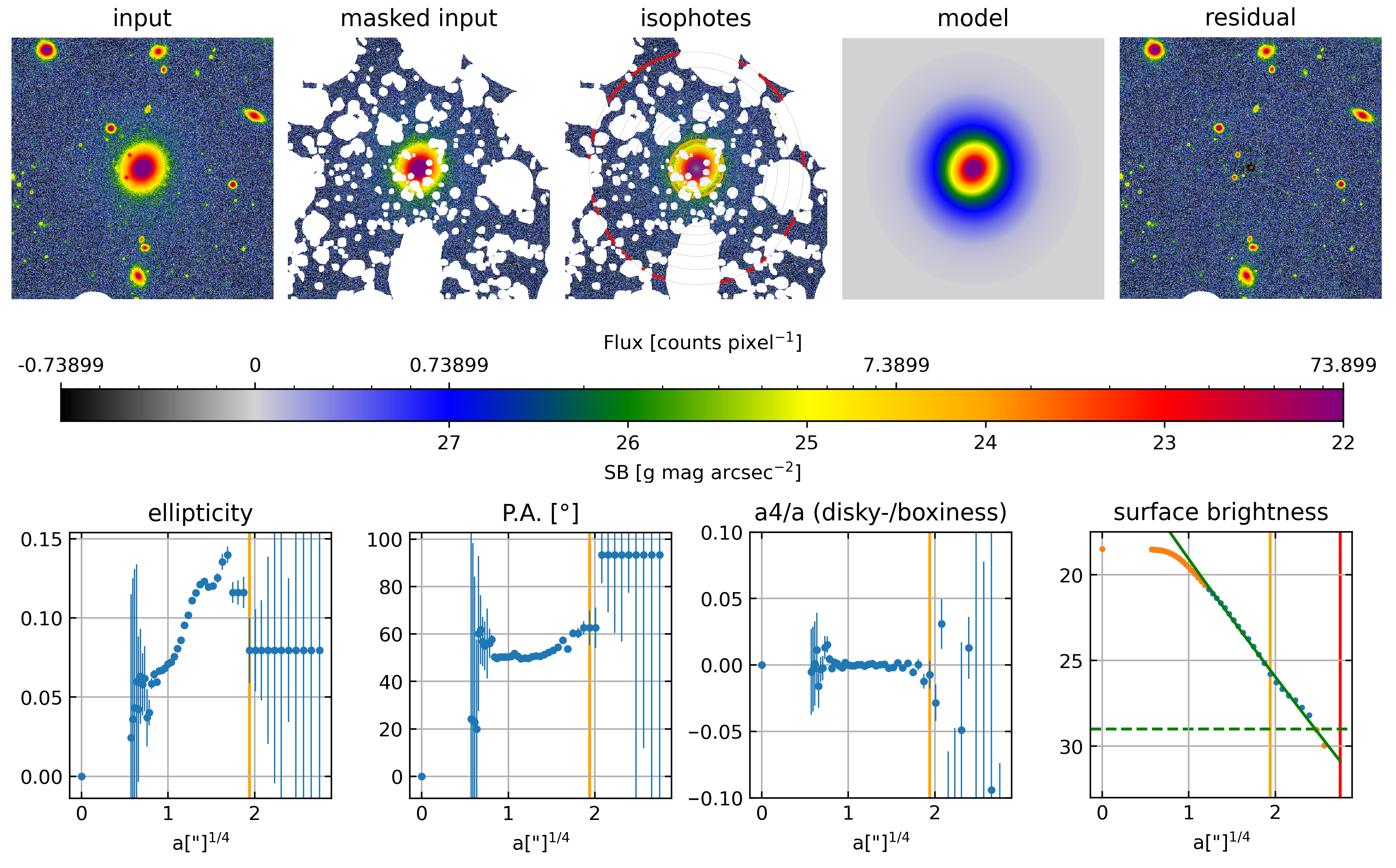}
    \caption{Selected outputs of the fitting routine for LEDA\,44562 in the $g'$ band with a cutout size of $\mathrm{2.1\arcmin\times2.1\arcmin}$. The ellipticity, position angle, and central coordinates are automatically fixed beyond their respective fixed radii; the orange marked isophote indicates the radius beyond which the ellipticity is fixed. In this case, the P.A. is fixed at a lower isophote than the ellipticity. The background is determined at the red-marked position. In the surface brightness profile plot, the blue dots are the data points used for the fit, the orange dots are excluded from the fit, the solid green line corresponds to the best-fit Sérsic profile (with $n=4.4\pm1.2$), and the dashed green line indicates the threshold from which the Sérsic profile is used as the model. The Sérsic model assumes perfect ellipses, i.e., $a_4/a=0$.}
    \label{fig:ELLexample}
\end{figure*}
Using \verb+isopy+, we measure the isophotes in the $g'$ band, while allowing the central coordinates of the isophotes, ellipticity, position angle, and disky-/boxyness ($a_4/a$) to vary. The surface brightness (SB) of each isophote is updated for the $u'$ and $r'$ bands while keeping the isophotal shapes fixed to those measured in the $g'$ band. The SB of each isophote is determined using the median flux, but in the case of an apparent spiral structure or star formation regions, the mean flux is used to avoid biasing the measurement. To the measured SB profiles, we fit a single Sérsic function to the outer regions, avoiding the seeing-dominated center ($a<2\,\mathrm{arcsec}$). In the case where multiple components would be required to fit the profile, we manually restrict the fit to the surface brightness range of the outer component, as this is only required to extrapolate the outer flux, as in \cite{klugeisophotespy}. 
The measured SB profile is replaced by the Sérsic profile at $29\,\mathrm{mag\,arcsec^{-2}}$ because of the strong scatter at low surface brightnesses. If necessary, the transition point was adjusted manually to a nearby surface brightness where both profiles are consistent.
The structural parameters are then directly derived from these models. To derive the colors, we use the total combined galaxy models in the respective filters. The model fluxes are integrated to infinity and converted to magnitudes, from which the colors are computed. The uncertainties are determined by adding in quadrature the Poisson uncertainty and the absolute difference between the total model flux integrated to infinity and the flux enclosed within the $g'$-band $30\,\mathrm{mag\,arcsec^{-2}}$ isophote. An example \verb+isopy+ output is shown in Figure \ref{fig:ELLexample}. For some galaxies, we manually set the blending threshold to a brighter SB, to avoid a scatter in the model due to the uncertainties in the flux measurements of individual isophotes or to avoid an impact due to variations in the background (here also the fit ranges are adjusted accordingly). The latter is especially relevant for galaxies close to the centers of the brightest cluster galaxies that were subtracted before, as those galaxies can be affected by residuals of the subtraction. Furthermore, 10 pairs of galaxies were modeled iteratively due to their strong overlap. Three authors (RZ, MK, and JNP) unanimously agreed that for each of the 403 bright galaxies, the models are as good as possible. None is discarded.

Finally, all models are subtracted from the images, and residuals are manually masked. After that, the object detection is run again on the bright-galaxy-subtracted images, and new automatic masks are created using the same parameters as before. This yields less conservative and more accurate masks, especially for objects that overlapped with the subtracted bright galaxies.

\subsection{Modeling of Faint Galaxies} \label{sec:udgmeasurements}
\label{sec:dwarfmoddelling}
We preselect faint galaxy candidates with an apparent magnitude between 17\,$g'$\,mag and 27.2\,$g'$\,mag and a mean surface brightness within the effective radius between $15\,g'\,\mathrm{mag\,arcsec^{-2}}$ and $29.4\,g'\,\mathrm{mag\,arcsec^{-2}}$. The faint limits are set to mitigate false detections. Further false detections are mitigated by removing objects with $\verb+FLAGS+\geq4$ or $\verb+PETRO_RADIUS+=0$.  Furthermore, we reject clear point sources from our sample with $\verb+S/G+>0.9$ or $R_e<2\,\mathrm{px}=0.4\,\mathrm{arcsec}$, where $R_e$ is the directly integrated half-light radius (FLUX\_RADIUS for PHOT\_FLUXFRAC =0.5), which is not corrected for the PSF. Hence, our galaxy sample is limited to resolved galaxies and does not include globular clusters, nor ultra-compact dwarfs.
\begin{figure*}[ht]
    \centering
    \includegraphics[width=\textwidth]{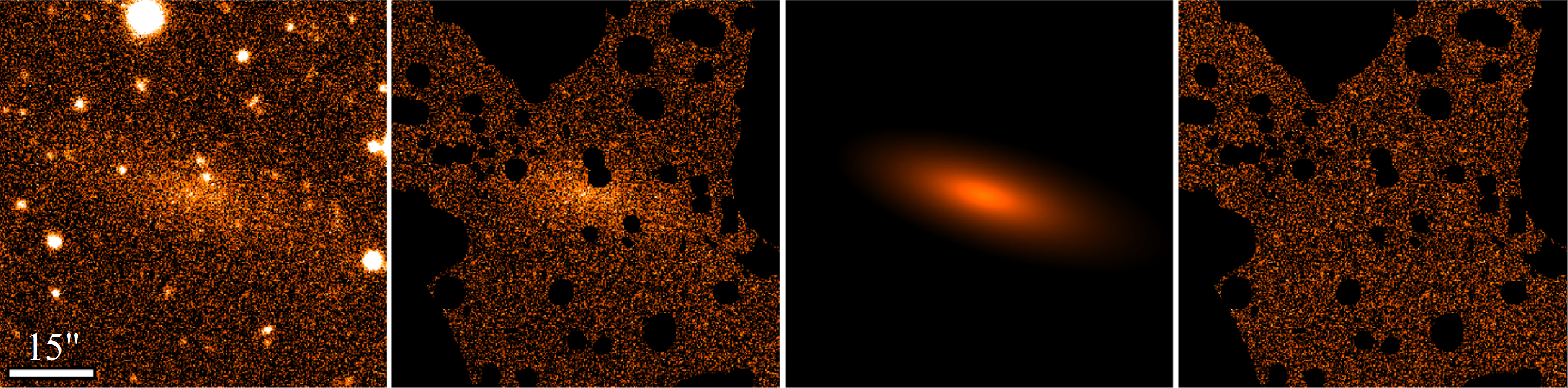}
    \includegraphics[width=\textwidth]{scale0to4g.png}
    \caption{Unmasked cutout around the UDG DF15, the automatically masked image, the best-fit model, and the residual image (from left to right)}
    \label{fig:dwarfexample}
\end{figure*}

To model the fainter galaxies, we fit single Sérsic models to all galaxy candidates. Unlike in \citet{Zoeller2024}, we do not perform a manual check and remasking of the fits. The main reason is to avoid an observer bias, i.e., unintentionally creating better manual masks for the galaxies in the Coma image than for the reference images or manually rejecting fewer fit results, which could potentially lead to an incorrect completeness correction. Furthermore, \cite{Zoeller2024} aimed for a clean sample, whereas we aim for a representative sample and statistically correct for contamination and completeness. 

The masking and fitting procedure is adopted from \cite{Zoeller2024} with some modifications. 
The cut-outs around the object of interest have side lengths of $12\,R_e$ (directly integrated) with a minimum of 101\,px and a maximum of 351\,px. 
Briefly summarized, for each individual mask, the central object and all overlapping objects are first unmasked. The resulting masks are then combined.
Afterward, all unmasked objects contained in the \verb+SExtractor+ catalog are modeled simultaneously using \verb+GALFIT+ either with a Sérsic or a PSF model.
For Sérsic fits, the values of $\mu_e$ and $\mu_0$ are obtained from additional \verb+GALFIT+ runs using the corresponding surface-brightness parameterizations, following \cite{Zoeller2024}; only if these runs fail does the pipeline fall back to an analytic conversion using $\mu_0=\mu_e-2.5\log_{10}(e)\,b_n$, where $b_n$ is defined by the Sérsic profile $I(r)=I_e\exp\{-b_n[(r/r_e)^{1/n}-1]\}$ such that $r_e$ encloses half of the total luminosity.
The colors of the objects are measured using elliptical apertures after subtracting the local background and convolving the cutouts in the different filter bands to the same target PSF, i.e., a Moffat profile with the worst local PSF FWHM of all filter bands. Here, we used as parameters of the elliptical aperture the \verb+GALFIT+ ellipse parameters, adopting $R_e$ as the semi-major axis, and required a minimum aperture area of 100\,px for sufficient S/N and a maximum semi-major axis of 15\,px to limit unmasked contamination. For all filters, we used the identical masks. Color uncertainties were computed by quadratically adding the \verb+photutils+ aperture-magnitude uncertainties.
In Figure \ref{fig:dwarfexample}, we show the extreme example of the UDG DF15 \citep{vanDokkum2015} which exhibits a very low surface brightness, extremely large extent and a lot of contamination that needs to be masked. Here, we show from left to right the unmasked cutout around DF15, the automatically masked image, the best-fit \verb+GALFIT+ model with $R_e=(21.1\pm1.3)\,\mathrm{arcsec}=(9.6\pm0.6)\,\mathrm{kpc}$, $m=(19.51\pm0.06)\,g'\,\mathrm{mag}$, $M=(-15.49\pm0.06)\,g'\,\mathrm{mag}$, $n=1.10\pm0.06$, $\mu_e=(27.54\pm 0.08)\,g'\,\mathrm{mag\,arcsec^{-2}}$, $\mu_0=(25.49\pm0.05)\,g'\,\mathrm{mag\,arcsec^{-2}}$, and $b/a=0.334\pm0.007$, as well as the residual image.

As we do not perform any additional manual remasking, we improved the automatic masking compared to \cite{Zoeller2024}, focusing on the main issues that required manual remasking. The first improvement is an additional mask for faint point sources as described in Section \ref{sec:masking}. Furthermore, we increased the mask sizes by convolving them with a tophat kernel with a diameter of 7\,px after the unmasking process. An issue that remains is objects that were not deblended by \verb+SExtractor+. This affects the galaxies that we inject for the injection and recovery test (Section \ref{sec:injrec}) equally, and therefore is statistically corrected. However, for individual objects, the results of \cite{Zoeller2024} can be more accurate.

The resulting catalogs are processed analogous to \cite{Zoeller2024}, rejecting objects with large uncertainties ($\Delta(u'-g')>0.2\,\mathrm{mag}$, $\Delta(g'-r')>0.2\,\mathrm{mag}$, $\Delta m > 1\,g'\,\mathrm{mag}$, $\mathrm{\Delta \mu_0} > 1\,g'\,\mathrm{mag\,arcsec^{-2}}$, $\mathrm{\Delta \mu_e} > 1\,g'\,\mathrm{mag\,arcsec^{-2}}$, or $\Delta R_e / R_e>0.5$). Furthermore, we correct for galactic extinction \citep{Schlafly2011} and, assuming that objects are at the distance of the Coma cluster, we apply K-correction \citep{Chilingarian2010,Kcorrection2012}, and calculate absolute magnitudes using the distance modulus, correct for cosmological dimming, and convert the $R_e$ values from arcsec to kpc using the values obtained from \cite{Scolnic2025}, see Section \ref{sec:intro}. 
K-corrections were computed at the fixed heliocentric redshift of the Coma cluster using the individual galaxy colors, with $g'-r'$ used for the $g'$ and $r'$ bands and $u'-r'$ for the $u'$ band. 
We adopt the mean Coma redshift so that the K-correction is appropriate for the cluster on average, rather than for each galaxy individually; the Coma redshift dispersion \citep[$\sigma_z=0.003$;][]{Scolnic2025} changes the K-corrections by only about $0.005\,\mathrm{mag}$ for $g'-r'=0.6$. At the low redshift of Coma, these corrections are small, and even for the maximum allowed color uncertainty of $0.2\,\mathrm{mag}$ in our sample, the induced uncertainty in the K-corrections is $\lesssim0.01\,\mathrm{mag}$ for typical dwarf colors; we therefore do not include it as a separate contribution to the color uncertainties.\\
We also apply the same corrections to the catalog of the bright galaxies, but without performing the automatic catalog cleaning, as we manually ensured the goodness of those fits.  
For more details, we refer to \cite{Zoeller2024}.

\subsection{Cluster Member Candidate Selection}
\label{sec:selection}
\begin{figure*}[t]
    \begin{center}
        \includegraphics[width=\textwidth]{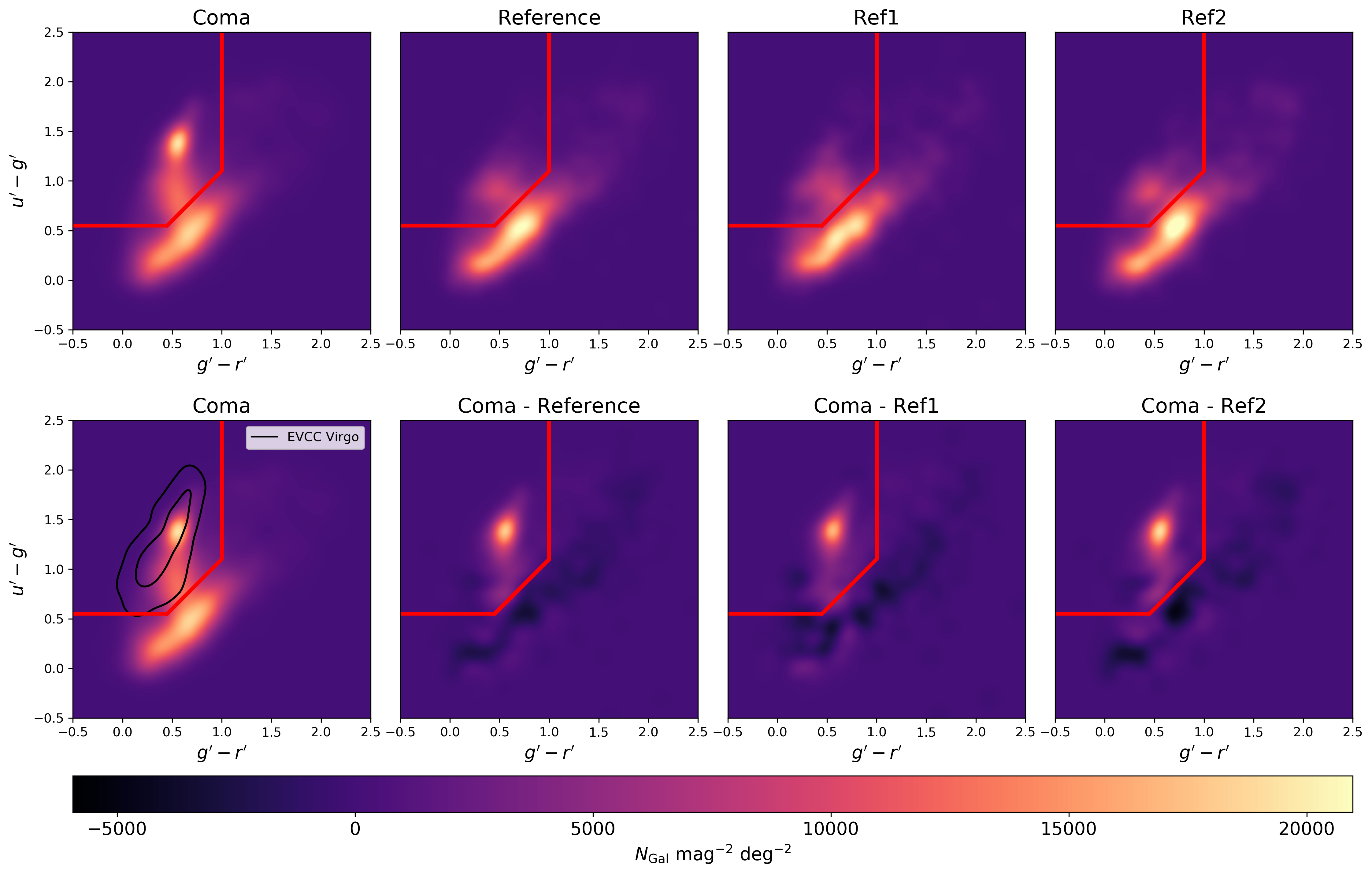}
        \caption{$u'-g'$ versus $g'-r'$ number density color-color diagrams for the faint Coma cluster and comparison samples.
        The top row shows the density distributions for Coma, the combined reference fields, and the individual reference fields Ref1 and Ref2, all without completeness correction. The bottom row shows Coma with EVCC Virgo contours overlaid, followed by residual maps obtained by subtracting each reference sample from the Coma distribution. Galaxies in the top left of the red visualized selection cutoffs are considered to be quenched. The samples are limited to galaxies with $m<24\,g'\,\mathrm{mag}$.}
    \label{fig:colorcolor}
    \end{center}
\end{figure*}
First, we preselect quiescent galaxies using a $u'-g'$ versus $g'-r'$ color--color selection.  Here, we use the effect that at the redshift of Coma, the 4000\,$\mathrm{\AA}$ break in quenched galaxies lies between the $u'$ and $g'$ bands, shifting them to redder $u'-g'$ colors and separating them from the star-forming sequence separates the star-forming galaxies and the quenched galaxies in color--color space \citep[for galaxy-population separation in color--color space, see e.g.,][]{Strateva2001,Williams,McIntosh}. 
We adjust the selection cutoffs from \citet{Zoeller2024} by shifting them by 0.2\,mag in $u'-g'$ to include also transitioning galaxies from the star-forming sequence to the quenched sequence and only exclude clear star-forming galaxies. The selection criteria are given by:
\begin{align}
    u'-g' &> g'-r + 0.1 \label{eq:diag}\\
    u'-g' &> 0.55\\
    g'-r' &< 1
\end{align}

In Figure \ref{fig:colorcolor}, we show the selection cuts as red lines overlaid on the galaxy number density in the color--color diagram. Galaxies in the upper-left region of the diagram are classified as quenched. The top row shows the distributions for the Coma cluster, the combined reference fields, and the individual reference fields. The bottom row shows the Coma distribution overlaid with contours from EVCC galaxies classified as possible Virgo cluster members based on their spectroscopic velocities \citep{Kim2014}, followed by residual maps formed by subtracting the reference-sample distributions from the Coma distribution. The EVCC contours occupy a similar region of color–color space as the adopted Coma quiescent selection, supporting the interpretation that this part of the diagram is populated by cluster galaxies with quenched or transitioning stellar populations. 
Despite the shallower apparent-magnitude limit of the EVCC, which reaches down to about $19\,g\,\mathrm{mag}$, the comparison still extends to about $-12.1\,g\,\mathrm{mag}$ at the Virgo distance modulus of 31.1 \citep{Kim2014}, only $\sim0.8$ mag brighter than our adopted faint limit of $-11.3\,g'\,\mathrm{mag}$ (see Section \ref{sec:LFfunction}). Moreover, since Virgo is less massive than Coma \citep{Kubo2007,Simionescu2017} and hosts a more prominent blue, star-forming population \citep{BoselliGavazzi2006}, the similar location of its galaxies in color--color space further supports that our selection is not overly restrictive.
In addition, the residual maps show that the positive Coma excess relative to the reference fields lies primarily above and to the left of the selection boundaries, while the reference-field populations are concentrated mainly along the bluer star-forming locus. \\
The residual maps also contain regions with negative values, where the reference-field number density exceeds that of the Coma sample. These negative residuals are typically at the level of $\sim10–20\%$ of the Coma-field galaxy number density in the corresponding region of color--color space, and we do not interpret them as a physical deficit of Coma galaxies. They likely arise from differences in the completeness functions of the Coma and reference-field samples, field-to-field variations due to cosmic variance, and Poisson noise. However, these localized negative residuals do not affect the main conclusion from the residual maps: the adopted selection encloses the color--color region where Coma shows a coherent excess relative to the reference fields. \\
The adopted selection therefore identifies the region dominated by quiescent and transitioning cluster galaxies, while effectively removing star-forming field interlopers and hence, the vast majority of contaminating sources from the sample. Based on the rejected fraction in the reference fields, we estimate that the color--color selection removes $\approx80\%$ of the contamination (see Section \ref{sec:referencefield}).\\
The sample of Coma cluster member candidates fulfilling the quiescent selection criterion consists of NGC 4889, NGC 4874, and NGC 4839, 403 bright galaxies (all bright galaxies fulfill the criterion) and 6335 faint galaxies.

\subsection{Contamination: Reference Fields} \label{sec:referencefield}
To probe the contamination of our Coma cluster galaxy sample, we run the same pipeline on the reference fields. Here, we only adjust the galactic extinction correction and use exactly the same color--color selection as for the Coma cluster. Furthermore, we assume that all found objects would be at the distance of the Coma cluster, regardless of their true distance, as also done for the contamination contained in the Coma sample.
For Ref1, the final faint galaxy sample contains 304 galaxies, and for Ref2, it contains 332 galaxies. Furthermore, we find three bright galaxies in Ref1 and eight in Ref2. We note that the selection of quiescent galaxies based on the color--color diagram removed 77\% of the objects in Ref1 and 81\% in Ref2, hence we conclude that this selection cut removes about 80\% of the contamination from the sample. 

\subsection{Red Sequence}
\label{sec:redsequence}
\begin{figure*}[t]
    \begin{center}
        \includegraphics[width=0.9\textwidth]{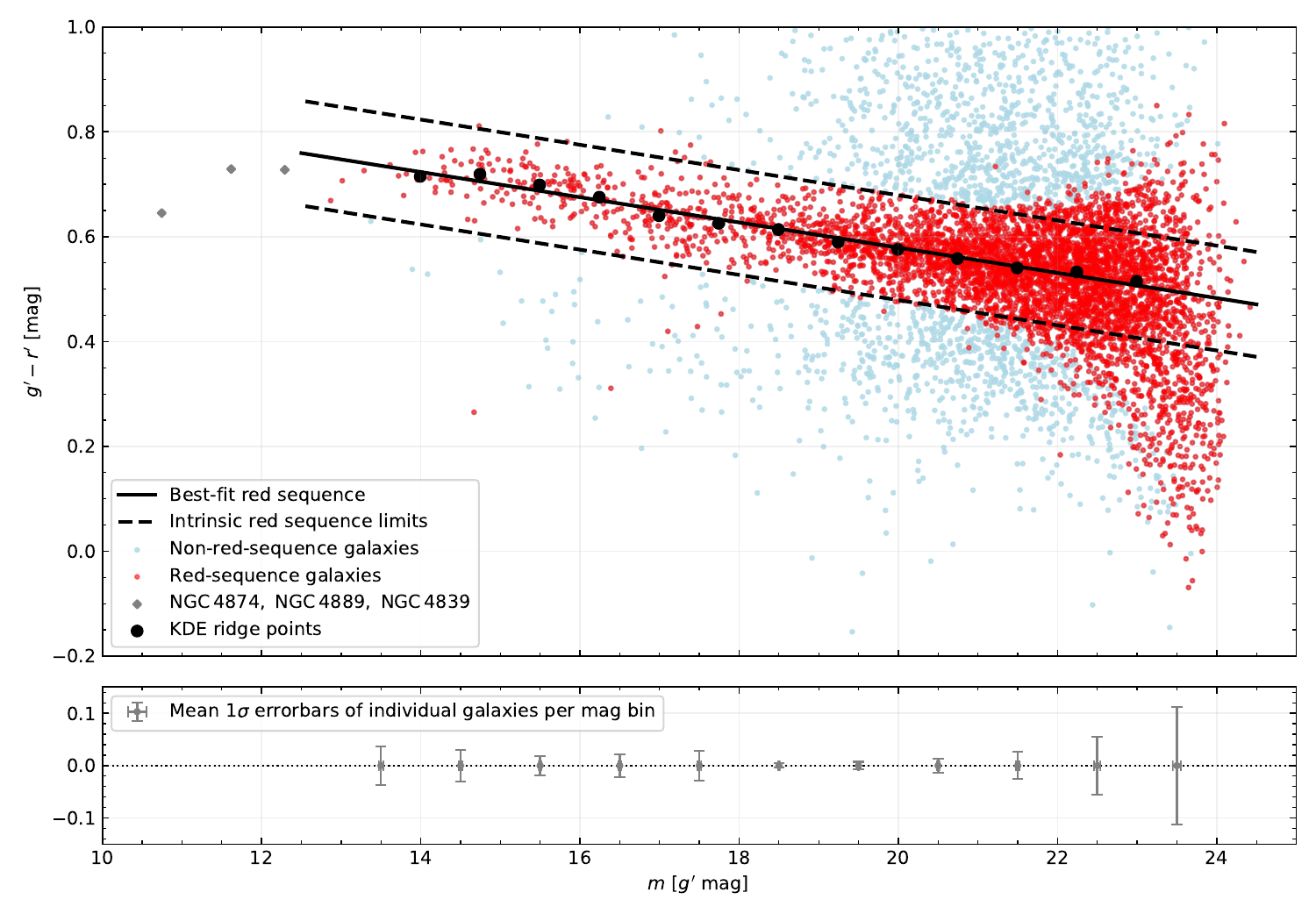} 
    \caption{$g'-r'$  color-magnitude diagram of the Coma cluster. The KDE ridge points are depicted as black dots, the best-fit red sequence as a solid black line, and the limits of the intrinsic red sequence as dashed black lines. Galaxies that deviate less than $3\times \Delta (g'-r')$ from these limits are accounted as red sequence members and are visualized in red, whereas non-red sequence members are depicted in blue. The three brightest galaxies, NGC 4874, NGC 4889, and NGC 4839, are shown separately as grey diamonds.
    In the bottom panel, we visualize the mean errorbars of the individual galaxies within a magnitude bin. Note here that the colors of the faint galaxy sample were measured using aperture magnitudes and hence their uncertainties only include photon noise and the uncertainty of the background subtraction, whereas the colors of the bright sample ($m<18\,g'\,\mathrm{mag}$) are based on the SB profile modeling of the whole galaxy and hence also includes the modeling uncertainties causing larger errorbars.}
    \label{fig:redseq}
    \end{center}
\end{figure*} 
Red-sequence selections are frequently used to identify cluster member candidates \citep[e.g.,][]{GladdersYee2000,Koester2007,Rykoff2014,Kluge2024}. We therefore fit the $g'-r'$ red sequence of the Coma cluster as a diagnostic reference and to assess how a red-sequence-based member selection would behave. However, we do not use this red-sequence selection for the final cluster-member candidate sample, because doing so would introduce a significant bias of the GLF as discussed in the following.
 
To determine the red sequence, we determine the ridge of a Gaussian Kernel Density Estimate (KDE)  in the $g'-r'$ vs. $m_{g'}$ space. For this, we calculate the KDE maxima in bins of ${0.75\,g'\,\mathrm{mag}}$. Here, we use all bright galaxies (except the three brightest ones) and the preselected quiescent galaxies with $m_{g'}<23.2\,g'\,\mathrm{mag}$, as well as only bins containing at least 10 objects. Using linear regression, we fit the red sequence to those maxima of the magnitude bins, giving a red sequence with:
\begin{equation}
    g'-r'=(-0.024\pm0.001) \,m_{g'} + (1.06\pm0.02).
    \label{eq:redsequence}
\end{equation}
To determine the red sequence members, we assume an intrinsic red sequence width of $3\sigma\approx 0.1\,\mathrm{mag}$ \citep[][priv. comm.]{Kluge2024}. Objects that deviate by less than $3\times \Delta (g'-r')$ from the intrinsic red sequence are considered red sequence members.  
The red sequence selection is visualized in Figure \ref{fig:redseq}, with the maxima of the KDE bins displayed as black dots, the best-fit red sequence as the solid black line, and the limits of the $3\sigma$ intrinsic red sequence as dashed black lines. Galaxies considered to be red sequence members 
are depicted in red, and those that are removed in the red sequence selection are shown in blue. Note that individual bright galaxies can have relatively large uncertainties and hence are still accounted to the red sequence. 
\begin{figure*}[t]
    \centering
    \includegraphics[width=\linewidth]{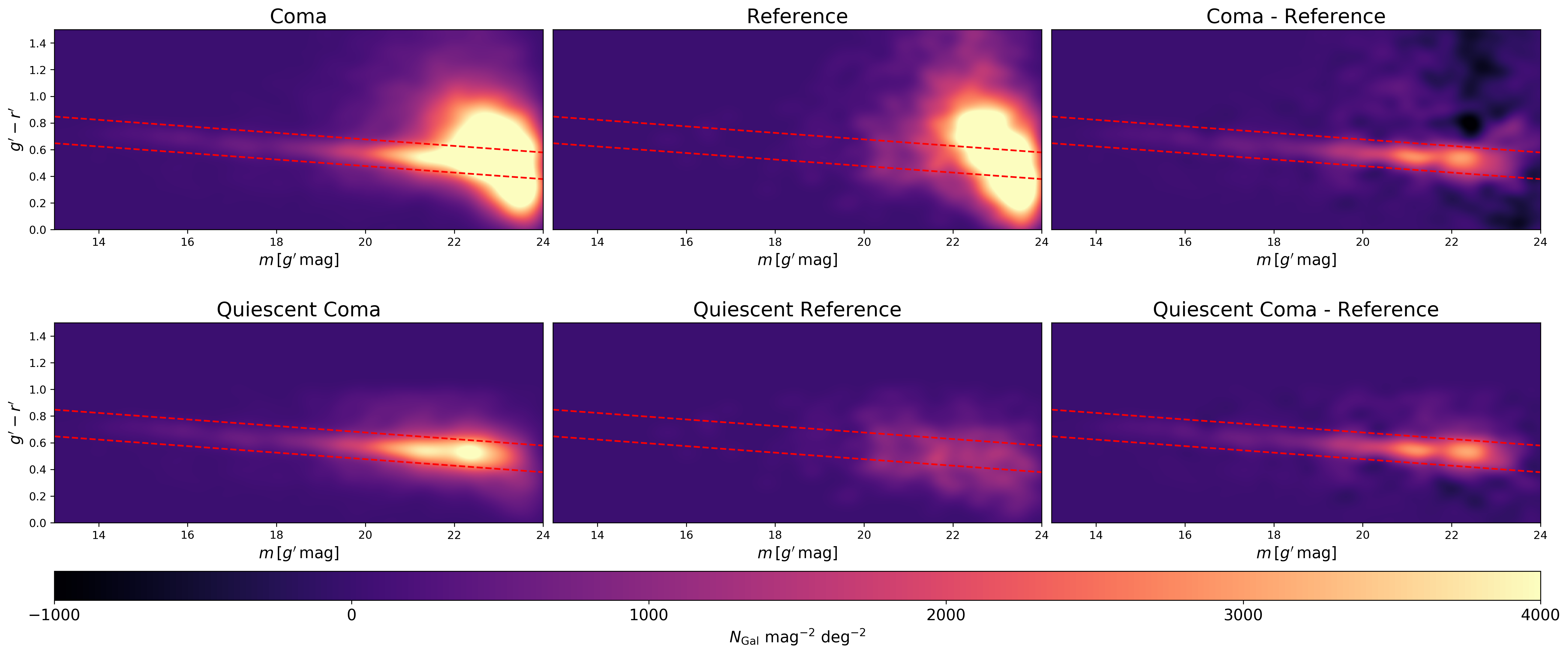}
    \caption{$g'-r'$ versus $m_{g'}$ number-density color-magnitude diagrams for the Coma cluster and reference fields. The top row shows the full galaxy samples for Coma, the combined reference fields, and the residual distribution obtained by subtracting the reference field distribution from the Coma distribution. The bottom row shows the corresponding distributions after applying the quiescent color–color preselection. The red dashed lines indicate the adopted $3\sigma$ intrinsic width around the fitted Coma red sequence.}
    \label{fig:rskde}
\end{figure*}
\begin{figure}[ht]
    \centering
    \includegraphics[width=\linewidth]{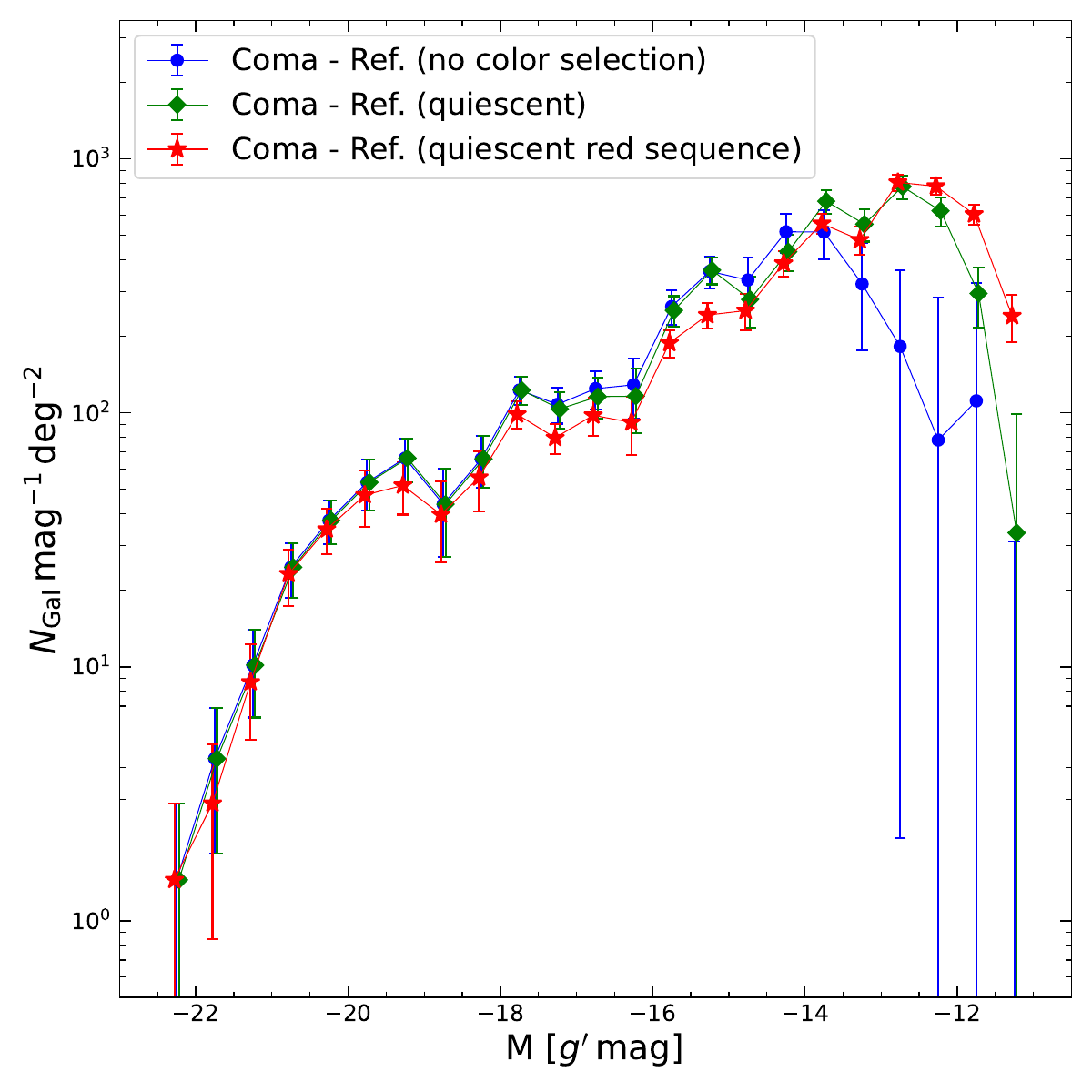}
    \caption{Background-subtracted galaxy number density counts for the Coma cluster field after subtracting the reference-field contribution without applying a completeness correction. The blue points show the luminosity function without any color selection, the green diamonds show the sample after applying the quiescent color–color selection, and the red stars show the quiescent sample further restricted to galaxies within the adopted red-sequence selection. Absolute magnitudes are expressed on a Coma-equivalent scale, used as a common reference scale rather than a physical distance assumption.}
    \label{fig:selectionbias}
\end{figure}

The effect of the color selections is illustrated in Figure \ref{fig:rskde}.  
In the top row, which shows the full sample before applying any color selection, both the Coma and combined reference-field color–magnitude distributions contain a broad galaxy population, including many objects outside the red-sequence region. After subtracting the reference fields, however, the Coma excess is already concentrated around the fitted red sequence. This indicates that the dominant cluster-associated population follows a well-defined red sequence in $g'-r'$ versus $m_{g'}$ space. The bottom row shows the same comparison after applying the $u'-g'$ versus $g'-r'$ quiescent preselection. After applying the quiescent color--color preselection, the reference field density is reduced, while the remaining Coma excess forms a coherent structure along the fitted red sequence and lies largely within the adopted intrinsic $3\sigma$ width. The residual signal within the red-sequence region is not significantly changed compared to the full sample, indicating that the color--color preselection removes mainly contaminating field galaxies while preserving the dominant Coma population. This demonstrates that the color--color preselection efficiently removes the majority of contaminating field galaxies while retaining the main Coma cluster population. The close correspondence between the quiescent Coma residual and the red sequence further supports the interpretation that the vast majority of the selected Coma cluster population is quiescent or transitioning and lies on or near the red sequence.

Figure \ref{fig:selectionbias} further illustrates the effect of the different selection choices on the background-subtracted galaxy luminosity function. We note that the GLFs shown here are based on raw counts and do not include a completeness correction; the error bars include only Poisson uncertainties. Without any color preselection, the luminosity function is difficult to constrain at faint magnitudes ($M \gtrsim -14\,g'\,\mathrm{mag}$) because the subtraction of the reference-field population introduces large Poisson uncertainties, as seen from the large error bars of the blue data points. At these magnitudes, the Poisson noise is dominated by the much more numerous background objects rather than by the Coma galaxies themselves. In this regime, the full-sample luminosity function also lies systematically below the quiescent-preselected luminosity function, suggesting that the full-sample subtraction becomes unstable when the reference-field contribution is large. 

For our dataset, this makes a purely statistical subtraction of the full photometric sample insufficient for a robust measurement of the faint-end GLF. The color--color preselection reduces the contaminating population by about 80\%, thereby substantially reducing both the Poisson noise and the sensitivity to uncertainties in the background estimate. This remains beneficial even when substantially larger reference fields are available.
At the same time, the fact that the quiescent-preselected GLF is not systematically lower than the full-sample GLF argues against the color–color selection being overly restrictive. 
Applying the $u'-g'$ versus $g'-r'$ quiescent preselection strongly reduces this uncertainty while preserving the overall shape and normalization of the Coma luminosity function over the magnitude range where the full-sample measurement is reliable.

Using the red-sequence selection would further reduce the contamination by about 50--60\%. However, this gain in purity comes at the cost of systematically reducing the recovered Coma galaxy population. The additional red-sequence selection produces a luminosity function that is systematically lower than both the full sample and the quiescent-preselected sample. This indicates that a red-sequence selection would not simply remove contaminants, but would also exclude genuine Coma galaxies that do not fulfill the red-sequence selection criterion, including objects in the blue cloud, transitioning systems, and possibly faint red-sequence galaxies due to a potentially larger intrinsic width at the faint end. Using the red-sequence selection for the final GLF would therefore bias the inferred luminosity function by artificially lowering the number density and potentially altering its shape. For this reason, we do not impose the red-sequence cut in the final cluster-member selection, but use the broader quiescent color–color preselection.

\subsection{Completeness Function} \label{sec:injrec}

Correcting for the completeness of the sample is crucial for determining the GLF. For the bright galaxy sample, we assume 100\% completeness in the unmasked areas, as such bright objects are easily detected by \verb+SExtractor+, and in the modeling process (Sec. \ref{sec:EllS0}),  we successfully measured their structural parameters for the entire bright sample.
To derive the completeness function of our faint galaxy sample, we perform an injection-recovery test. For this, we randomly draw the structural parameters from our final Coma faint galaxy sample undoing the corrections of the catalogue processing (corrections for galactic extinction, color term of the zero-point, K-correction, and cosmic dimming), generate single Sérsic mock galaxies, convolve them with the local PSF, add the photon noise of the galaxy, and inject them onto a grid with a step size of 500\,px into the science frame. Here, we do not add the photon noise of the sky, as it is already in the science stack. For the colors of the injected galaxies, we use the $g'-r'$ color predicted from the best-fit red sequence and a $u'-g'=1.4$, corresponding to the maximum density of the Coma cluster faint galaxy population in the $u'-g'$ versus $g'-r'$ color-color space.\\
Then, we run the pipeline again, starting at the \verb+SExtractor+ run for faint objects. This includes all subsequent parts of the pipeline, e.g., creation of masks, running \verb+GALFIT+, rejection of bad fits, large color uncertainties, and the color selection itself.

We count all galaxies that we find within 1\,arcsec that are retained in the final catalog as being recovered. We also classify as unrecovered any galaxies whose recovered magnitude is more than $0.75\,g'\mathrm{mag}$ brighter than the injected value (i.e., a factor of $\approx2$ in total flux), as these are likely mismatches with another object.

This procedure is repeated twice, injecting different random sets of galaxies on grids shifted by 250\,px in x-direction and in x- and y-direction. Additionally, we perform one run injecting only galaxies with $M<-14\,g'\,\mathrm{mag}$ to improve the sampling at the bright end.

The first injection sample is used to estimate the completeness as a function of magnitude $C(M)$. However, because this injection sample is drawn from the observed galaxy population, it inherits the selection biases of the observed catalog. We therefore construct an independent 2D completeness model, $C_{\rm 2D}(M,\mu_e)$, from injections that uniformly sample the $M-\mu_e$ plane. This model is then used to reweight the original injection sample before fitting the final 1D completeness function.

\begin{figure*}[t]
    \begin{center}
        \includegraphics[width=\textwidth]{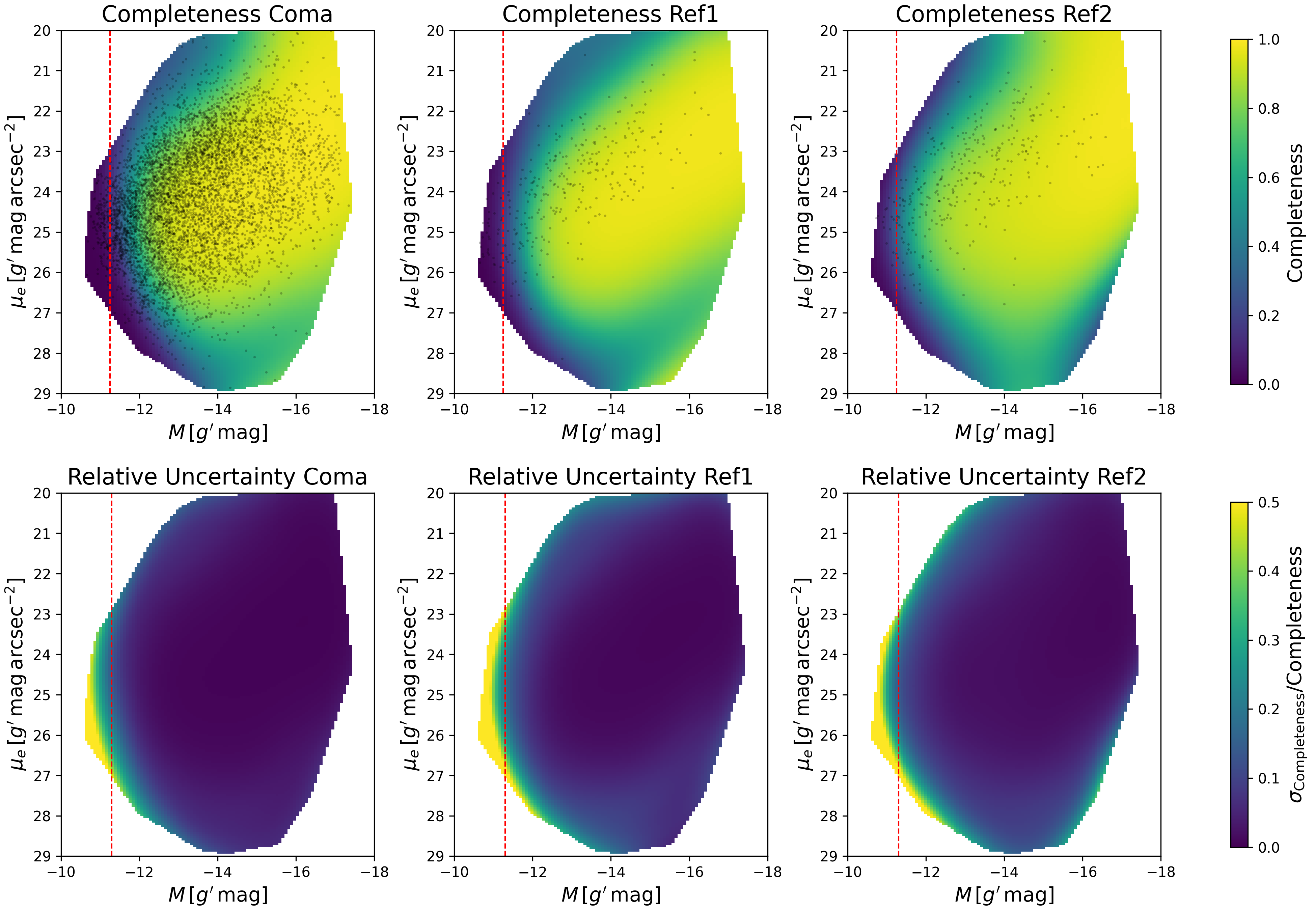}
    \caption{2D Completeness function ($C_{\rm 2D}(M,\mu_e)$) for the Coma field and the reference fields (top) and its relative uncertainty (bottom). 
    The dashed red line indicates $M=-11.3\,g'\,\mathrm{mag}$, the limiting magnitude adopted for the GLF analysis. At this magnitude, the corresponding 1D completeness of the Coma cluster is approximately 20\%. Overlayed in black are the individual quiescent galaxies, detected in the individual fields. Absolute magnitudes are expressed on a Coma-equivalent scale, used as a common reference scale rather than a physical distance assumption.}
    \label{fig:2DCompletenessA1656}
    \end{center}
\end{figure*}

To derive $C_{\rm 2D}(M,\mu_e)$, we run additional sets of injection-recovery tests. Here, we draw galaxies from the quiescent Coma cluster sample, uniformly sampling the $M-\mu_e$ parameter space. For this, we perform two runs for the Coma cluster and eight runs for each reference field. We model $C_{\rm 2D}(M,\mu_e)$ using a 2D 4th order polynomial, where we clip the completeness polynomial to values between $10^{-9}$ and $1-10^{-9}$ to avoid undefined values in the log-likelihood computation.

To derive the best-fit parameters of the 2D polynomial, we derive the Bayesian posterior distribution using the dynamic nested sampling package \verb+DYNESTY+ \citep{Speagle2020}. We use the unbinned data from the  injection-recovery test that evenly sampled the $M-\mu_e$ plane and model the recovery of each injected galaxy as a Bernoulli process,
\begin{equation}
    \mathcal{L}_i =
    p_i^{k_i}
    \left(1-p_i\right)^{1-k_i},
\end{equation}
where $k_i=1$ means that the object is recovered, $k_i=0$ means that it is not recovered, compute a log-likelihood for each galaxy magnitude -- surface brightness pair ($M_i,\mu_{e,i}$), and derive the joint log-likelihood:

\begin{align}
    \ln(\mathcal{L}) = \sum_i &k_i\ln\{ C_{2D}(M_i,\mu_{e,i})\} \nonumber \\
    &+ (1-k_i)\ln\{1-C_{2D}(M_i,\mu_{e,i})\}.
\end{align}

The resulting 2D completeness maps for the Coma cluster field and both reference fields individually are shown in Figure~\ref{fig:2DCompletenessA1656}. The top row shows the median completeness, $C_{\rm 2D}(M,\mu_e)$, while the bottom row shows the corresponding relative uncertainty, $\sigma_C/C$. The maps are only evaluated within the region of parameter space covered by the source injections. The black points indicate the observed faint quiescent galaxies found in the respective fields. In all three fields, the completeness is highest for bright galaxies with intermediate $\mu_e$. It decreases both toward fainter magnitudes and toward the low-surface-brightness regime. 
At low $\mu_e$, the completeness additionally decreases toward brighter magnitudes. This may be related to the structural properties of these objects, since at fixed $\mu_e$, a brighter total magnitude generally implies a larger $R_e$ and/or more extended profiles depending on $n$, increasing the probability of overlap with other sources and making masking and simultaneous profile fitting more challenging.
A drop in completeness is also visible toward high surface brightnesses, presumably because they can be misclassified as point sources or attributed to more extended sources.

We then apply this two-dimensional completeness correction $C_{\rm 2D}(M,\mu_e)$ to the original injection catalog. For each posterior draw $s$ of the $C_{\rm 2D}(M,\mu_e)$ model, we assign
\begin{equation}
    \tilde{w}_{s,i} =
    \frac{1}{C_{\rm 2D}^{(s)}(M_i,\mu_{e,i})} .
\end{equation}

To avoid changing the relative importance of different magnitude bins due to faint galaxies' lower completeness, we normalize the weights separately in each bin; we denote these normalized weights by $w_{s,i}$.

We model the final one-dimensional completeness function $C(M)$ using a 3rd order polynomial, where we clip the completeness polynomial to values between $10^{-9}$ and $1-10^{-9}$ to avoid undefined values in the log-likelihood computation. Note that the completeness for galaxies fainter than the clip-off magnitude is not exactly zero, since a few are detected in the science image. However, as the completeness is very low in this regime and the completeness correction is not trustworthy, we decide to limit our sample to brighter magnitudes than this clip-off magnitude.

For a given realization of the 2D completeness correction, the log-likelihood is the weighted sum of the Bernoulli log-likelihoods of all injected galaxies,
\begin{align}
    \ln(\mathcal{L}_s)
    =\sum_i &w_{s,i} \left[k_i \ln\{C(M_i)\} \right. \nonumber \\
    &\left.
    + (1-k_i)\ln\left\{1-C(M_i)\right\}\right] .
\end{align}

To propagate the uncertainty of the 2D completeness correction into the polynomial fit, we draw several posterior realizations of the 2D completeness model before running the nested sampler. Each realization gives a different set of weights for the injected galaxies. For each set of polynomial parameters sampled by \verb+DYNESTY+, we compute the weighted log-likelihood for every one of these weight realizations and marginalize over the uncertainty in the 2D completeness correction by averaging the corresponding likelihoods \citep{Trotta2008}.
\begin{equation}
    \mathcal{L}
    =
    \frac{1}{S}
    \sum_{s=1}^{S}
    \mathcal{L}_s .
    \label{eq:sumloglike}
\end{equation}
Equivalently, the log-likelihood evaluated during nested sampling is
\begin{equation}
    \ln(\mathcal{L})
    =
    \ln
    \left[
    \frac{1}{S}
    \sum_{s=1}^{S}
    \exp\left\{
    \ln(\mathcal{L}_s)
    \right\}
    \right] .
    \label{eq:log-mean-exp}
\end{equation}

In practice, this is done using a numerically stable log-mean-exp calculation. The weight realizations are kept fixed during the nested-sampling run, so that the likelihood remains deterministic for a given set of polynomial parameters. In this way, the posterior uncertainty of the 1D completeness function includes both the injection-recovery statistics and the uncertainty of the 2D completeness correction.

\begin{figure}[ht]
    \begin{center}
        \includegraphics[width=0.47\textwidth]{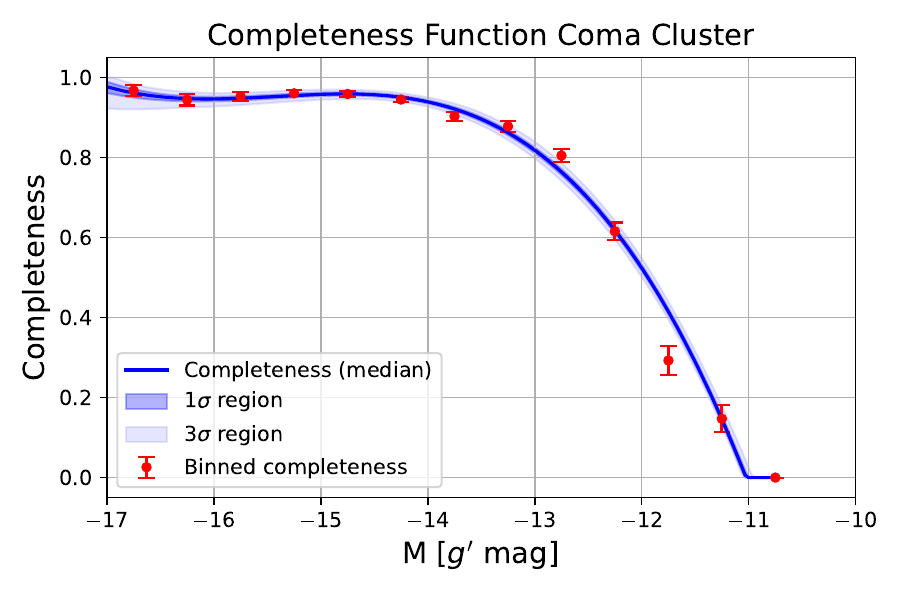}
    \caption{Completeness function for the Coma field. The red points show the weighted binned injection-recovery results, while the blue curve and shaded regions show the median and uncertainty of the unbinned nested-sampling fit. Absolute magnitudes are expressed on a Coma-equivalent scale, used as a common reference scale rather than a physical distance assumption.}
    \label{fig:CompletenessA1656}
    \end{center}
\end{figure}

To inform the priors for the nested sampling, we use the results from a least-squares fit to the weighted binned data points with a bin width of $0.5\,\mathrm{mag}$. The binned points are computed as the weighted recovered fraction in each bin, using weights from the maximum-likelihood 2D completeness model and normalized to have a mean of unity within each magnitude bin. Their error bars combine the corresponding weighted-binomial uncertainty with the scatter obtained by recomputing the binned completeness for posterior draws of the 2D completeness model.

\begin{figure}[t]
    \begin{center}
        \includegraphics[width=0.47\textwidth]{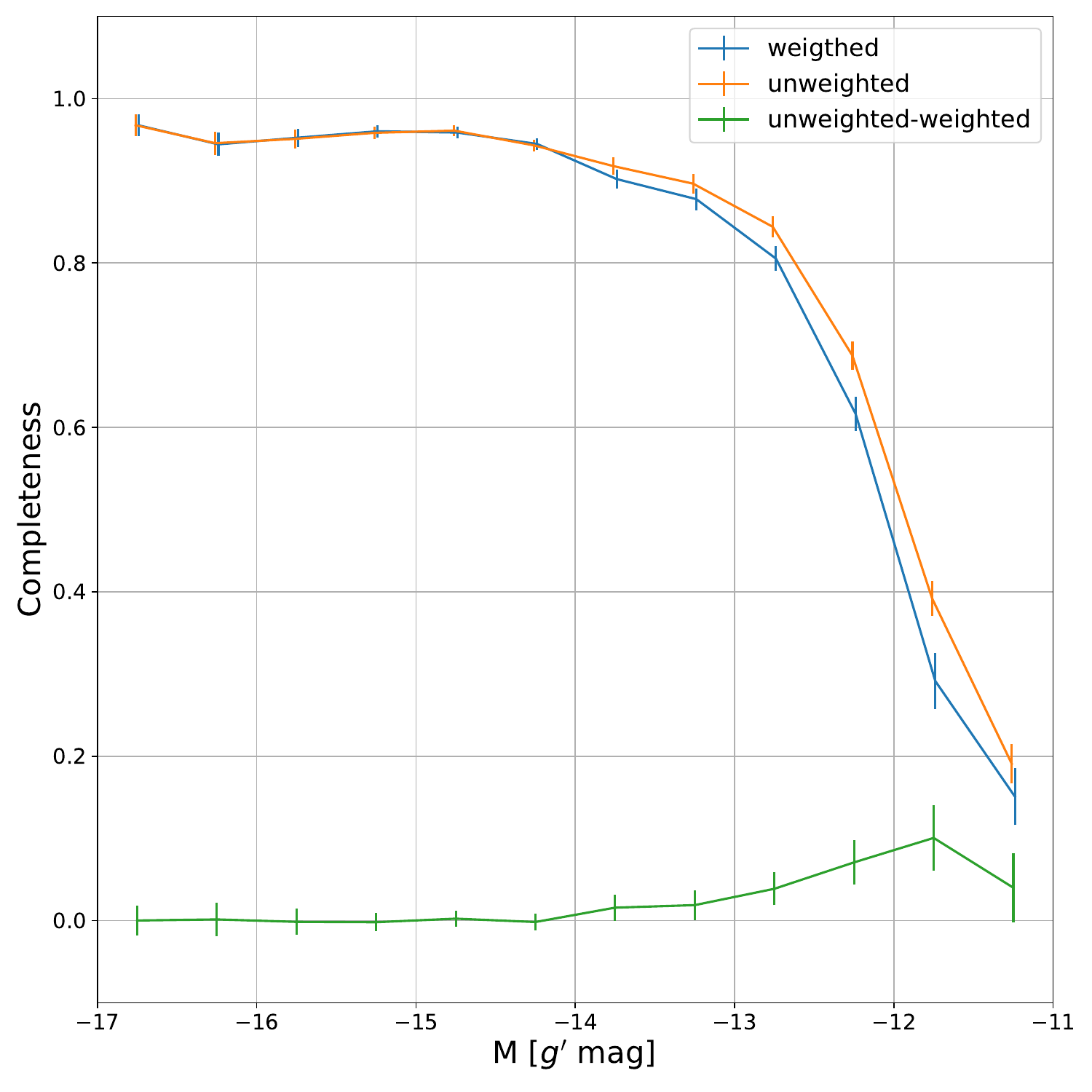}
    \caption{Comparison of the binned weighted (blue) and unweighted (orange) 1D completeness as a function of injected magnitude for the Coma cluster. The difference between the two curves shows the effect of applying the 2D completeness-based reweighting, while the green curve shows the residual difference between the unweighted and weighted estimates. Absolute magnitudes are expressed on a Coma-equivalent scale, used as a common reference scale rather than a physical distance assumption.}
    \label{fig:Completenessweightvsunweight}
    \end{center}
\end{figure}

In Figure \ref{fig:CompletenessA1656}, we show the derived completeness function for the Coma cluster, with the weighted binned data points overlaid for visualization. We find a relatively flat completeness function in the range of $-16\,g'\,\mathrm{mag}\lessapprox M\lessapprox-14\,g'\,\mathrm{mag}$, with a slight upturn at the bright end, and dropping fast for $ M\gtrapprox-13\,g'\,\mathrm{mag}$. We note that the weighting based on $C_{\rm 2D}(M,\mu_e)$ does mainly affect the completeness function at $ M\gtrapprox-14\,g'\,\mathrm{mag}$, whereas at brighter magnitudes it does not significantly affect the completeness function as visualized in Figure \ref{fig:Completenessweightvsunweight}.

\begin{figure}[t]
    \begin{center}
        \includegraphics[width=0.47\textwidth]{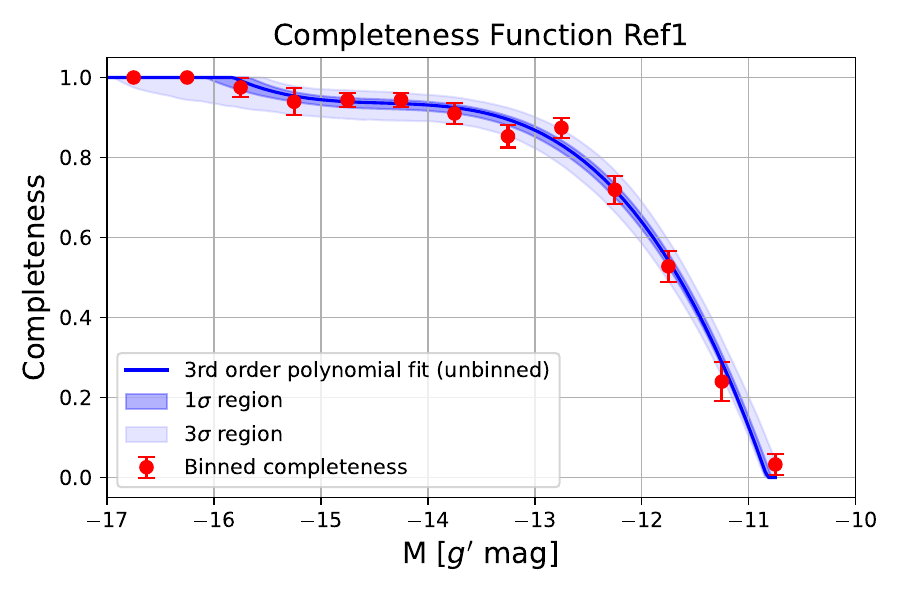}
        \includegraphics[width=0.47\textwidth]{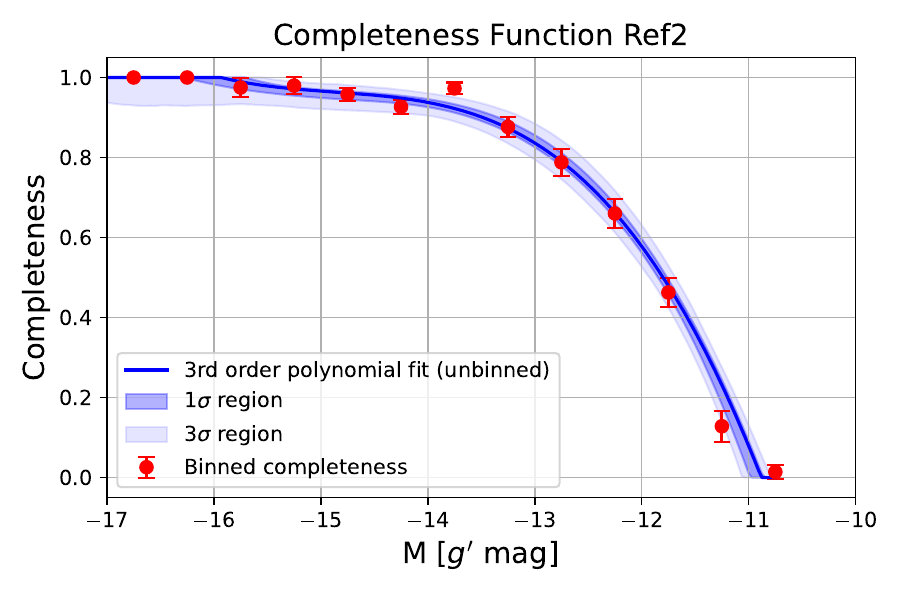}
    \caption{Completeness function for the reference fields. The red points show the weighted binned injection-recovery results, while the blue curve and shaded regions show the median and uncertainty of the unbinned nested-sampling fit. Absolute magnitudes are expressed on a Coma-equivalent scale, used as a common reference scale rather than a physical distance assumption.}
    \label{fig:Completenessref}
    \end{center}
\end{figure}

To derive the completeness function for the reference fields, we follow the approach presented before. Here, we draw random galaxies from the catalogs of both reference fields and reweight them with $C_{\rm 2D}(M,\mu_e)$ for the individual reference fields.

For Ref1, we perform the injection recovery test nine times and for Ref2 seven times (due to the larger area and the resulting more objects per run), drawing from the whole reference field sample and injecting them onto a grid with a step size of 500\,px, with the grid starting at different random x- and y-positions.

Due to the low number of relatively bright galaxies in the reference field catalogs, we did an additional run, randomly drawing only galaxies with $M<-14\,g'\,\mathrm{mag}$ to improve the statistics for the completeness function at the bright end. In total, we injected 1445 galaxies in Ref1 and 1565 in Ref2.

The resulting completeness functions are displayed in Figure \ref{fig:Completenessref}. For the reference fields, we find a similar shape of the completeness function as for the Coma field, but with the reference fields reaching slightly fainter magnitudes. Both reference fields reach 100\% completeness for their bright galaxies, unlike the completeness for the Coma cluster, and also the completeness function of the Coma cluster is dropping faster than both reference fields. This could be explained by the higher crowding in the Coma cluster field, and also due to the higher fraction of diffuse galaxies in the Coma cluster, leading to a lower completeness. That Ref1 reaches fainter magnitudes than Ref2, could be due to its greater depth in the $g'$ band.

\section{Results and Discussion} \label{sec:resultsdiscussion}

\subsection{Luminosity Function}
\label{sec:LFfunction}

\begin{figure*}[ht]
    \begin{center}
        \includegraphics[width=0.95\textwidth]{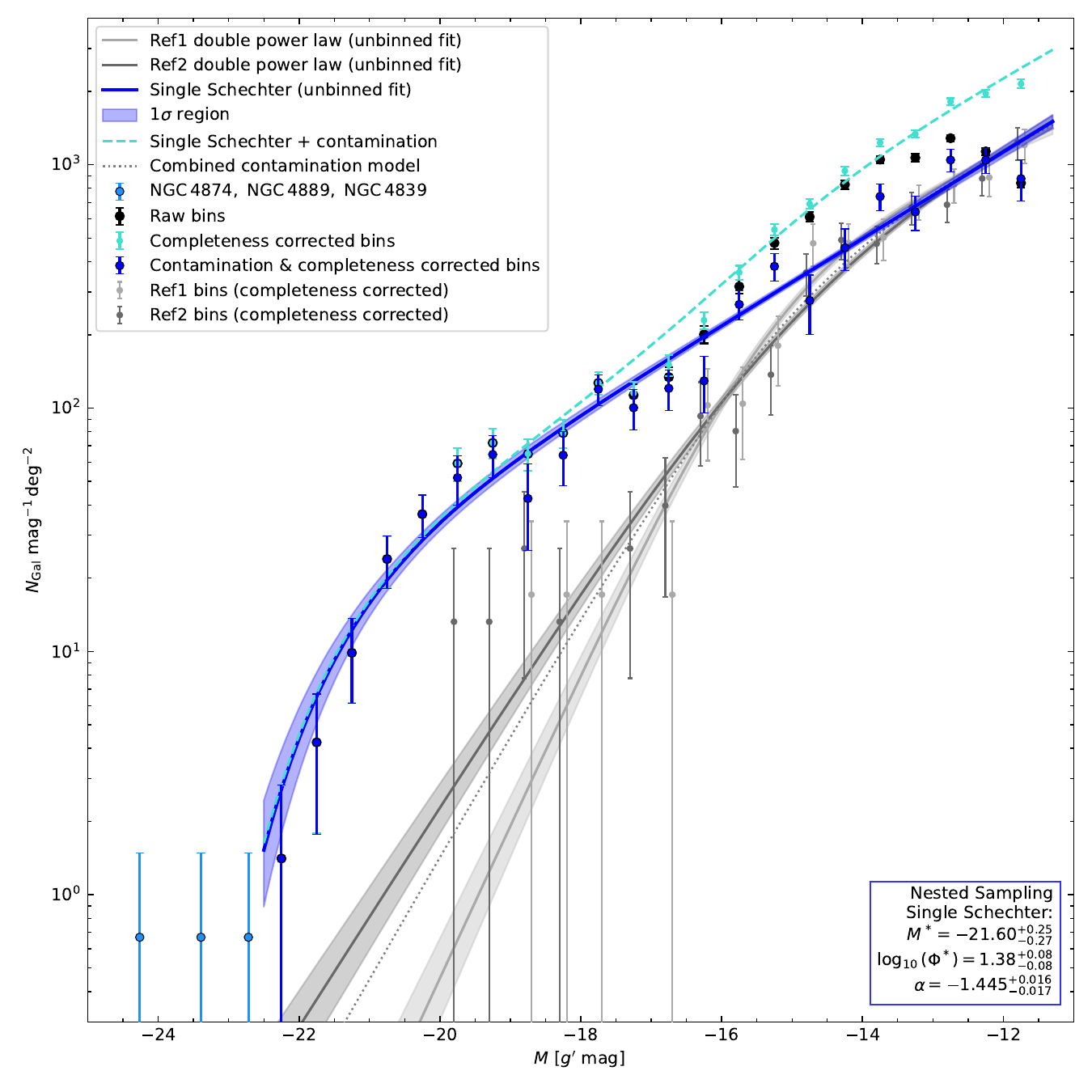}
        \caption{
        The fiducial single Schechter GLF of the Coma cluster is visualized as a blue solid line with its 16th--84th percentile posterior interval in shaded blue. This fit marginalizes over the individual galaxy magnitude uncertainties, completeness-function uncertainty, and background contamination.
        The completeness corrected counts per bin of galaxies from the reference fields are shown as light-grey data points for Ref1 and as dark-grey data points for Ref2 with their final double power-law functions as solid grey lines with their 16th--84th percentile posterior intervals in shaded. These BCF fits likewise marginalize over the individual galaxy magnitude uncertainties and completeness-function uncertainty.
        The errorbars are given by $\sigma_{\mathrm{Poisson}}$ for the raw counts, $\sqrt{\sigma_{\mathrm{Poisson}}^2+\sigma_{\mathrm{compl.}}^2}$ for the completeness corrected counts, and by $\sqrt{\sigma_{\mathrm{Poisson}}^2+\sigma_{compl.}^2+\sigma_{\mathrm{BG}}^2}$ for the completeness and contamination corrected counts. Here, $\sigma_{\mathrm{BG}}=\sqrt{\sigma_{\mathrm{Poisson,BG}}^2+\sigma_{\mathrm{model,BG}}^2}$ is the combined uncertainty from the Poisson noise of the combined reference field bins and the uncertainty due to cosmic variance estimated by the standard deviation of the two BCFs.
        The grey dotted and turquoise dashed lines are shown for visualization only and represent the combined-reference-field BCF, derived analogously to the individual reference fields, and the corresponding total GLF in the Coma field, $\Phi_{\mathrm{C}}+\Phi_{\mathrm{BG}}$, respectively; neither is used in the fitting.
        The three brightest galaxies, NGC 4889, NGC 4874, and NGC 4839, are shown in light blue and excluded from the Schechter fit. Their data points are not binned but shown at their actual positions, normalized by the total Coma footprint of $1.49\,\mathrm{deg}^2$. Absolute magnitudes are expressed on a Coma-equivalent scale, used as a common reference scale rather than a physical distance assumption.}
        \label{fig:LF}
    \end{center}
\end{figure*}

The goal of this work is to determine the GLF of the Coma cluster.  To derive the GLF, we want to use the unbinned data points, as the best-fit GLF derived by a fit to binned data points is sensitive to the chosen bin width and bin centers. This is especially affecting the bright end of the GLF due to the expected steep decline, and relatively low number of counts per bin. Hence, we want to use a maximum likelihood estimation using nested sampling based on the individual data points.
However, we first measure the GLF using binned data points to inform the priors for the nested sampling. Here, we correct the raw bin counts (black data points in Figure \ref{fig:LF}) with the best-fit completeness function and subtract the binned completeness corrected data points from the reference fields. For this, we use only galaxies with $M<-11.3\,g'\,\mathrm{mag}$ , i.e., a detection probability larger than $\approx 20\%$. This cut also avoids the regime where $C_{\rm 2D}(M,\mu_e)$ becomes poorly constrained, as indicated in Figure \ref{fig:2DCompletenessA1656} by the red dashed line. Furthermore, we exclude the three brightest galaxies NGC 4889, NGC 4874, and NGC 4839 (light blue), which is often done in the literature because the GLF flattens in this regime and is not well described by a Schechter nor double Schechter function \citep[e.g.,][]{Lin2004,CuillandreLF2025}. The completeness corrected bins are overlayed in turquoise, and the completeness and contamination corrected bins in blue, where we subtracted the combined completeness corrected binned data points of the reference fields. We note here that at the magnitude ($M\approx-17\,g'\,\mathrm{mag}$) where the sample is split into a bright and a faint galaxy sample, we do not find any indication of an offset, implying that the two methods are consistent. 

For deriving the GLF of the Coma cluster using an unbinned maximum likelihood estimator, we require an analytic description for the contamination. 
For this, we fit a double power law to the unbinned reference field data, motivated by \citet{Marr2023}, who showed that the sum over many redshifts of galaxy number counts exhibits a transition from a Euclidean bright-end slope to an $\alpha$-dependent faint-end slope, we use a double power law for the BCF. 
Due to our sample selection, the exact form will not apply directly; we therefore allow the bright- and faint-end slopes ($\beta_{\mathrm{BG}}$ and $\alpha_{\mathrm{BG}}$) as well as the turnover magnitude ($M^*_{\mathrm{BG}}$) and normalization ($\Phi^*_{\mathrm{BG}}$) to vary freely. Here, we fit the double power law to each of the reference fields individually to get an estimate for the cosmic variance. 

\begin{equation}
    \begin{aligned}
    \Phi_{\mathrm{BG}}(M) =
    &\Phi^{*}_{\mathrm{BG}}
    \bigg[
    10^{0.4(\alpha_{\mathrm{BG}}+1)(M-M^{*}_{\mathrm{BG}})} \\
    &+10^{0.4(\beta_{\mathrm{BG}}+1)(M-M^{*}_{\mathrm{BG}})}
    \bigg]^{-1}.
    \end{aligned}
    \label{eq:dplref}
\end{equation}

We use data-informed priors based on a binned double-power-law fit to the combined number counts from both reference fields combined. Specifically, we adopt Gaussian priors on $M^*_{\mathrm{BG}}$ and $\log_{10}(\Phi^*_{\mathrm{BG}})$, centered on the best-fit values from this combined binned reference-field fit and with standard deviations given by their corresponding $1\sigma$ uncertainties. For the slopes, we use uniform priors, $\alpha_{\mathrm{BG}} \sim \mathcal{U}(-1.5,-0.5)$ and $\beta_{\mathrm{BG}}\sim\mathcal{U}(-3.0,-1.0)$.

To perform the maximum likelihood estimation, we again use the dynamic nested sampling package \verb+DYNESTY+ \citep{Speagle2020}. We use a Poisson log-likelihood that is, following \cite{Cash1979}, given by:
\begin{equation}
    \ln(\mathcal{L}) =  -\mu + \sum_{i} n_i\ln\{f(M_i)\},
\end{equation}
where $\mu$ is the total expected number of counts and $f(M_i)$ is the finite expected number of counts at $M_i$. $n_i$ is the number of counts per bin, but as we want to perform the fit using the unbinned data and get the luminosity function normalized by the area in units per $\mathrm{deg^2}$, we use $n_i=1/A$, where $A$ is the respective covered area.

For deriving the BCF $\Phi_{\mathrm{BG}}(M)$, we use the following log-likelihood for a given draw $s$ of the completeness function:

\begin{equation}
\begin{aligned}
\ln(\mathcal{L}_s)=
&-\int
C_{\mathrm{BG},s}(M)
\Phi_{\mathrm{BG}}(M)\,dM\\
&+\sum_{i=1}^{N_{\mathrm{ref}}}
\frac{
\ln\{C_{\mathrm{BG},s}(M^\mathrm{BG}_i)
\Phi_{\mathrm{BG}}(M^{\mathrm{BG}}_i)\}
}{A_{\mathrm{ref}}}.
\end{aligned}
\label{eq}
\end{equation}

To marginalize over the uncertainty of the completeness function, we average the likelihood over $S=1000$ draws from the posterior distribution of the completeness-function fit and evaluate the log-likelihood during nested sampling as described before for the completeness function (Equations \ref{eq:sumloglike} and \ref{eq:log-mean-exp}).


The posterior distribution is then derived using \verb+DYNESTY+ for each reference field. The posterior distributions of these model-selection-stage reference field fits are shown in Appendix~\ref{sec:GLFposteriors} in Figure~\ref{fig:PosteriorLFref_combined}. 
Their faint-end slopes $\alpha_{\mathrm{BG}}$ agree well, and the two BCFs also agree well at the faint end, whereas they deviate significantly in the bright and intermediate regimes. The differences between the two BCFs therefore trace the field-to-field variation of the background population, i.e., cosmic variance, but may also partly result from low-number statistics, especially at the bright end, where only a few reference-field galaxies are detected.

For the GLF of the Coma cluster, we follow a similar procedure.  
The joint log-likelihood for the nested sampling maximum likelihood estimate is given in Equation \ref{eq:LLDS}. Here, $f(M_i)$ consists of the expected counts of the Coma cluster itself and the expected counts of interloping background galaxies given the completeness function of the Coma cluster image $C(M)$. Because the faint galaxies are detected on the images after subtracting the models of the bright galaxies and masking their central residuals, the effective area in which the faint galaxies are detected ($A_\mathrm{f}=1.39\,\mathrm{deg^2}$) is smaller than the area of the bright sample ($A_\mathrm{b}=1.41\,\mathrm{deg^2}$). Hence, we split the sum in the joint log-likelihood into our bright galaxy sample and our faint galaxy sample, and divide them by their respective effective area. 
For a given realization $s$ of the nuisance parameters, i.e. the completeness function $C_s(M)$ and the background GLF $\Phi_{\mathrm{BG},s}(M)$, the joint log-likelihood used to determine the GLF of the Coma cluster $\Phi_{\mathrm{C}}(M)$ is given by:

\begin{equation}
    \begin{aligned}    
        \ln(&\mathcal{L}_s)= -\int C_{s}(M)(\Phi_{\mathrm{C}}(M)+\Phi_{\mathrm{BG},s}(M))\,dM\\
                      &+\sum_{i=1}^{N_{\mathrm f}}\frac{\ln\{C_{s}(M^\mathrm{f}_i)(\Phi_{\mathrm{C}}(M^\mathrm{f}_i)+\Phi_{\mathrm{BG},s}(M^\mathrm{f}_i)) \}}{A_\mathrm{f}}\\
                      &+\sum_{j=1}^{N_{\mathrm b}}\frac{\ln\{C_{s}(M^\mathrm{b}_j)(\Phi_{\mathrm{C}}(M^\mathrm{b}_j)+\Phi_{\mathrm{BG},s}(M^\mathrm{b}_j)) \}}{A_\mathrm{b}}.
                      \label{eq:LLDS}
    \end{aligned}   
\end{equation}

\subsection{GLF Model Comparison}
\label{sec:GLFmodelcomparison}
For the purpose of selecting the functional form of the Coma GLF, we first perform the following model comparison using the measured best-estimate magnitudes of the individual galaxies. The magnitude uncertainties are not included as nuisance quantities at this stage and are considered subsequently in Section~\ref{sec:EddingtonBias}. To account for the posterior uncertainty in these nuisance quantities during the model-selection fits, we marginalize the likelihood over $S=1000$ realizations of the completeness function $C(M)$ and the BCF $\Phi_{\mathrm{BG}}(M)$. The galaxy magnitudes are held fixed at their measured best-estimate values at this stage.
Each realization consists of one posterior draw of $C(M)$ and one posterior draw of $\Phi_{\mathrm{BG}}(M)$, which are treated as independent. The log-likelihood during nested sampling is then evaluated using the same log-mean-exp form as described above for the completeness-function fit (Equations \ref{eq:sumloglike} and \ref{eq:log-mean-exp}).

To derive the GLF of the Coma cluster, we test four different models: a single Schechter function \citep[e.g.,][]{Mobasher2003,dePropris2003,CuillandreLF2025}
\begin{equation}
    \begin{aligned}
       \Phi_{\mathrm{C}}(M)=&0.4\ln(10)\,\Phi^{*}
       \times \exp\Bigl[-10^{-0.4(M-M^{*})}\Bigr] \\
        &\times 10^{-0.4(M-M^{*})(\alpha+1)},
    \end{aligned}
    \label{eq:singleschechter}
\end{equation}
a double Schechter function \citep[e.g.,][]{Blanton2005,Agulli2016}
\begin{equation}
    \begin{aligned}
       \Phi_{\mathrm{C}}(M)=&0.4\ln(10) \times \exp\Bigl[-10^{-0.4(M-M^{*})}\Bigr] \\
        &\times \Bigl[\Phi^{*}_1 10^{-0.4(M-M^{*})(\alpha_1+1)}\\
        &+\Phi^{*}_2 10^{-0.4(M-M^{*})(\alpha_2+1)}\Bigr],
    \end{aligned}     
    \label{eq:doubleschechter}
\end{equation}
a double Schechter function with two independent $M^*$ \citep[e.g.,][]{Mercurio2003,Lan2016}
\begin{equation}
\begin{aligned}
\Phi_{\rm C}(M) =&\,
0.4\ln(10)\,\Phi^{*}_1
\exp\left[-10^{-0.4(M-M^{*}_1)}\right]
\\
&\times
10^{-0.4(M-M^{*}_1)(\alpha_1+1)}
\\
&+
0.4\ln(10)\,\Phi^{*}_2
\exp\left[-10^{-0.4(M-M^{*}_2)}\right]
\\
&\times
10^{-0.4(M-M^{*}_2)(\alpha_2+1)} ,
\end{aligned}
\label{eq:doubletwomstarschechter}
\end{equation}
and a Gaussian + Schechter function \citep[e.g.,][]{Biviano1995,Secker1996,CuillandreLF2025}
\begin{equation}
\begin{aligned}
\Phi_{\rm C}(M) =&\,
\Phi_{\rm G}
\exp\left[
-\frac{1}{2}
\left(\frac{M-M_{\rm G}}{\sigma_{\rm G}}\right)^2
\right]
\\
&+0.4\ln(10)\,\Phi^{*}
\exp\left[-10^{-0.4(M-M^{*})}\right]
\\
&\times
10^{-0.4(M-M^{*})(\alpha+1)} .
\end{aligned}
\label{eq:gaussianschechter}
\end{equation}

To inform the priors for the nested sampling, we use the results from the least-squares fits of the different models to the binned data points and center the priors around the best-fit values. For normal priors, we use $\sigma_{\mathrm{prior}}=2\times\sigma_{\mathrm{binned\,fit}}$. The priors, best-fit values, and the Bayesian Information Criterion (BIC) of the four nested sampling runs are summarized in Table \ref{tab:schechter-constraints}. The BIC favors the single Schechter function due to its simplicity; hence, we use the single Schechter model as the best description of the Coma GLF.

\input{schechter_model_comparison_table.tex}



The posterior distributions of all four model-selection fits are shown in Appendix \ref{sec:GLFposteriors} for the single Schechter function (Figure \ref{fig:PosteriorLF}), for double Schechter function (Figure \ref{fig:PosteriorLF_DS}), the double Schechter function with two independent $M^*$ (Figure \ref{fig:PosteriorLF_2MDS}), and the Gaussian + Schechter function (Figure \ref{fig:PosteriorLF_GS}). 

\begin{figure*}[ht]
    \begin{center}
        \includegraphics[width=\textwidth]{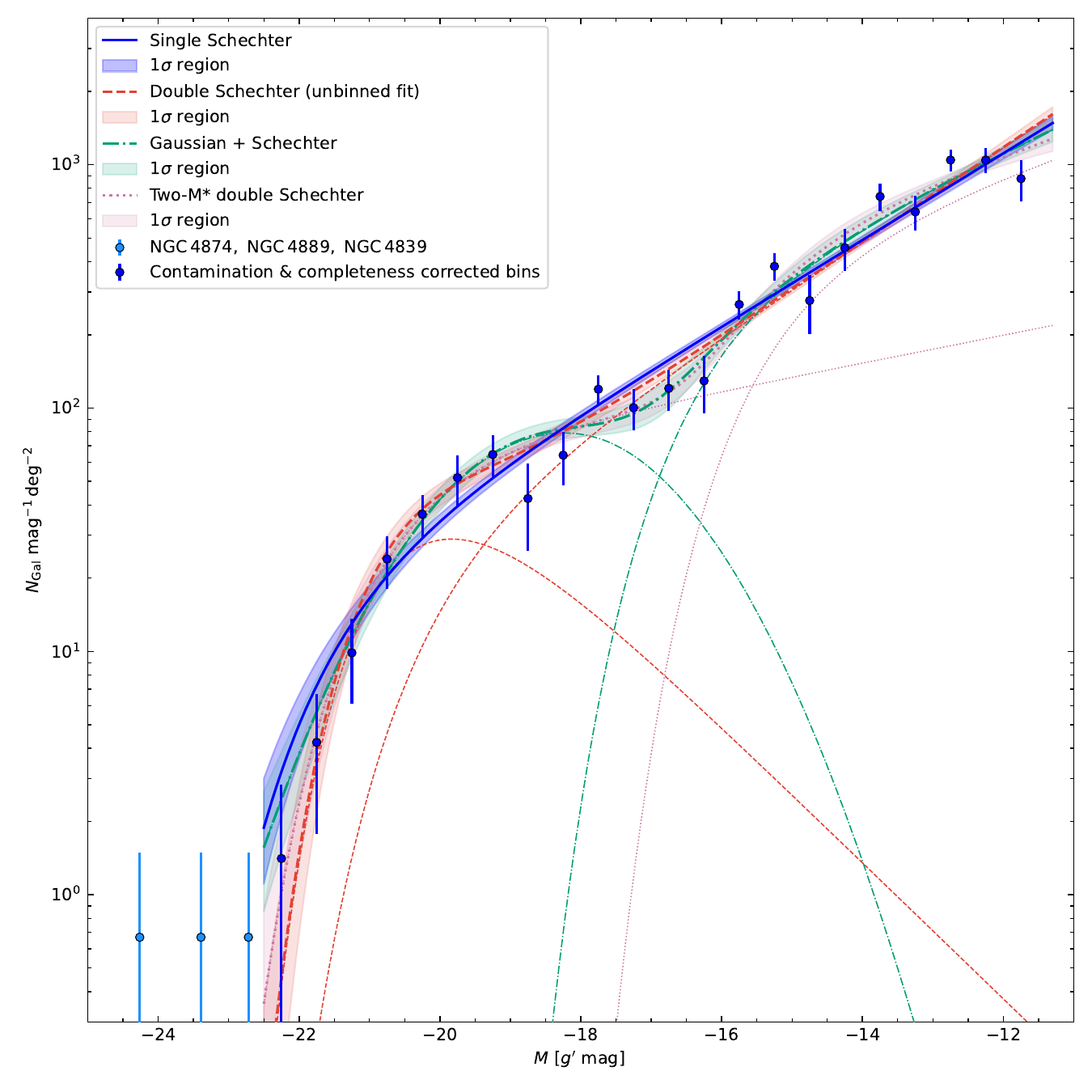}
        \caption{Model comparison for the Coma GLF. Blue circles indicate the binned Coma GLF data points after contamination and completeness correction, and the light-blue circles indicate the three brightest cluster galaxies, NGC~4874, NGC~4889, and NGC~4839.  
        The solid blue, dashed orange-red, dotted purple, and dash-dotted green-blue curves show the posterior median predictions of the nominal-magnitude model-selection fits for the single Schechter, double Schechter, two-$M^*$ double-Schechter, and Gaussian+Schechter models, respectively, with shaded bands enclosing the $16$th--$84$th percentile posterior intervals.
        For the multi-component models, the thin curves with the same color and linestyle show the individual components contributing to the total GLF.}
        \label{fig:GLFmodelcomp}
    \end{center}
\end{figure*}

A comparison of the different fit functions is shown in Figure~\ref{fig:GLFmodelcomp}. The nominal-magnitude single Schechter fit is depicted as a solid blue line, the double Schechter as a dashed orange-red line, the double Schechter with two independent $M^*$ as a dotted purple line, and the Gaussian+Schechter model as a green-blue dash-dotted line. The shaded regions indicate the corresponding $16$th--$84$th percentile posterior intervals. For the multiple-component models, we visualize the individual components as thin lines in the same color and linestyle as the total model. The nominal-magnitude single Schechter fit follows the completeness and contamination corrected binned GLF well over the full fitted magnitude range. This fit is used here for the model comparison only; the fiducial single Schechter GLF, including the individual magnitude uncertainties as nuisance quantities, is derived in Section~\ref{sec:EddingtonBias} and shown in Figure~\ref{fig:LF}.

Figure \ref{fig:GLFmodelcomp} shows that all four models provide broadly similar
descriptions of the GLF over the magnitude range best constrained by
the data. The main differences appear at the bright end and over the intermediate
magnitude range, approximately $-20\,g'\,\mathrm{mag} \lesssim M \lesssim -16\,g'\,\mathrm{mag}$.
Despite the limited flexibility of the single Schechter function, it follows the completeness and contamination corrected binned GLF well over the full fitted range.
For the nominal-magnitude single Schechter fit, the faint-end slope is $\alpha=-1.444^{+0.015}_{-0.015}$, which is
is also consistent within $2\sigma$ to the slope of the faint component of the double Schechter model, for which $\alpha_2=-1.485^{+0.026}_{-0.029}$. This indicates that the faint-end behavior is already well captured by the simpler single Schechter description.

The double Schechter model adds flexibility to the single Schechter form by adding a second component with the same characteristic magnitude $M^*$ but an independent slope and normalization. Relative to the single Schechter function, this mainly changes the bright end of the GLF, $M\lesssim-19\,g'\,\mathrm{mag} $. At fainter magnitudes, the single and double Schechter fit is nearly equivalent. However, the improvement in flexibility does not translate into a lower BIC. The double Schechter model has $\Delta{\rm BIC}={\rm BIC}_{\rm model}-{\rm BIC}_{\rm single\,Schechter}=9.7$ relative to the single Schechter function, indicating that the additional parameters are not justified by the likelihood improvement.

The two-$M^*$ double Schechter model adds flexibility by allowing the two components to have independent characteristic magnitudes. In Figure~\ref{fig:GLFmodelcomp}, this mainly affects the bright and intermediate-magnitude regime. Nevertheless, the model remains disfavored by the BIC with $\Delta{\rm BIC}=6.3$ with respect to the single Schechter, and thus, the data does not require this additional complexity. Considering the BIC, the double Schechter with two independent $M^*$ is favored with respect the double Schechter with the same $M^*$. As both Schechter components of the best-fit two-$M^*$ double Schechter model contribute significantly at the faint end of the GLF, the parameter $\alpha_2$ can not be interpreted as the faint-end slope of the GLF anymore, unlike for the single Schechter, and the double Schechter function with one $M^*$.

The Gaussian+Schechter model provides an alternative empirical description by adding a Gaussian term to the Schechter function. This extra component mainly affects the intermediate-magnitude regime, but its parameters, especially  $M_{\rm G}$ and $\sigma_{\rm G}$, are strongly degenerate with the Schechter component, as reflected in the posterior distributions in Figure \ref{fig:PosteriorLF_GS} and the large uncertainties, especially of $M^*$.  Additionally, the added flexibility is not sufficient to overcome the BIC penalty, the model remains disfavored relative to the single Schechter fit ($\Delta{\rm BIC}=6.9$), but is also disfavored compared to the two-$M^*$ double Schechter model with the same number of free parameters.

Overall, the model comparison indicates that the Coma GLF does not require a multi-component functional form. Although the double Schechter, two-$M^*$ double Schechter, and Gaussian+Schechter models can reproduce small changes in curvature in the bright to intermediate magnitude range indicated by the binned datapoints, none of them is preferred once the BIC penalty for the additional degrees of freedom is included. We therefore adopt the single Schechter function as the fiducial functional form for the Coma GLF. The final parameter constraints for this model, including the uncertainties in the individual galaxy magnitudes, are derived in the following Section~\ref{sec:EddingtonBias}. 

\subsection{Eddington Bias and Fiducial GLF Constraints}
\label{sec:EddingtonBias}

The model comparison in Section~\ref{sec:GLFmodelcomparison} favors a single Schechter function as the simplest description of the Coma cluster GLF. The model comparison was performed using the measured best-estimate magnitudes of the individual galaxies. However, measurement uncertainties in the magnitudes can scatter galaxies between different parts of the GLF. In particular, for a steeply varying number-count distribution, symmetric magnitude uncertainties can result in an asymmetric redistribution of galaxies, commonly referred to as Eddington bias \citep{Eddington1913}. We therefore repeat the inference for the preferred single Schechter model while explicitly treating the individual galaxy magnitudes as nuisance quantities. The results of this analysis are adopted as our final constraints on the Coma cluster GLF.

For each galaxy $i$ with measured absolute magnitude $M_i$ and magnitude uncertainty $\sigma_{M_i}$, we generate a realization $s$ according to
\begin{equation}
M_{i,s}\sim\mathcal{N}\left(M_i,\sigma_{M_i}^2\right).
\label{eq:mag_nuisance_draw}
\end{equation}
This procedure is applied independently to every galaxy in the faint and bright Coma samples and to the galaxies in both reference fields. After drawing the magnitudes, we apply the same magnitude selection as in the nominal GLF analysis and retain galaxies with $M<-11.3\,g'\,\mathrm{mag}$. Consequently, a galaxy close to this magnitude limit can scatter into or out of the fitted sample in an individual realization. The number of galaxies entering the likelihood is therefore allowed to vary between realizations rather than being artificially fixed to the number obtained from the nominal magnitudes. The catalog used for these realizations extends to $M\approx-10.7\,g'\,\mathrm{mag}$, about $0.6\,\mathrm{mag}$ fainter than the adopted GLF limit, while the median $1\sigma$ magnitude uncertainty around the limit is only $\approx0.04\,g'\,\mathrm{mag}$. Thus, galaxies can scatter across the limit in both directions without being affected by catalog truncation.

We use $S=10000$ independent magnitude realizations. This is an increase by a factor of ten relative to the $S=1000$ nuisance realizations used in the preceding analysis. The larger number of realizations is adopted to obtain a sufficiently stable Monte Carlo marginalization once the individual magnitude uncertainties are included in addition to the completeness and background uncertainties.

\begin{figure*}[ht]
\begin{center}
\includegraphics[width=0.49\textwidth]{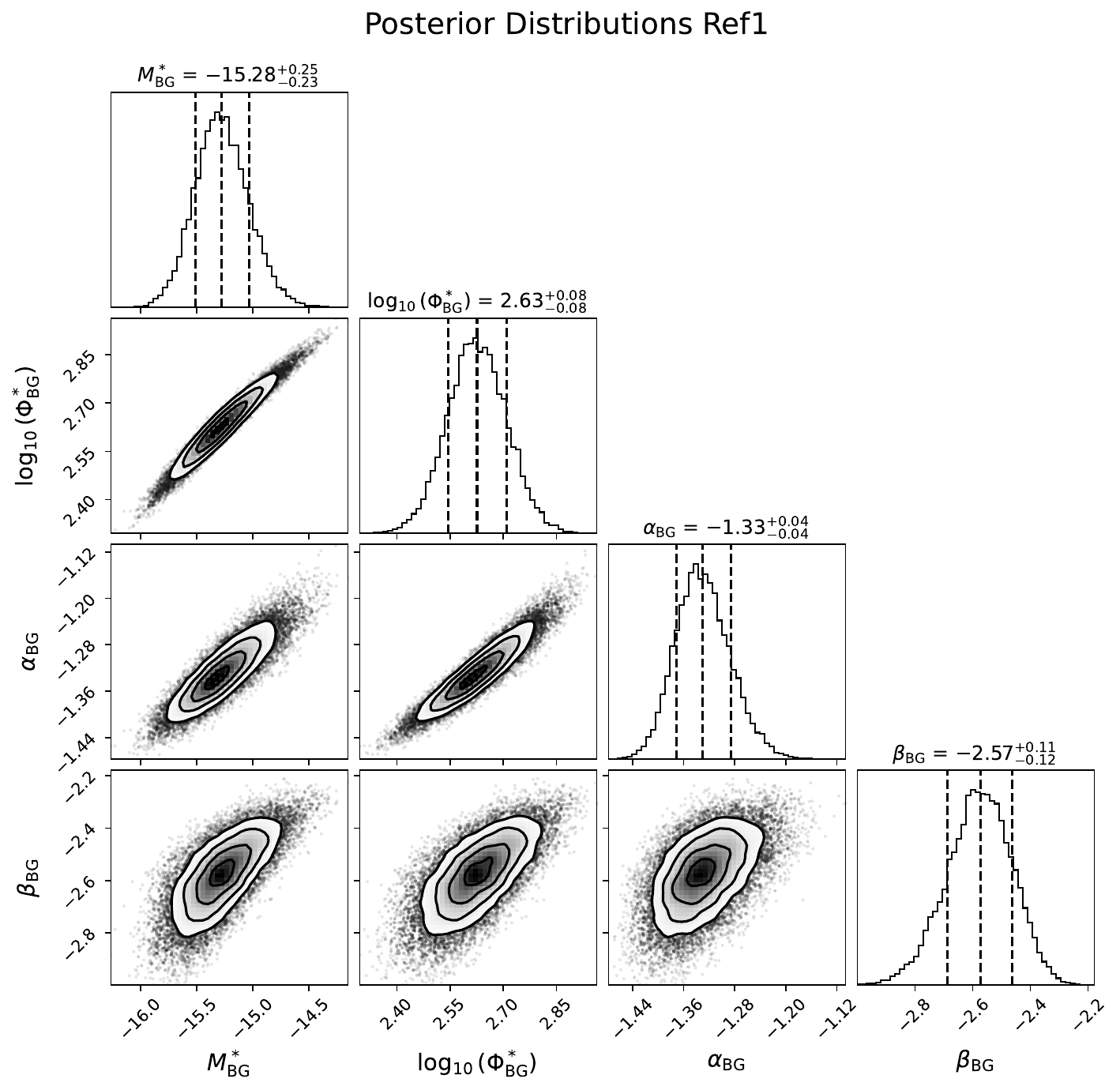}
\includegraphics[width=0.49\textwidth]{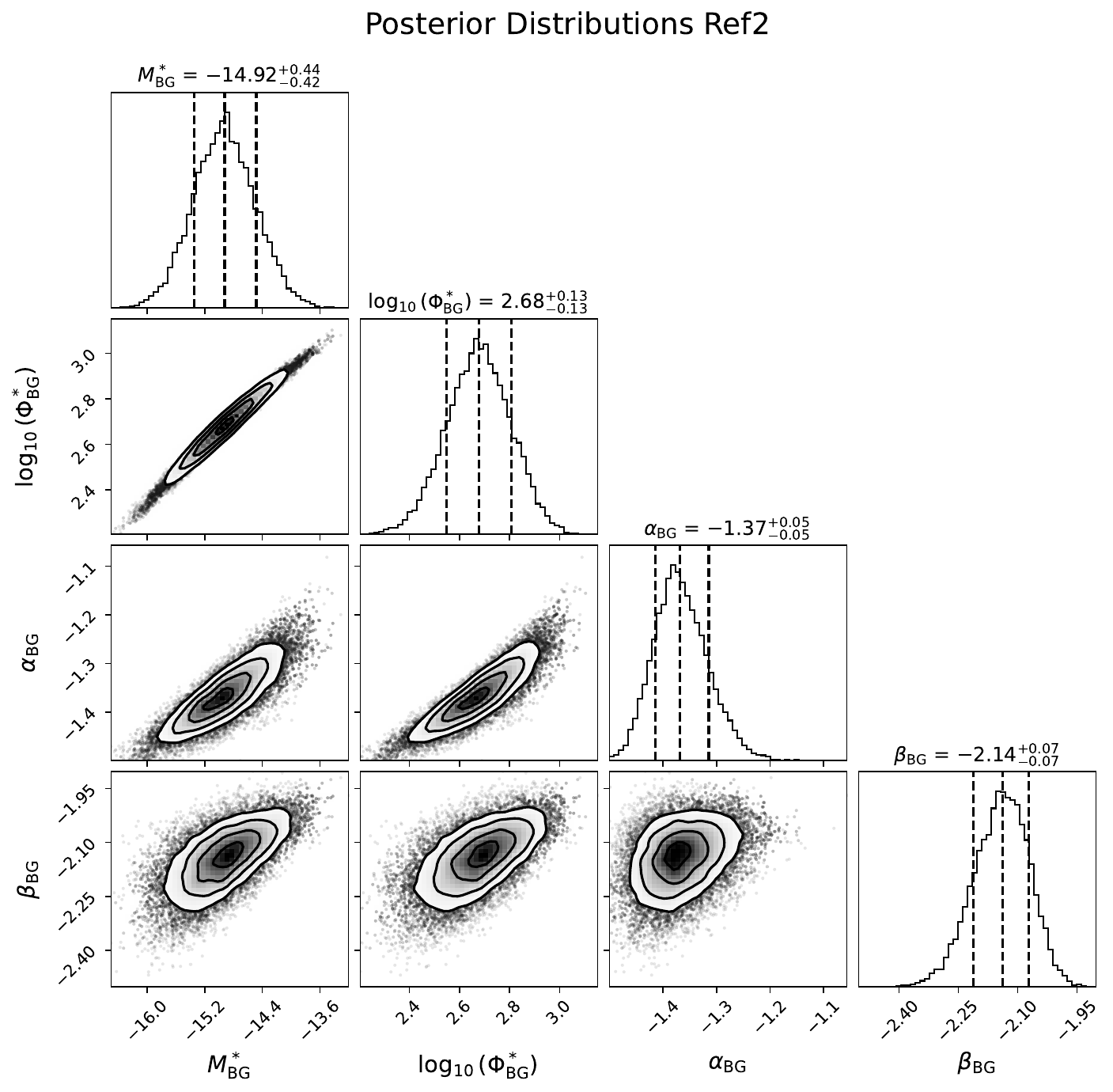}
\caption{Final posterior distributions of the double-power-law background-count functions for the two reference fields, Ref1 (left) and Ref2 (right). In addition to the uncertainty of the completeness function, the likelihood is marginalized over $10000$ realizations of the individual galaxy magnitudes drawn from their measurement uncertainties. The contours correspond to the $0.5\sigma$, $1\sigma$, $2\sigma$, and $3\sigma$ levels. These posteriors are used to propagate the background uncertainty into the final Coma cluster GLF fit. }
\label{fig:PosteriorLFref_final}
\end{center}
\end{figure*}


We first repeat the fits of the BCFs to both reference fields while also accounting for the individual galaxy magnitude uncertainties. For each reference-field realization, the magnitudes are drawn according to Equation~\ref{eq:mag_nuisance_draw}, and the same magnitude selection of $M<-11.3\,g'\,\mathrm{mag}$ as in the nominal analysis is applied. As before, the uncertainty of the completeness function is included by drawing a realization from its posterior distribution. The corresponding likelihood for realization $s$ is

\begin{equation}
\begin{aligned}
\ln(\mathcal{L}_s)={}&-\int C_{\mathrm{BG},s}(M)\Phi_{\mathrm{BG}}(M)\,dM\\
&+\sum_{i=1}^{N_{\mathrm{ref},s}}
\frac{
\ln\left\{
C_{\mathrm{BG},s}(M^{\mathrm{BG}}_{i,s})
\Phi_{\mathrm{BG}}(M^{\mathrm{BG}}_{i,s})
\right\}
}{A_{\mathrm{ref}}}.
\end{aligned}
\label{eq:LL_BCF_eddington}
\end{equation}

Here, $N_{\mathrm{ref},s}$ denotes the number of reference-field galaxies remaining in the fitted sample after applying the magnitude selection in realization $s$. The likelihood is marginalized over all $S=10000$ joint realizations of the galaxy magnitudes and completeness function using the same log-mean-exp formulation as in the preceding analysis. The resulting posterior distributions are shown in Figure~\ref{fig:PosteriorLFref_final}.

For the final Coma GLF fit, we retain the single Schechter functional form selected by the BIC analysis in Section~\ref{sec:LFfunction}. We do not repeat the model comparison after introducing the magnitude nuisance quantities; rather, the purpose of the present analysis is to determine whether the inferred parameters of the preferred model are affected by Eddington bias and to obtain our final parameter constraints.

For every realization $s$, one realization of the Coma completeness function and one realization of the background-count function are drawn from their respective posterior distributions. These are combined with the corresponding realization of the individual Coma galaxy magnitudes. The completeness and background draws are treated as independent. The likelihood for realization $s$ has the same form as Equation~\ref{eq:LLDS}, but is evaluated at the realization-dependent magnitudes $M_{i,s}$:
\begin{equation}
\begin{aligned}
&\ln(\mathcal{L}_s)={}-\int C_s(M)\left[\Phi_{\mathrm{C}}(M)+\Phi_{\mathrm{BG},s}(M)\right]\,dM\\
&+\sum_{i=1}^{N_{\mathrm{f},s}}\frac{\ln\left\{C_s(M^{\mathrm{f}}_{i,s})\left[\Phi_{\mathrm{C}}(M^{\mathrm{f}}_{i,s})+\Phi_{\mathrm{BG},s}(M^{\mathrm{f}}_{i,s})\right]\right\}}{A_{\mathrm{f}}}\\
&+\sum_{j=1}^{N_{\mathrm{b},s}}\frac{\ln\left\{C_s(M^{\mathrm{b}}_{j,s})\left[\Phi_{\mathrm{C}}(M^{\mathrm{b}}_{j,s})+\Phi_{\mathrm{BG},s}(M^{\mathrm{b}}_{j,s})\right]\right\}}{A_{\mathrm{b}}}.
\end{aligned}
\label{eq:LLDS_eddington}
\end{equation}
Here, $N_{\mathrm{f},s}$ and $N_{\mathrm{b},s}$ denote the numbers of faint and bright galaxies remaining in the fitted sample after applying the magnitude selection in realization $s$. The total number of galaxies entering the fit therefore varies between realizations, with a median of $6540^{+5}_{-4}$ galaxies, where the uncertainties denote the $16$th--$84$th percentile interval. The likelihood entering the nested sampling is then marginalized over the $S=10000$ joint nuisance realizations using the same log-mean-exp formulation as before.

\begin{figure}[ht]
\begin{center}
\includegraphics[width=0.47\textwidth]{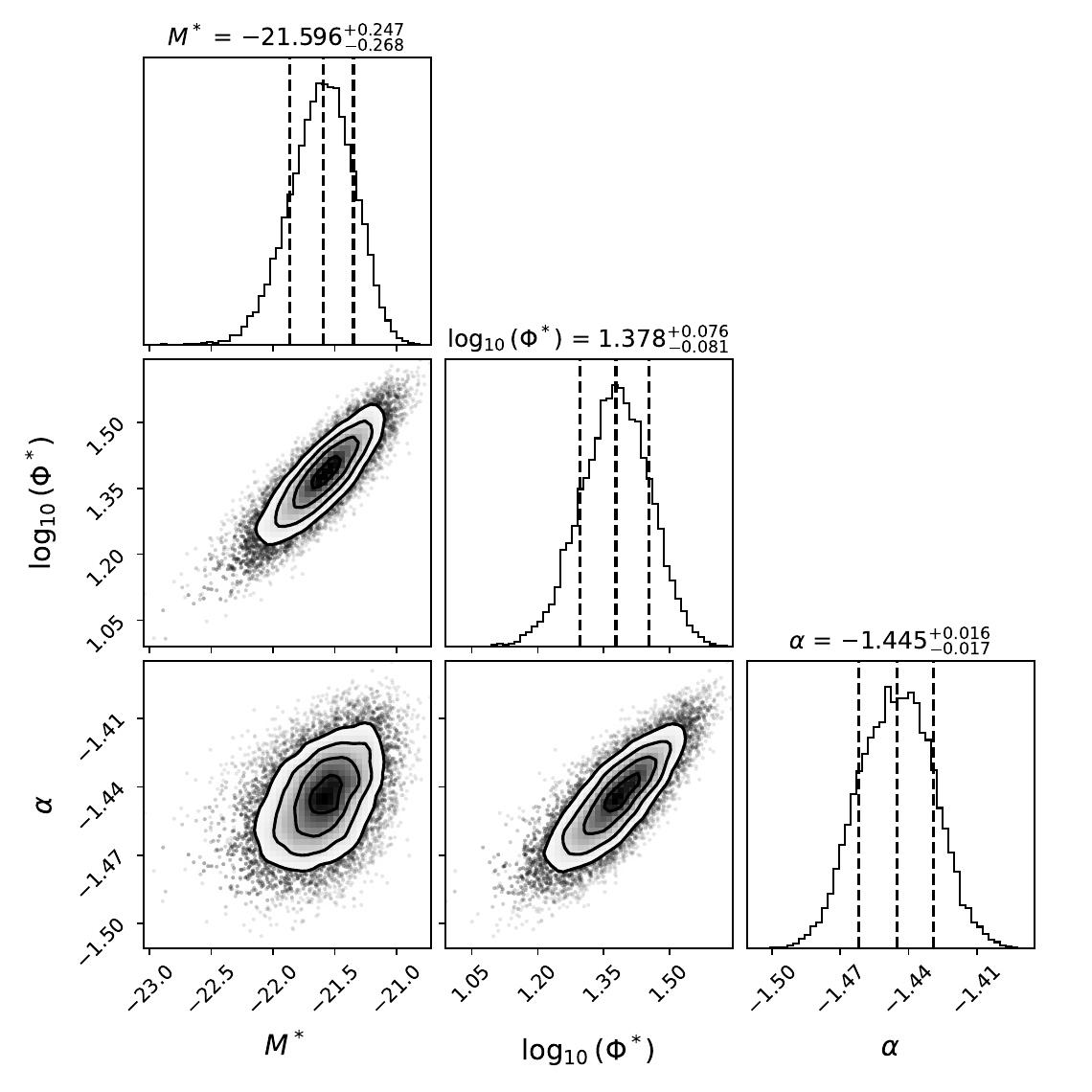}
\caption{Final posterior distribution of the fiducial single Schechter fit to the Coma cluster GLF. The likelihood is marginalized over $10000$ joint realizations of the individual galaxy magnitudes, completeness function, and background contamination. The contours correspond to the $0.5\sigma$, $1\sigma$, $2\sigma$, and $3\sigma$ levels. These posterior constraints constitute the final GLF results adopted throughout this work.}
\label{fig:PosteriorLF_final}
\end{center}
\end{figure}

The posterior distribution obtained from this fit is shown in Figure~\ref{fig:PosteriorLF_final}. Since this analysis simultaneously propagates the uncertainties associated with the measured galaxy magnitudes, completeness function, and background contamination, we adopt this posterior as our fiducial GLF posterior. Accordingly, all final single Schechter parameters quoted below, and the GLF shown in Figure~\ref{fig:LF}, are derived from this nuisance-marginalized fit rather than from the nominal-magnitude fit used for the model comparison.

The resulting final single Schechter parameters are
\begin{equation}
\begin{aligned}
M^* &=-21.60^{+0.25}_{-0.27}\,g'\,\mathrm{mag},\\
\log_{10}\left(\Phi^*\,[\mathrm{deg}^{-2}\,\mathrm{mag}^{-1}]\right)
&= 1.378^{+0.076}_{-0.081},\\
\text{and}\qquad
\alpha &= -1.445^{+0.016}_{-0.017}.
\end{aligned}
\end{equation}
where the quoted uncertainties correspond to the $16$th--$84$th percentile interval of the marginalized posterior. These values constitute our final constraints on the Coma cluster GLF.



For comparison, the nominal-magnitude single Schechter fit used during the model-selection analysis yielded $M^*=-21.71^{+0.26}_{-0.29}\,g'\,\mathrm{mag}$, $\log_{10}(\Phi^*\,[\mathrm{deg}^{-2}\,\mathrm{mag}^{-1}])=1.355^{+0.076}_{-0.079}$, and $\alpha=-1.444^{+0.015}_{-0.015}$. Including the individual galaxy magnitudes as nuisance quantities changes these parameters only slightly, with $\Delta M^*=0.11\,g'\,\mathrm{mag}$, $\Delta\log_{10}(\Phi^*)=0.023$, and $\Delta\alpha=0.001$. All three parameters remain well within their respective posterior uncertainties, indicating that the inferred single Schechter GLF is only weakly affected by Eddington bias over the fitted magnitude range.


The resulting fiducial single Schechter Coma cluster GLF is shown in Figure~\ref{fig:LF} as a solid blue line, with its $16$th--$84$th percentile posterior interval indicated by the shaded blue region. In contrast to the single Schechter curve shown in the model comparison in Figure~\ref{fig:GLFmodelcomp}, this GLF is derived from the final fit in which the individual galaxy magnitudes, completeness function, and background contamination are treated as nuisance quantities and marginalized over. Furthermore, we show the corresponding total GLF in the Coma field, $\Phi_{\mathrm{C}}+\Phi_{\mathrm{BG}}$, as a turquoise dashed line. At the faint end, the background contribution remains comparable to the Coma signal even after the color--color preselection, corresponding to a contamination level of approximately 50\%. For the visualization, we use a $\Phi_{\mathrm{BG}}$ model derived analogously to the individual reference fields but using the combined data of both reference fields, shown as the grey dotted line; this combined-reference-field BCF is not used in the Coma GLF fit and is shown for visualization only. The fiducial $\Phi_{\mathrm{C}}$ agrees well with the completeness and contamination corrected binned data points, while $\Phi_{\mathrm{C}}+\Phi_{\mathrm{BG}}$ provides a good description of the completeness corrected binned counts. We therefore adopt the fiducial GLF shown in Figure~\ref{fig:LF} as our final estimate of the Coma cluster GLF. 

\subsection{Comparison with Other Works on the Coma Cluster GLF}
\begin{figure*}[ht]
    \begin{center}
        \includegraphics[width=\textwidth]{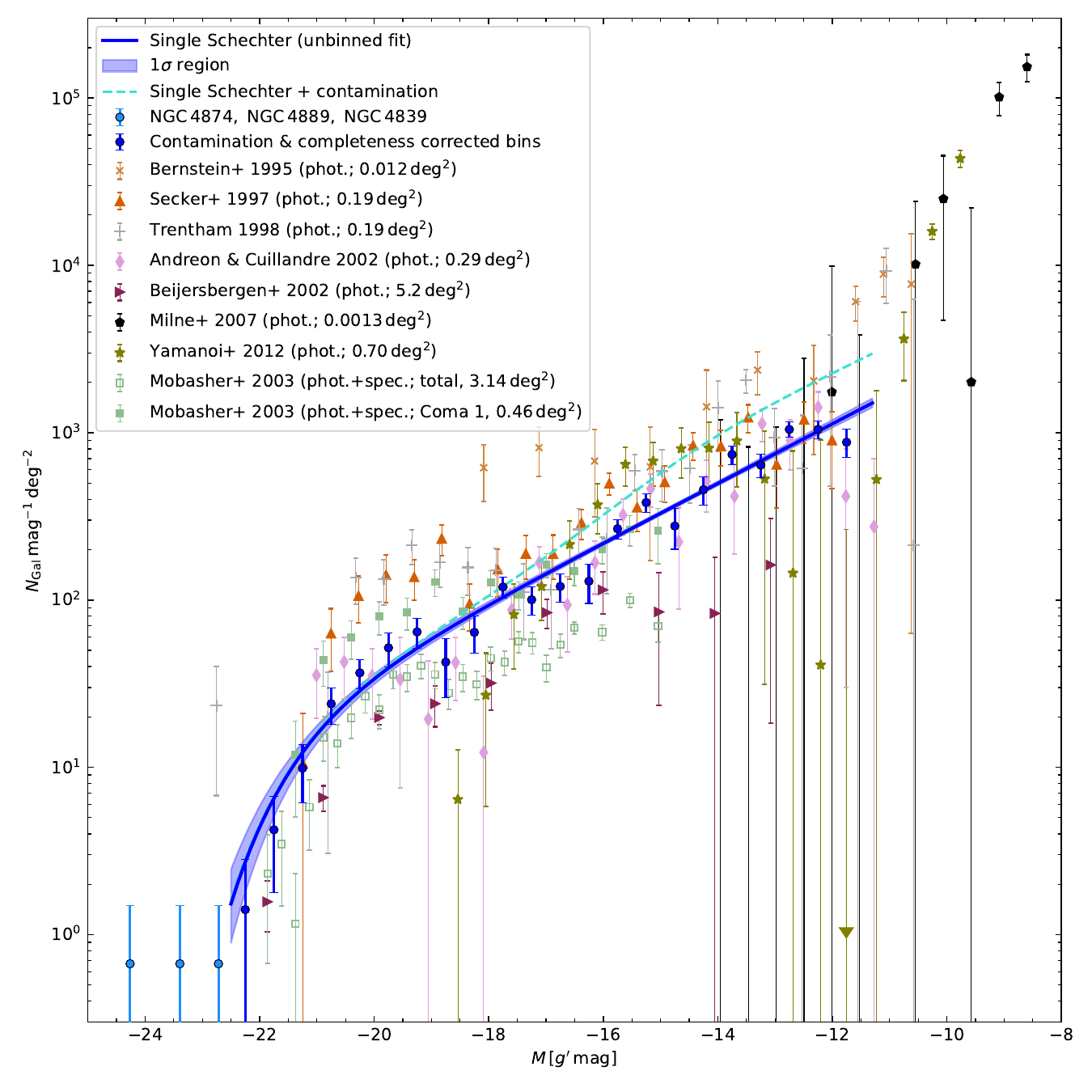}
        \caption{Comparison with GLF of the Coma cluster from the literature. Our best-fit GLF of the Coma cluster is visualized as a blue solid line with its $1\sigma$ uncertainty interval in shaded blue. The completeness and contamination corrected bin counts representing the galaxies used for the unbinned GLF fitting procedure are shown in blue, and the three brightest galaxies, NGC 4889, NGC 4874, and NGC 4839, which were not included for the fit, are shown in light blue. The observed GLF from \cite{Bernstein1995} is shown in brown, those from \cite{Secker1997} in orange, those from \cite{Trentham1998} in dark grey, those from \cite{Andreon2002} in pink, those from \cite{Beijersbergen2002} in dark red, those from \cite{Milne2007} in black, those from \cite{Yamanoi2012} in olive, and the total and Coma~1 GLFs from \cite{Mobasher2003} are shown as open and filled light-green squares, respectively. Note that one data point from \cite{Yamanoi2012} is outside their plotting range and was therefore not extracted by us. We indicate this point with an arrow marking its maximum possible value.
        The data points of the three brightest cluster galaxies from our sample are not binned but shown at their actual positions, normalized by the total Coma footprint of $1.49\,\mathrm{deg}^2$.}
        \label{fig:LFliterature}
    \end{center}
\end{figure*}
In Figure \ref{fig:LFliterature}, we compare our derived Coma cluster GLF to previous GLF measurements in the Coma cluster. Direct comparisons of different surveys are problematic, as the shape and normalization of the GLF depend on the observed filter band \citep{Beijersbergen2002}, but also on the survey region \citep{Adami2007}, e.g., the galaxy number density in the cluster center is inherently higher than in the outer region. We designed our survey to cover a large region ($\approx1.5\,\mathrm{deg^2}$) around the Coma cluster center and measured the structural parameters of all galaxies spanning the range from the BCG to dwarf galaxies ($-24.5\,g'\,\mathrm{mag} \lessapprox M\lessapprox-11.3\,g'\,\mathrm{mag}$) and containing a large variety of dwarf galaxies from compact dwarf galaxies to UDGs \citep{Zoellerinprep}. This allows us to derive a representative GLF that also probes the faint end well.

To enable a consistent comparison, we homogenize the literature measurements before comparing them with our GLF. Table \ref{tab:literature_glf_conventions} summarizes the relevant conventions of each study, including the adopted magnitude and photometric system, distance modulus, survey area, and the form and units in which the GLF is reported. Table \ref{tab:literature_glf_conversions} lists the corresponding transformations used to convert these measurements to our $g'$-band system and to surface densities in units of deg$^{-2}$ mag$^{-1}$ at our adopted distance modulus of $\mu_0=34.97$. These transformations account, where required, for differences in magnitude system, bin width, survey area, logarithmic versus linear representation, and the change in differential number density associated with the filter transformation. We do not attempt to correct for differences in the region of the Coma cluster sampled by the individual surveys.

\begin{deluxetable*}{lllclll}[ht]
\tabletypesize{\scriptsize}
\tablewidth{0pt}
\tablecaption{Published Conventions of the Comparison Coma Cluster Luminosity Functions
\label{tab:literature_glf_conventions}}
\tablehead{
\colhead{Reference} &
\colhead{Source} &
\colhead{Magnitude} &
\colhead{System} &
\colhead{$\mu_{\rm lit}$} &
\colhead{Area} &
\colhead{Published GLF quantity ($Q_{\rm lit}$)}
}
\startdata
\cite{Yamanoi2012} &
Figure 13 &
$M_R$ &
AB &
35.00 &
$0.698\,{\rm deg}^{2}$ &
$\Phi\,[{\rm deg}^{-2}(0.5\,{\rm mag})^{-1}]$ \\[2pt]
\cite{Trentham1998} &
Table 5 &
$M_R$ &
Vega &
34.83 &
$0.188\,{\rm deg}^{2}$ &
$\log_{10}\!\left(\Phi\,[{\rm deg}^{-2}(0.5\,{\rm mag})^{-1}]\right)$ \\[2pt]
\cite{Bernstein1995} &
Figure 5 &
$M_R$ &
Vega &
34.90 &
$0.0117\,{\rm deg}^{2}$ &
$N\,[{\rm mag}^{-1}]$ \\[2pt]
\cite{Mobasher2003}, total &
Table 5 &
$M_R$ &
Vega &
35.13 &
$3.14\,{\rm deg}^{2}$ &
$\Phi\,[{\rm Mpc}^{-2}\,{\rm mag}^{-1}]$ \\[2pt]
\cite{Mobasher2003}, Coma~1 &
Table 3 &
$M_R$ &
Vega &
35.13 &
$0.459\,{\rm deg}^{2}$ &
$\Phi\,[{\rm Mpc}^{-2}\,{\rm mag}^{-1}]$ \\[2pt]
\cite{Andreon2002} &
Figure 3 &
$M_R$ &
Vega &
35.10 &
$0.29\,{\rm deg}^{2}$ &
$N\,[{\rm mag}^{-1}]$ \\[2pt]
\cite{Secker1997} &
Figure 9 &
$m_R$ &
Vega &
\ldots &
$0.194\,{\rm deg}^{2}$  &
$\log_{10}\!\left(N\,[(0.5\,{\rm mag})^{-1}]\right)$ \\[2pt]
\cite{Milne2007} &
Table 4 &
$m_R$ &
Vega &
\ldots &
$0.00128\,{\rm deg}^{2}$ &
$N\,[(0.5\,{\rm mag})^{-1}]$ \\[2pt]
\cite{Beijersbergen2002} &
Figure 9 &
$M_r-5\log h_{100}$ &
Vega\tablenotemark{a} &
34.20\tablenotemark{b} &
$5.2\,{\rm deg}^{2}$  &
$\log_{10}\!\left(\Phi\,[h_{100}^{2}\,{\rm Mpc}^{-2}\,{\rm mag}^{-1}]\right)$ \\
\enddata
\tablenotetext{a}{The AB/Vega convention is not stated explicitly. Because the photometry was
calibrated against Landolt standards, and the authors converted SDSS magnitudes to their own photometric system for comparison, we treat the published $r$ magnitudes as Vega-based.} 
\tablenotetext{b}{Effective distance modulus for $h_{100}=1$, inferred from
the relation between apparent and absolute magnitude adopted by
Beijersbergen et al.\ (2002).}
\tablecomments{Uppercase $R$ denotes Johnson--Cousins/Bessel--Cousins $R$,
whereas lowercase $r$ denotes Sloan $r$. Absolute and apparent magnitudes are
denoted by uppercase $M$ and lowercase $m$, respectively.  Here, $N$ denotes
counts over the quoted survey area, whereas $\Phi$ denotes a surface-density
luminosity function.}
\end{deluxetable*}

\begin{deluxetable*}{lll}[ht]
\tabletypesize{\scriptsize}
\tablewidth{0pt}
\tablecaption{Conversions of the Comparison Luminosity Functions to the System of This Work
\label{tab:literature_glf_conversions}}
\tablehead{
\colhead{Reference} &
\colhead{Magnitude conversion} &
\colhead{Surface-density conversion}
}
\startdata
\cite{Yamanoi2012} &
$M_{g'}=\mathcal{T}_R(M_R-0.21+\Delta\mu)$ &
$\Phi_{g'}=\mathcal{J}_R(Q_{\rm lit}/0.5)$ \\
\cite{Trentham1998} &
$M_{g'}=\mathcal{T}_R(M_R+\Delta\mu)$ &
$\Phi_{g'}=\mathcal{J}_R(10^{Q_{\rm lit}}/0.5)$ \\
\cite{Bernstein1995} &
$M_{g'}=\mathcal{T}_R(M_R+\Delta\mu)$ &
$\Phi_{g'}=\mathcal{J}_R(Q_{\rm lit}/A)$ \\
\cite{Mobasher2003}, total &
$M_{g'}=\mathcal{T}_R(M_R+\Delta\mu)$ &
$\Phi_{g'}=\mathcal{J}_R(Q_{\rm lit}(1.77)^2)$ \\
\cite{Mobasher2003}, Coma~1 &
$M_{g'}=\mathcal{T}_R(M_R+\Delta\mu)$ &
$\Phi_{g'}=\mathcal{J}_R(Q_{\rm lit}(1.77)^2)$ \\
\cite{Andreon2002} &
$M_{g'}=\mathcal{T}_R(M_R+\Delta\mu)$ &
$\Phi_{g'}=\mathcal{J}_R(Q_{\rm lit}/A)$ \\
\cite{Secker1997} &
$M_{g'}=\mathcal{T}_R(m_R-\mu_0)$ &
$\Phi_{g'}=\mathcal{J}_R(10^{Q_{\rm lit}}/(0.5A))$ \\
\cite{Milne2007} &
$M_{g'}=\mathcal{T}_R(m_R-\mu_0)$ &
$\Phi_{g'}=\mathcal{J}_R(Q_{\rm lit}/(0.5A))$ \\
\cite{Beijersbergen2002} &
$M_{g'}=\mathcal{T}_r(M_r'+0.16+\mu_{\rm lit})-\mu_0$ &
$\Phi_{g'}=\mathcal{J}_r(10^{Q_{\rm lit}}(1.22)^2)$ \\
\enddata
\tablecomments{
All measurements are converted to the distance modulus adopted in this work, $\mu_0=34.97$, with $\Delta\mu=\mu_{\rm lit}-\mu_0$. 
Here, $Q_{\rm lit}$ denotes the numerical GLF quantity reported by each study, as specified in Table~\ref{tab:literature_glf_conventions}.
The area $A$ denotes the survey area in deg$^2$ listed in
Table~\ref{tab:literature_glf_conventions}.
The operator $\mathcal{T}_R$ denotes the adopted $R$-to-$g'$ magnitude transformation given by Equation~\ref{eq:Rtog}, while $\mathcal{T}_r$ denotes the corresponding $r$-to-$g'$ transformation derived from the Coma cluster red sequence in Equation~\ref{eq:redsequence}.
The operators $\mathcal{J}_R$ and $\mathcal{J}_r$ account for the change in
differential number density under the corresponding magnitude transformations, such that $\mathcal{J}_R(\Phi_R)=\Phi_R/0.9736$ and $\mathcal{J}_r(\Phi_r)=1.024\,\Phi_r$.
For \cite{Yamanoi2012}, we first convert the published AB magnitudes using $R_{\rm Vega}=R_{\rm AB}-0.21$ \citep{Blanton2007}.
For \cite{Mobasher2003}, the published GLFs are given in
Mpc$^{-2}$ mag$^{-1}$; their adopted angular scale,
$1^\circ=1.77\,{\rm Mpc}$, converts these to
deg$^{-2}$ mag$^{-1}$ through multiplication by $(1.77)^2$.
For \cite{Beijersbergen2002}, we define $M_r'\equiv M_r-5\log h_{100}$ and adopt $r_{\rm AB}=r_{\rm Vega}+0.16$ \citep{Blanton2007}. Their published absolute magnitudes are first converted to apparent $r$-band AB magnitudes using $\mu_{\rm lit}=34.20$, after which the $r$-to-$g'$ transformation is applied; the resulting apparent $g'$ magnitudes are then converted to absolute magnitudes using $\mu_0$. Their published GLF is given in $h_{100}^{2}\,{\rm Mpc}^{-2}\,{\rm mag}^{-1}$; using their adopted angular scale, $1^\circ=1.22h_{100}^{-1}\,{\rm Mpc}$, converts it to deg$^{-2}$ mag$^{-1}$ through multiplication by $(1.22)^2$. 
}
\end{deluxetable*}

As most of the literature GLFs were derived using the $R$ band, we stick for consistency between the literature datasets to their $R$ band GLFs. To convert these measurements to the $g'$ band, we use $R_C-r'=-0.138-0.131(g'-r')$ \citep{Tonry2012} for $R_C$ in Vega magnitudes, and for surveys measured in $R$ band AB magnitudes, we use $m_{AB}-m_{Vega}=0.21$ \citep{Blanton2007} for the conversion with $M_{R_C}$ in the following equations referring to Vega magnitudes:

\begin{align}
    M_{R_c}-M_{r'}&=-0.138-0.131(g'-r')\nonumber\\
                    &-(g'-r')+(M_{g'}-M_{r'})\\
    M_{R_c}&=-0.138-1.131(g'-r')+M_{g'}\\
    M_{g'}&=M_{R_C}+0.138+1.131(g'-r')
\end{align}

Given the red sequence of the Coma cluster (Equation \ref{eq:redsequence}) and distance modulus ($m_{g'}-M_{g'}=34.97$), we approximate the conversion from the $R$ band to the $g'$ band as:

\begin{align}
    M_{g'}&=M_{R_C}+0.138+1.131(-0.024 m_{g'} + 1.06)\\
        &=M_{R_C}-0.0271(m_{g'}-M_{g'}+M_{g'})+1.337\\
        &= M_{R_C}-0.0271M_{g'}+0.3877\\
     M_{g'}&\approx0.974M_{R_C}+0.378
     \label{eq:Rtog}
\end{align}

We stress here that the comparison with the literature should be considered with caution, especially keeping in mind that the GLF can vary from filter to filter and from field to field \citep{Beijersbergen2002,Adami2007}.

Because our $R$-to-$g'$ transformation depends partly on the $g'-r'$ color derived from the Coma cluster red sequence, it is not a simple constant magnitude offset. The inclination of the red sequence implies that fainter galaxies are systematically bluer, which can in principle change the faint-end slope of the GLF when transforming between filters. For our linear transformation, $M_{g'}=0.974M_R+0.378$, the corresponding Schechter faint-end slopes are related by
\begin{equation}
    \alpha_{g'}=\frac{\alpha_R+1}{0.974}-1,
    \label{eq:alpha_conversion}
\end{equation}
as derived in Appendix~\ref{app:alpha_conversion}. For the range of published $R$-band slopes considered here, $-1.7\lesssim\alpha_R\lesssim-1.18$, this correction changes $\alpha$ by only $\approx0.005$--$0.019$. We therefore neglect this small correction when quoting the published faint-end slopes, particularly because differences caused by the different regions of the Coma cluster sampled by the individual surveys are likely substantially larger.

The \cite{Mobasher2003} measurements are based on a combined photometric and spectroscopic survey, which allows them to estimate cluster membership without relying solely on statistical background subtraction. We show both their total Coma cluster GLF (open light-green squares) and the GLF of their central Coma~1 field (filled light-green squares). For Coma~1, we use the measurements derived from their combined Deep and Shallow Spectroscopic Surveys (DSS+SSS), rather than the extended-SSS measurements used to constrain the bright end. The Coma~1 field covers only $0.46\,\mathrm{deg^2}$ and is centered on the dense cluster core. Its surface densities are therefore significantly higher than ours at bright and intermediate magnitudes, while the measurements become consistent with our GLF toward the faint end. Their total GLF covers a substantially larger region of $3.14\,\mathrm{deg^2}$ and generally lies below our measurement at bright and intermediate magnitudes. Our $1.49\,\mathrm{deg^2}$ footprint lies between these two spatial scales and includes the Coma~1 region while extending farther into lower-density regions. It is therefore consistent that our surface densities lie between the Coma~1 and total \cite{Mobasher2003} measurements over much of the bright-to-intermediate magnitude range.
A caveat is that their total GLF combines the faint end measured in the deeper DSS+SSS fields with the bright end from the shallower extended SSS, after normalizing the two over their common magnitude range. This effectively assumes that the LF shape measured in the deeper fields is representative of the larger surveyed region, which \cite{Mobasher2003} note may not hold if the GLF varies with local density. For their total GLF, they report a relatively shallow faint-end slope of $\alpha=-1.18^{+0.04}_{-0.02}$. Their GLF reaches only to approximately $M\approx-15\,g'\,\mathrm{mag}$ when converted to our magnitude system, and therefore does not probe the substantially fainter dwarf-galaxy regime covered by our sample. A direct comparison of the fitted faint-end slopes is consequently complicated by the different magnitude ranges, in addition to the different spatial coverage of the cluster.

The \cite{Beijersbergen2002} sample (dark-red triangles) covers $5.2\,\mathrm{deg^2}$, making it the widest-area literature measurement included in our comparison and sampling the Coma cluster from its dense central region into the outskirts. Their GLF lies below our measurement over most of the common magnitude range. A lower average surface density compared to our measurement is expected from their substantially larger spatial coverage, since a larger fraction of lower-density cluster regions is included. Their GLF does not reach as faint magnitudes as our sample, and the uncertainties of their faintest data points are large, limiting the constraints on the behavior of the GLF at the faint end.

The \cite{Trentham1998} sample (dark grey plus signs) has also measured the Coma GLF ranging from bright galaxies to dwarf galaxies but focused only on a small region ($0.19\,\mathrm{deg^2}$) in the cluster center. At the bright end, our measurement deviates significantly, but at the faint end, their data points and our best-fit GLF match reasonably well within their uncertainties. However, they find a significantly steeper faint-end slope with $\alpha=-1.7$. The bright-end deviation might be due to them observing only the cluster center and hence finding a higher number density.  Similarly, \cite{Bernstein1995} (brown crosses) analyzed a very small region ($0.012\,\mathrm{deg^2}$) in the cluster center, reporting higher number densities. They find a slightly shallower but consistent faint-end slope compared to our measurement, with $\alpha=-1.42\pm0.05$. The \cite{Secker1997} sample (orange-red triangles) likewise covers only a small central region ($0.19\,\mathrm{deg^2}$) and shows systematically higher surface densities than our GLF throughout the overlapping magnitude range. They derive a faint-end slope of $\alpha=-1.41\pm0.05$, although their faintest measurements have relatively large uncertainties. The differences in normalization among these studies and our GLF are therefore likely affected by their substantially smaller and more centrally concentrated survey regions.


\cite{Yamanoi2012} observed very deep data (olive stars) reaching $\approx-10\,\mathrm{R \,mag}$. They are missing the bright end of the GLF as they masked the bright galaxies. At the faint end, they observed a strong decrease in the GLF, followed by a steep increase in their last four data points. 
Their uncertainties in the range where their GLF strongly decreases are large, but they find the same trend for three different fields. We note that \cite{Yamanoi2012} did not apply a comparable color--color preselection before statistical background subtraction. In our sample, the color--color selection reduces the overall contamination by about 80\%, while at the faint end the remaining contamination is still approximately 50\%. Without this preselection, the contaminating population would therefore be about five times larger than the Coma population. In such a regime, even a 10\% uncertainty in the background would translate into an approximately 50\% uncertainty in the recovered Coma signal, compared with about 10\% after preselection. 
This effect becomes particularly important in lower-density cluster regions, where the cluster-to-background contrast is reduced. The decline reported by \cite{Yamanoi2012} is less pronounced in their central Coma~1 field than in the more external fields, although it is present in all three. Differences in cluster-to-background contrast may therefore enhance, but cannot by themselves explain, this feature. Moreover, the larger contaminating population increases the Poisson noise in the cluster field itself, even if the background is constrained using much larger reference fields. The absence of such a preselection may therefore contribute to the large uncertainties in the \cite{Yamanoi2012} GLF and potentially to the apparent decline in this magnitude range, while their analysis with several independent control fields indicates that cosmic variance alone is unlikely to explain it.
Also \cite{Andreon2002} (pink diamonds) report a decrease of the GLF in their last two bins, however, again with large uncertainties. 
\cite{Milne2007} report in this magnitude range mainly negative values for their GLF (black pentagons). However, due to their tiny covered region ($\approx0.0013\,\mathrm{deg^2}$), their uncertainties (we quadratically added their Poisson noise and cosmic variance uncertainties) are very large, and even most of the negative data points are consistent within $1\sigma$ with our measurements. We note that for one of their datapoints, even the upper limit is negative and is therefore not shown.
Our faint end of the GLF overlaps with the range where their GLFs are dropping $(-14\,g'\,\mathrm{mag}\lessapprox M \lessapprox-11.3\,g'\,\mathrm{mag})$. We do not find a significant decrease of the GLF in this regime, neither in the best-fit GLF nor in the individual bins that have a significantly smaller uncertainty than their measurements. We conclude that there is no decline of the GLF and no evidence of two distinct dwarf galaxy components in the GLF as suggested by \citet{Yamanoi2012}, but that the GLF is steadily increasing. \cite{Yamanoi2012} interpreted their strong increase at the faint end as the contribution by unresolved galaxies and partly by globular clusters. We do not include unresolved objects in our catalog but focus on resolved galaxies. Hence, we can not find this upturn by design of our sample but also due to our magnitude limit. Generally, \cite{Bernstein1995}, \cite{Trentham1998}, and \cite{Milne2007} focused on small regions but reached fainter magnitudes than our sample, but at the cost of not getting a large and representative sample, which also leads to larger uncertainties than in our survey. Furthermore, we restricted our GLF measurement to a clean sample with good Sérsic fits and accurate $u'-g'$ and $g'-r'$ colors to also derive structural parameter number densities \citep{Zoellerinprep}. 
This restrictive selection limits the completeness at the very faint end but yields a high-confidence measurement of the structural parameters and colors of galaxies, extending to the faintest galaxies used to determine the GLF. The resulting incompleteness is automatically taken into account by the completeness function (see Section \ref{sec:injrec}).

Our derived $g'$ band GLF function of quenched galaxies in the Coma cluster is based on 
a large covered area ($1.5\,\mathrm{deg^2}$) and spans a wide luminosity range
from the BCG to faint dwarf galaxies ($M\lessapprox-11.3\,g'\,\mathrm{mag}$), providing a high-precision measurement of the GLF through an unbinned single Schechter fit, which is also reflected in the small uncertainties of the binned data points across the entire probed luminosity range. With this, we hope to provide a new benchmark for numerical simulations. Considering the discussion above, we emphasize that for a proper comparison of our GLF with other observations or simulations, a directly matched comparison is crucial. 
The chosen filter band can affect the measured faint-end slope, since the color conversion between different filters is not a constant offset. In particular, the $g'-r'$ color depends on $M$ through the slope of the red sequence (Equation \ref{eq:redsequence}), such that fainter galaxies are bluer. This leads to a slightly steeper faint-end slope in the $g'$ band than in the $r'$ band. As shown above, however, converting the literature $R$ band slopes to the $g'$ band changes $\alpha$ by only $\approx0.005$--$0.019$. This is small compared to the much larger variation among the literature slopes by $\approx0.5$, which is presumably caused by differences in the probed area, limiting magnitude, and other potential systematic effects. For more closely matched comparisons, however, such a color-dependent slope conversion may become relevant. In particular, choosing a comparable cluster region is crucial, as the galaxy number density and the GLF vary strongly with cluster-centric radius.

\subsection{Comparison with the Coma Cluster Counterpart in the SLOW Simulations}
We now compare our observed GLF with predictions from cosmological simulations. Such a comparison can provide useful context for interpreting the shape and normalization of the GLF, but it is not straightforward. Simulations usually do not provide a direct analogue of the observed cluster, and GLFs or GSMFs are typically measured per volume rather than per projected area, as in observations. For example, \citet{Negri2022} present the GLF of massive clusters in the Cluster-EAGLE simulation within a sphere of the virial radius. They report a significantly flatter faint-end slope in the $g$ band, with $\alpha=-1.32\pm0.01$.

While this selection is physically meaningful, it is not directly comparable to observations, as the distances of individual galaxies cannot be determined accurately enough to apply the same spherical selection cut. 
Furthermore, despite having one of the largest area coverages among deep Coma cluster GLF surveys, our footprint remains significantly smaller than the region enclosed by the virial radius. Full coverage out to the virial radius is observationally challenging for nearby clusters because of their large angular extent and requires exceptionally wide surveys. For a direct comparison of simulations and observations, the simulated GLF therefore has to be restricted to the observed survey area and include the galaxies along the corresponding line of sight.
We therefore compare our GLF to a simulation selected to match our observational conditions as closely as possible, namely by using the same projected region through a constrained simulated counterpart of the Coma cluster.

\begin{figure}[t]
    \includegraphics[width=0.47\textwidth]{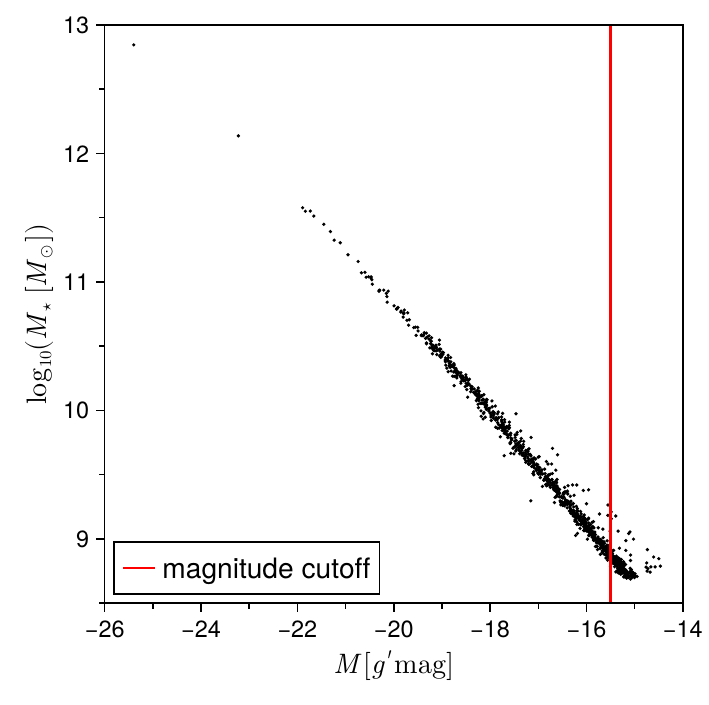}
    \caption{$M_\ast$ -- $M$ scaling for the simulated Coma cluster. The magnitude cutoff at $M=-15.5\,g'\,\mathrm{mag}$ is highlighted by the red vertical line.}
    \label{stelrel}
\end{figure}
\begin{figure}[t]
    \begin{center}
        \includegraphics[width=0.47\textwidth]{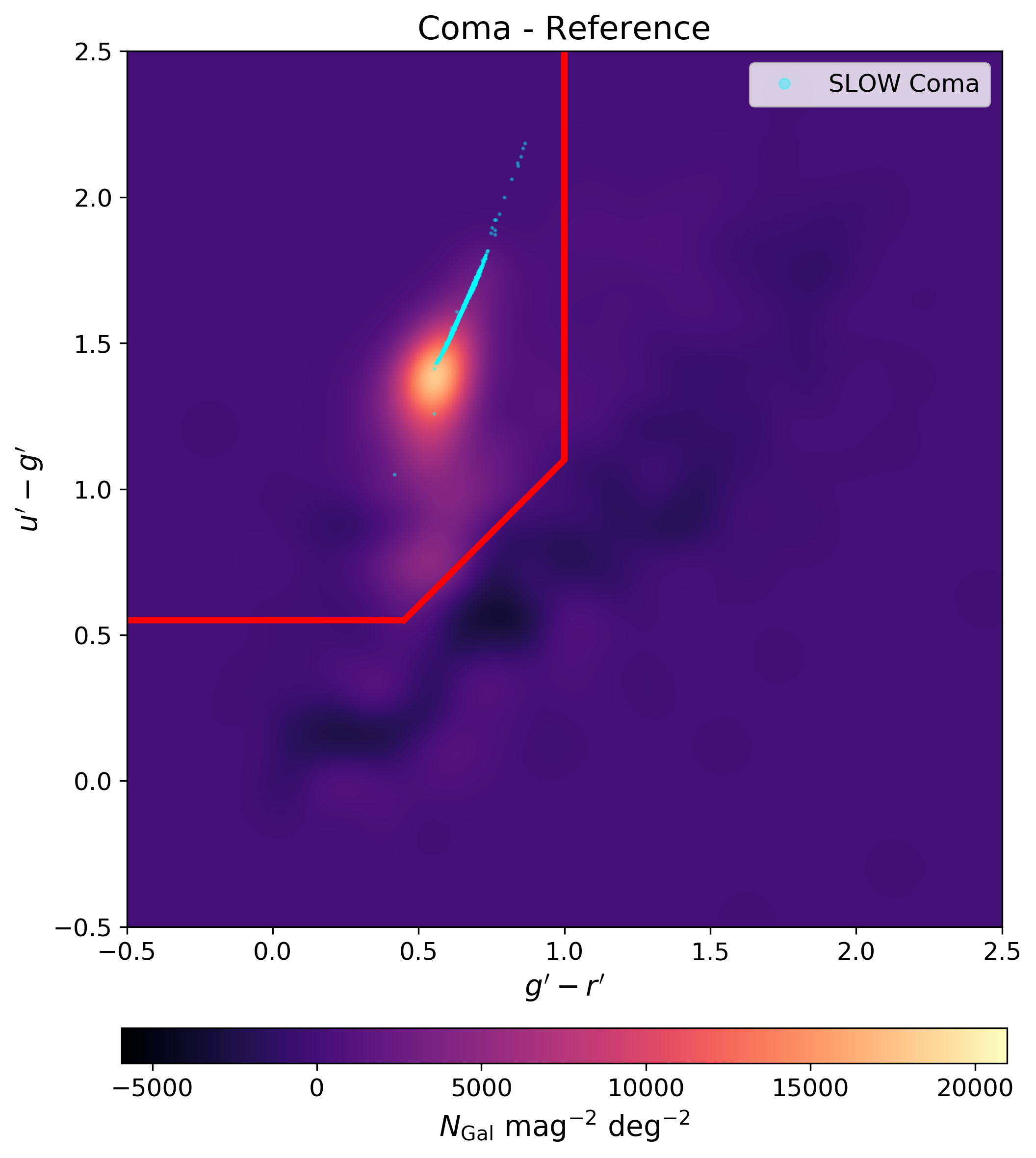}
        \caption{Background-subtracted Coma color–color density distribution for the faint galaxy sample, shown as Coma minus the combined reference fields in $g'-r'$ versus $u'-g'$. The color scale gives the residual galaxy density, while the red line indicates the adopted color-selection boundary. Cyan markers show individual SLOW Coma galaxies.}
        \label{fig:colorcolorSLOW}
    \end{center}
\end{figure}
\begin{figure}[t]
    \begin{center}
        \includegraphics[width=0.47\textwidth]{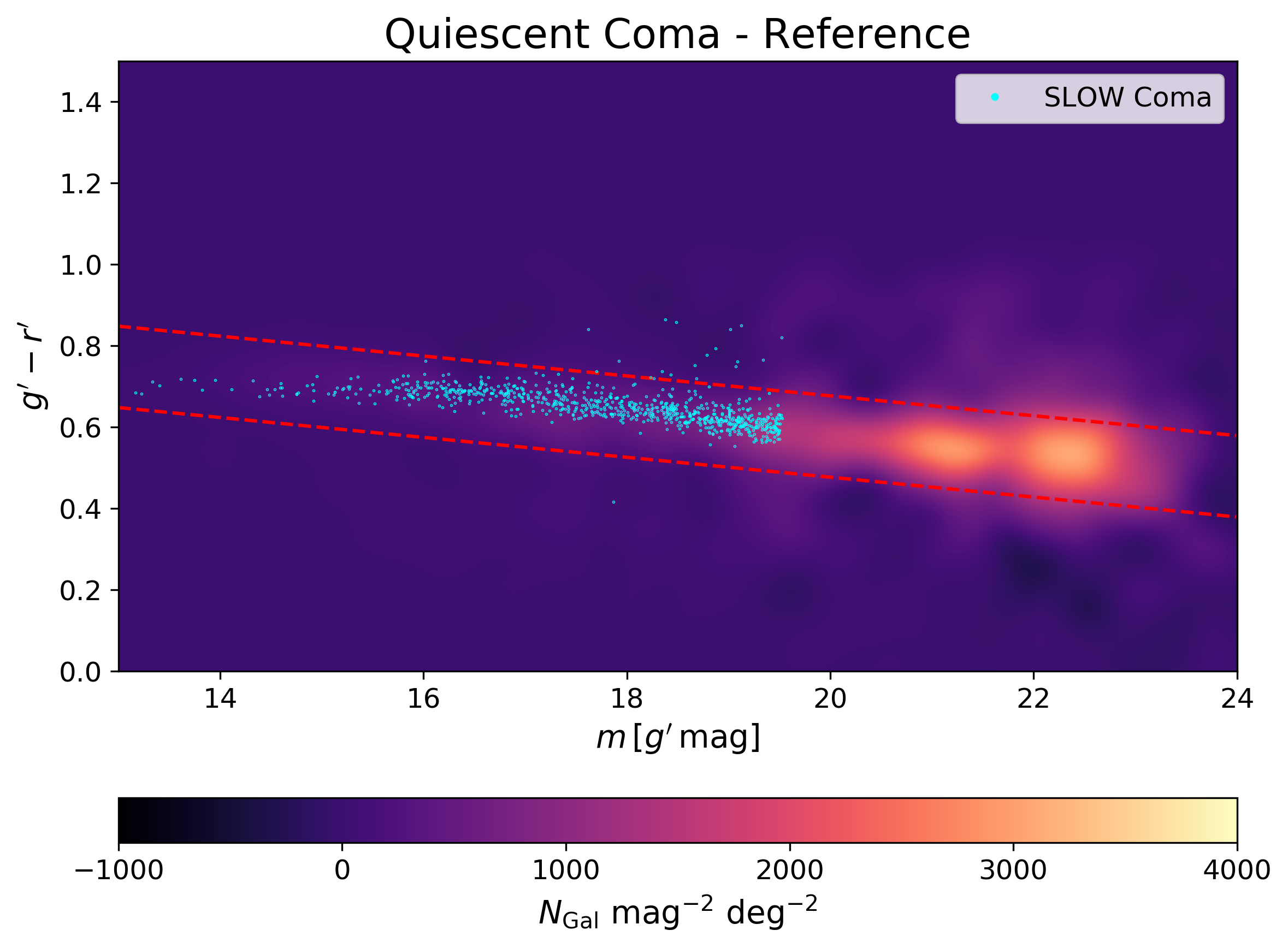}
        \caption{Background-subtracted red-sequence density distribution, shown as quiescent Coma minus the combined reference fields. The density map includes all galaxies with $m<24\,g'\,\mathrm{mag}$ except the three brightest objects. The red dashed lines indicate the adopted intrinsic red-sequence. Cyan markers show individual SLOW Coma galaxies, except the brightest two.}
        \label{fig:redsequenceSLOW}
    \end{center}
\end{figure}

As a direct reference for our results from numerical simulations, we employ a set of constrained zoom-in simulations 
\citep[\textbf{LO}cal \textbf{WE}b \textbf{R}esimulations with \textbf{D}ynamical friction and \textbf{E}xtended bla\textbf{CK} hole \textbf{S}pin model, LOWERDECKS; B. A. Seidel et al., in prep.;][]{sorce2015,sorce2018}
obtained from the \textbf{S}imulating the \textbf{L}Ocal \textbf{W}eb (\textbf{SLOW}) constrained simulations of the Local Universe \citep{Dolag2023,Hernandez2024}.
These zoom-in simulations use a modernized version of the Magneticum model \citep{Dolag2025,steinborn2015}, adding spin-coupled AGN feedback \citep{sala2024,salathesis} and an improved prescription for dynamical friction \citep{damiano2024}. The high resolution zoom-in region ($m_\ast=2.6\times10^6M_\odot$) used in this work is constructed from a stable Lagrangian region \citep{Seidel2025} around the well-constrained Coma\footnote{For an alternative re-simulation of the Coma cluster see \citet{malavasi2023}, based on the same constraints, Nr. 8 from the CLONES set \citep{sorce2015}.} counterpart in the SLOW simulation. We place a mock observer at the optimized observer position from \citet{Dolag2023}. Since the positions of the constrained clusters do not exactly match the positions of their real world counterparts \citep{Hernandez2024,Seidel2025}, we choose the FOV in the simulation in such a way, that the physical size of the observed region ($\mathrm{2.00Mpc \times 2.00Mpc}$) is reproduced.

To ensure completeness for the simulated GLF, we adapt a lower magnitude cutoff of $M=-15.5\,g'\,\mathrm{mag}$ that is set slightly above the stellar mass cut of $5\times10^8M_\odot$ we applied due to the mass resolution of the simulation. This can be seen in Figure \ref{stelrel}. 

All galaxies selected like this from the SLOW simulation fulfill the quiescent galaxy selection adopted for the observation (see Figure \ref{fig:colorcolorSLOW}), and the vast majority follows the observed red sequence and lies within the adopted intrinsic red sequence as shown in Figure \ref{fig:redsequenceSLOW}.

\begin{figure*}[ht]
    \begin{center}
        \includegraphics[width=\textwidth]{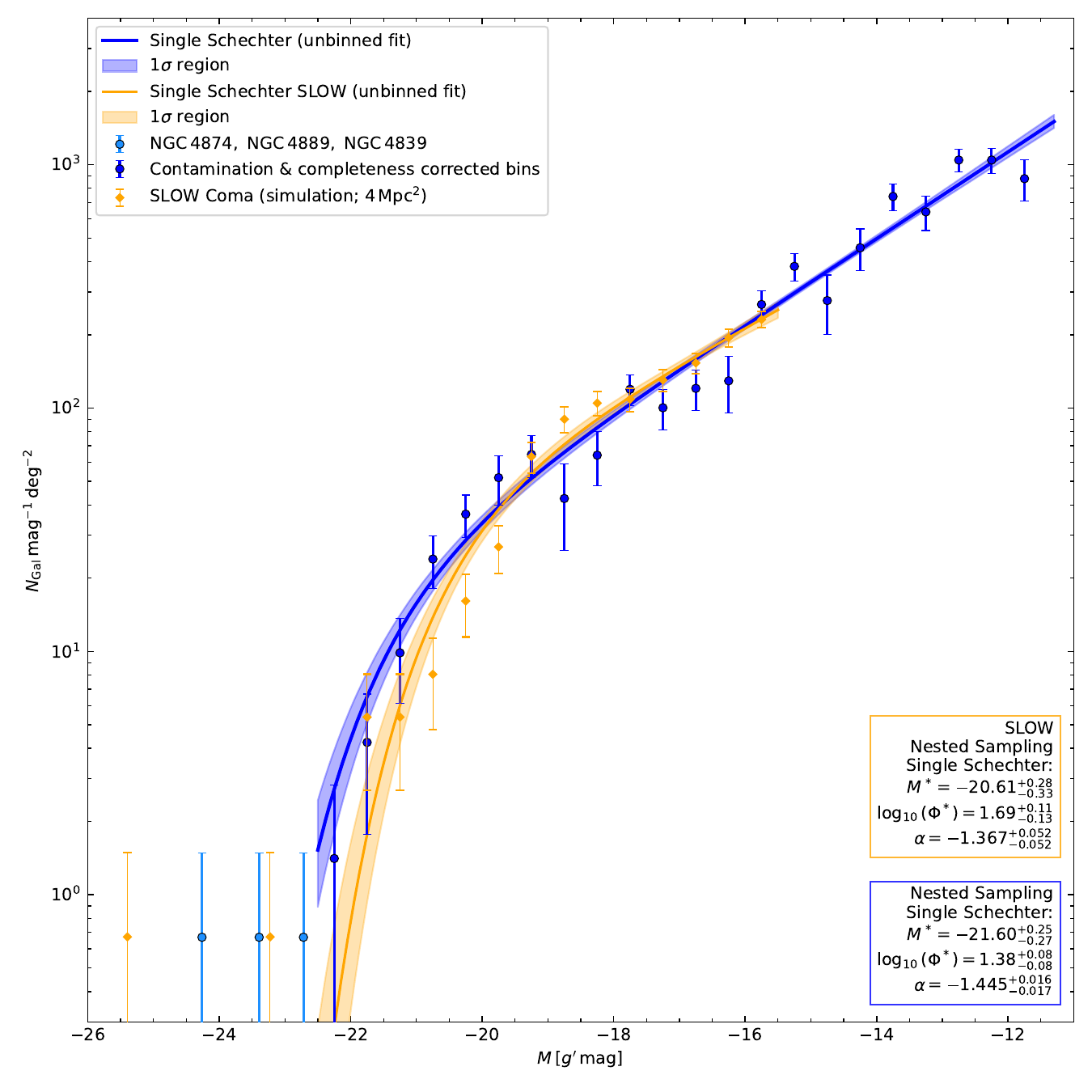}
        \caption{Comparison with the GLF of the Coma cluster counterpart in the SLOW simulation. Our best-fit GLF of the Coma cluster is visualized as a blue solid line with its $1\sigma$ uncertainty interval in shaded blue. The completeness and contamination corrected bin counts representing the galaxies used for the unbinned GLF fitting procedure are shown in blue, and the three brightest galaxies, NGC 4889, NGC 4874, and NGC 4839, which were not included for the fit, are shown in light blue. 
        The simulated GLF from the Coma cluster SLOW simulation \citep{Seidelinprep} is shown in orange. The data points of the three brightest cluster galaxies from our sample and the two brightest from the SLOW simulation are not binned but shown at their actual positions, normalized by the total Coma footprint of $1.49\,\mathrm{deg}^2$.}
        \label{fig:LFSLOW}
    \end{center}
\end{figure*}

The resulting GLF is represented by the orange diamonds in Figure \ref{fig:LFSLOW}, with the error bars given by the Poisson noise in each bin. Overall, there is good agreement between the simulated and observed GLF, with deviations in the BCG luminosities at the high-luminosity end and a slight overabundance of simulated galaxies in the intermediate to low-luminosity range compared to the real Coma cluster. We expect the largest uncertainty factor in the simulated luminosity function to be the subgrid model, which can suffer from over-cooling as well as over-quenching. This can impact the slope of the stellar mass functions \citep[e.g.,][]{Dolag2025}. The over-prediction of galaxy abundance at $\approx -18\,g'\,\mathrm{mag}$ is possibly a result of the stellar winds  and AGN feedback lowering the stellar mass of giant galaxies, in turn also causing the steeper drop-off of the luminosity function at the bright end. Additionally, slight deviations with regard to the dynamical state and assembly history of the Coma counterpart due to uncertainties in the constraints might impact the simulated GLF. We would expect the high-luminosity end to be most sensitive to these large-scale effects, as the mass (and stellar mass) in the BCG and ICL in the cluster can be shown to directly trace its assembly history \citep{kimmig2025}. However, the underlying population appears to be closely matched between simulation and observation, implying CDM simulations to be generally capable of reproducing the galaxy populations of local clusters. 

To compare the luminosity functions more quantitatively, we fit a single Schechter function to the unbinned data points from the SLOW Coma counterpart down to $M<-15.5\,g'\,\mathrm{mag}$ using the same procedure as outlined before for the actual Coma cluster, also excluding the galaxies in the magnitude range of the three brightest cluster galaxies (the 2nd brightest cluster galaxy is in the simulation actually outside of the FoV), but without the need for completeness and contamination correction. We find best-fit parameters of $M^{*}=-20.61^{+0.28}_{-0.33}\,g'\,\mathrm{mag}$, $\log_{10}(\Phi^{*}\,[\mathrm{deg}^{-2}\,\mathrm{mag}^{-1}])=1.69^{+0.11}_{-0.13}$, $\alpha=-1.367^{+0.052}_{-0.052}$.  The simulated faint-end slope is shallower than our observed one, but consistent within less than $2\sigma$. Given the limited magnitude range, it is not possible to determine, whether this shallower slope is caused by the slight bump at intermediate magnitude or reflects a general trend in the simulated galaxies given the overall agreement between the two datasets. Generally, the best-fit full GLF of the observation and the SLOW simulation agree well in the $-20\,g'\,\mathrm{mag} \lesssim M \lesssim -15.5\,g'\,\mathrm{mag}$ range.\\
To fit the observed and simulated GLFs over the same magnitude range and thereby enable an even more direct comparison, we would need either a substantially higher-resolution simulation that extends to the faint end of the observed GLF (which is not feasible with the current generation of zoom-ins) or to restrict the observed GLF to brighter magnitudes. However, limiting the observed sample to galaxies with $M<-15.5\,g'\,\mathrm{mag}$ would require additional reference fields, because the small number of galaxies in this magnitude range prevents the BCF from being robustly constrained using this magnitude range solely.

\section{Summary and Conclusion} \label{sec:summaryconclusion}
We have obtained deep $u'$-, $g'$-, and $r'$ band data for the Coma cluster using the 2.1\,m Fraunhofer Wendelstein Telescope with the final survey region covering $1.5\,\mathrm{deg^2}$ and two reference fields with comparable depth and resolution. The Coma cluster data reaches median $3\sigma$ surface brightness limits in $10\arcsec\times10\arcsec$ boxes of $\mathrm{ (30.0\,u',\,\,29.6\,g',\,\,28.7\,r')\,mag\,arcsec^{-2}}$.

We have created isophote models and measured the structural parameters, $u'-g'$, and $g'-r'$ colors of the three brightest cluster galaxies NGC\,4874 (including the ICL), NGC\,4889, and NGC\,4839 using \verb+isopy+. We subtracted these models from the images to allow an accurate measurement of the fainter galaxies. The parameters of the other 403 bright galaxies with $m\leq18\,g'$\,mag were also measured using isophotal fitting. Again, we subtracted these models to allow for a reliable detection and measurement of the structural parameters of dwarf galaxies even in the vicinity of those bright galaxies. Furthermore, we presented a method to fully automatically create masks and to measure the structural parameters using \verb+GALFIT+ of more than $10^5$ galaxies in the Coma cluster field. Additionally, we obtained $u'-g'$ and $g'-r'$ of those galaxies using elliptical apertures. Afterward, we selected cluster member candidates based on their quiescent sequence membership in the $u'-g'$ vs $g'-r'$ color-color diagram and restricted our sample to galaxies with high-confidence measurements of their structural parameters. In total, our final Coma cluster sample comprises more than 6000 cluster member candidates. 

Furthermore, we identically analyzed the two reference fields to quantify the contamination of our Coma cluster sample. Moreover, we have performed injection-recovery tests for the Coma cluster and the reference fields to derive the completeness functions of our sample selection, also taking into account the dependence of the completeness on $\mu_e$. For both reference fields, we fitted a double-power-law using the unbinned data points employing the dynamic nested sampling package \verb+DYNESTY+ for an analytical description of the contamination, taking into account the completeness function of the respective reference field, also propagating the uncertainties and probing the uncertainty due to cosmic variance. %

Finally, we derived the $g'$ band GLF of the Coma cluster using the unbinned data points of quiescent galaxies down to $M\approx-11.3\,g'\,\mathrm{mag}$, excluding the three brightest cluster galaxies. For this, we fitted a single Schechter function using \verb+DYNESTY+, taking into account the completeness function and the background contamination, and propagated their uncertainties, including the effect of the cosmic variance.  We derived the best-fit single Schechter parameters of $M^{*}=-21.60^{+0.25}_{-0.27}\,g'\,\mathrm{mag}$, $\log_{10}(\Phi^{*}\,[\mathrm{deg}^{-2}\,\mathrm{mag}^{-1}])=1.378^{+0.076}_{-0.081}$, $\alpha=-1.445^{+0.016}_{-0.017}$. 

This work combines deep wide-field imaging with careful cluster member candidate selection and galaxy profile fitting, along with corrections for incompleteness and contamination, to improve constraints on the Coma cluster GLF. With the resulting GLF we hope to provide a new reference measurement for future observational and theoretical studies of galaxy clusters. 
The high precision of our approach is also reflected in the small uncertainties of the binned GLF data points across the full luminosity range. The remaining observational uncertainties are dominated primarily by the limited number of available reference fields, which affects the precision with which background contamination can be statistically corrected. Our measured faint-end slope, $\alpha=-1.445^{+0.016}_{-0.017}$, is steeper than that predicted for massive galaxy clusters by the Cluster-EAGLE simulation \citep{Negri2022}, which finds $\alpha=-1.32\pm0.01$. However, as the differences in the construction of our GLF and this literature-based GLF form the EAGLE simulation do not allow a fair comparison, we constructed the comparison with the Coma cluster counterpart in the SLOW simulation \citep{Seidelinprep} specifically to match the observational conditions and analysis choices of our study as closely as possible. This provides a directly matched comparison between the simulated and observed GLFs.
The SLOW faint-end slope, $\alpha=-1.367^{+0.052}_{-0.052}$, is slightly shallower than our best-fitting value. Nevertheless, the two measurements are consistent within less than $2\sigma$. Moreover, the SLOW GLF agrees well with the observed GLF over the overlapping magnitude range, apart from the few brightest galaxies, indicating that the agreement is not limited to the fitted faint-end slope alone.\\
We note, however, that our observations reach a significantly deeper faint-end limit of $M=-11.3\,g'\,\mathrm{mag}$ compared to the simulation of $M=-15.5\,g'\,\mathrm{mag}$. The comparison, therefore, does not test the simulated GLF down to the full depth reached by the observations. 
Since SLOW is based on a cold dark matter model, this close agreement over the accessible magnitude range indicates that a CDM-based simulation can reproduce both the shape and the overall galaxy abundance of the observed Coma GLF under closely matched conditions. The steep observed faint-end slope is therefore consistent with a CDM framework.\\
Nevertheless, the GLF alone does not allow firm conclusions about the nature of dark matter, as the predicted dwarf-galaxy population also depends on baryonic physics and its implementation in simulations. Moreover, the full GLF should be considered rather than the faint-end slope $\alpha$ alone, since its shape and normalization together determine the distribution and total abundance of dwarf galaxies. Future studies should therefore compare observations with simulations based on different dark matter models under closely matched conditions. Ideally, future simulations should also reach substantially fainter magnitudes, where differences between the GLFs predicted by different dark matter models may become detectable.\\
Despite these limitations, we attempted to provide a robust observational benchmark for future simulations to probe baryonic feedback and possible signatures of different dark matter scenarios in galaxy clusters.

\begin{acknowledgments}

We thank the anonymous referee for the detailed and constructive comments throughout the review process, which helped to significantly improve the manuscript.

The Wendelstein 2.1m telescope project was funded by the Bavarian government and by the German Federal government through a common funding process. Part of the 2.1m instrumentation, including some of the upgrades for the infrastructure, were funded by the Cluster of Excellence “Origin of the Universe” of the German Science Foundation DFG.

This work would not have been practical without extensive use of NASA's Astrophysics Data System Bibliographic Services and the SIMBAD database, operated at CDS, Strasbourg, France.

We also used the image display tool SAOImage DS9, developed by the Smithsonian Astrophysical Observatory, and the image display tool Fitsedit, developed by Johannes Koppenhoefer.

We thank Lucas Valenzuela for providing the CosmoLuminosities julia package for the simulation post-processing.

The authors used OpenAI’s ChatGPT (various GPT models, from GPT-4 through GPT-5.6 Sol) and Grammarly to assist with grammar and language polishing of the manuscript. ChatGPT was also used for prototyping parts of the analysis code and for debugging purposes. All code and text generated with these tools were thoroughly reviewed, verified, and adapted by the authors.

The Pan-STARRS1 Surveys (PS1) and the PS1 public science archive have been made possible through contributions by the Institute for Astronomy, the University of Hawaii, the Pan-STARRS Project Office, the Max-Planck Society and its participating institutes, the Max Planck Institute for Astronomy, Heidelberg and the Max Planck Institute for Extraterrestrial Physics, Garching, The Johns Hopkins University, Durham University, the University of Edinburgh, the Queen's University Belfast, the Harvard-Smithsonian Center for Astrophysics, the Las Cumbres Observatory Global Telescope Network Incorporated, the National Central University of Taiwan, the Space Telescope Science Institute, the National Aeronautics and Space Administration under Grant No. NNX08AR22G issued through the Planetary Science Division of the NASA Science Mission Directorate, the National Science Foundation Grant No. AST–1238877, the University of Maryland, Eotvos Lorand University (ELTE), the Los Alamos National Laboratory, and the Gordon and Betty Moore Foundation.\\
Funding for SDSS-III has been provided by the Alfred P. Sloan Foundation, the Participating Institutions, the National Science Foundation, and the U.S. Department of Energy Office of Science. The SDSS-III web site is \url{http://www.sdss3.org/}. \\
SDSS-III is managed by the Astrophysical Research Consortium for the Participating Institutions of the SDSS-III Collaboration including the University of Arizona, the Brazilian Participation Group, Brookhaven National Laboratory, Carnegie Mellon University, University of Florida, the French Participation Group, the German Participation Group, Harvard University, the Instituto de Astrofisica de Canarias, the Michigan State/Notre Dame/JINA Participation Group, Johns Hopkins University, Lawrence Berkeley National Laboratory, Max Planck Institute for Astrophysics, Max Planck Institute for Extraterrestrial Physics, New Mexico State University, New York University, Ohio State University, Pennsylvania State University, University of Portsmouth, Princeton University, the Spanish Participation Group, University of Tokyo, University of Utah, Vanderbilt University, University of Virginia, University of Washington, and Yale University.\\
This work has made use of data from the European Space Agency (ESA) mission Gaia (\url{https://www.cosmos.esa.int/gaia}), processed by the Gaia Data Processing and Analysis Consortium (DPAC, \url{https://www.cosmos.esa.int/web/gaia/dpac/consortium}). Funding for the DPAC has been provided by national institutions, in particular the institutions participating in the Gaia Multilateral Agreement.

\end{acknowledgments}

\software{ Astropy \citep{astropy2018}, Photutils v1.3.0 and v1.11.0 \citep{photutils}, DYNESTY v2.1.5 \citep{Speagle2020}, NumPy \citep{numpy}, SciPy \citep{scipy}, Matplotlib \citep{matplotlib}, GALFIT v3.0.5 \citep{galfit}, SExtractor v2.19.5 \citep{Sextractor}, SCAMP v2.10.0 \citep{scamp}, SWarp v2.38.0 \citep{swarp}, PSFEx v3.17.1 \citep{PSFEx}, SAOImage DS9 v8.2.1.}

\bibliography{ComaLFpaper}{}
\bibliographystyle{aasjournalv7.1}

\appendix\section{Zero-point Color Terms}
\label{appendix:colorterm}

We determine the color correction terms in the $g'$ and $r'$ bands relative to the zero-point calibration of the Coma cluster stack. The zero-point calibration itself was restricted to stars with $0.25 < g'-r' < 1.0$. For the $u'$ band, we do not measure a color term because point sources that are sufficiently bright in SDSS are saturated in our $u'$-band images. However, this does not affect our color selection, which is based only on WWFI colors.

For the point sources, we measured the zero-point residual $\Delta ZP$ relative to the Pan-STARRS reference catalog as a function of the $g'-r'$ color measured from our zero-point-calibrated images. We restricted the fit to unsaturated, well-measured sources with $14 < g' < 22$ in the Pan-STARRS catalog, $|\Delta ZP_{g'}| < 0.2$, and $0.25 < g'-r' < 1.0$.

\begin{figure}[ht]
    \centering
    \includegraphics[width=\linewidth]{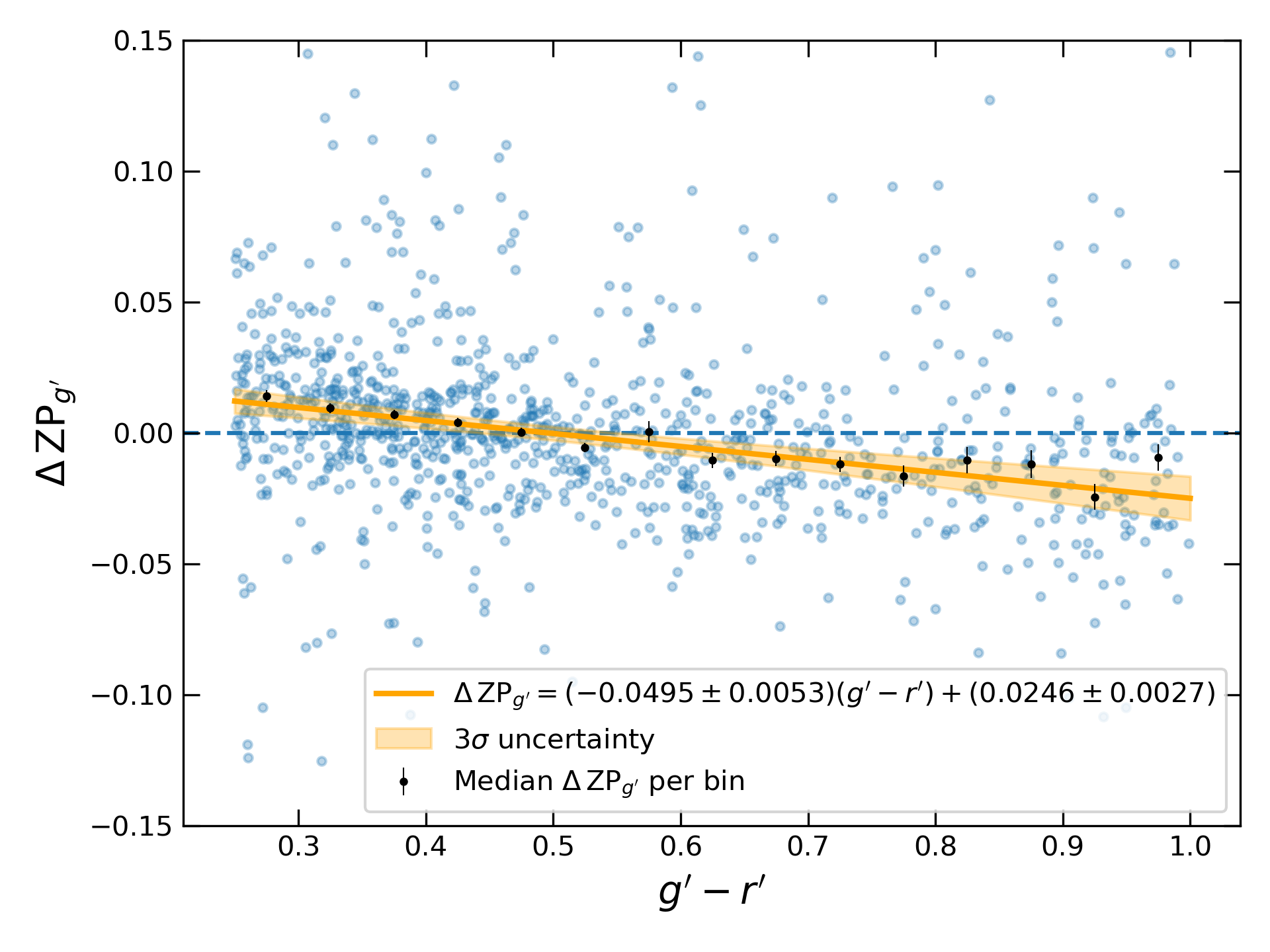}\\
    \includegraphics[width=\linewidth]{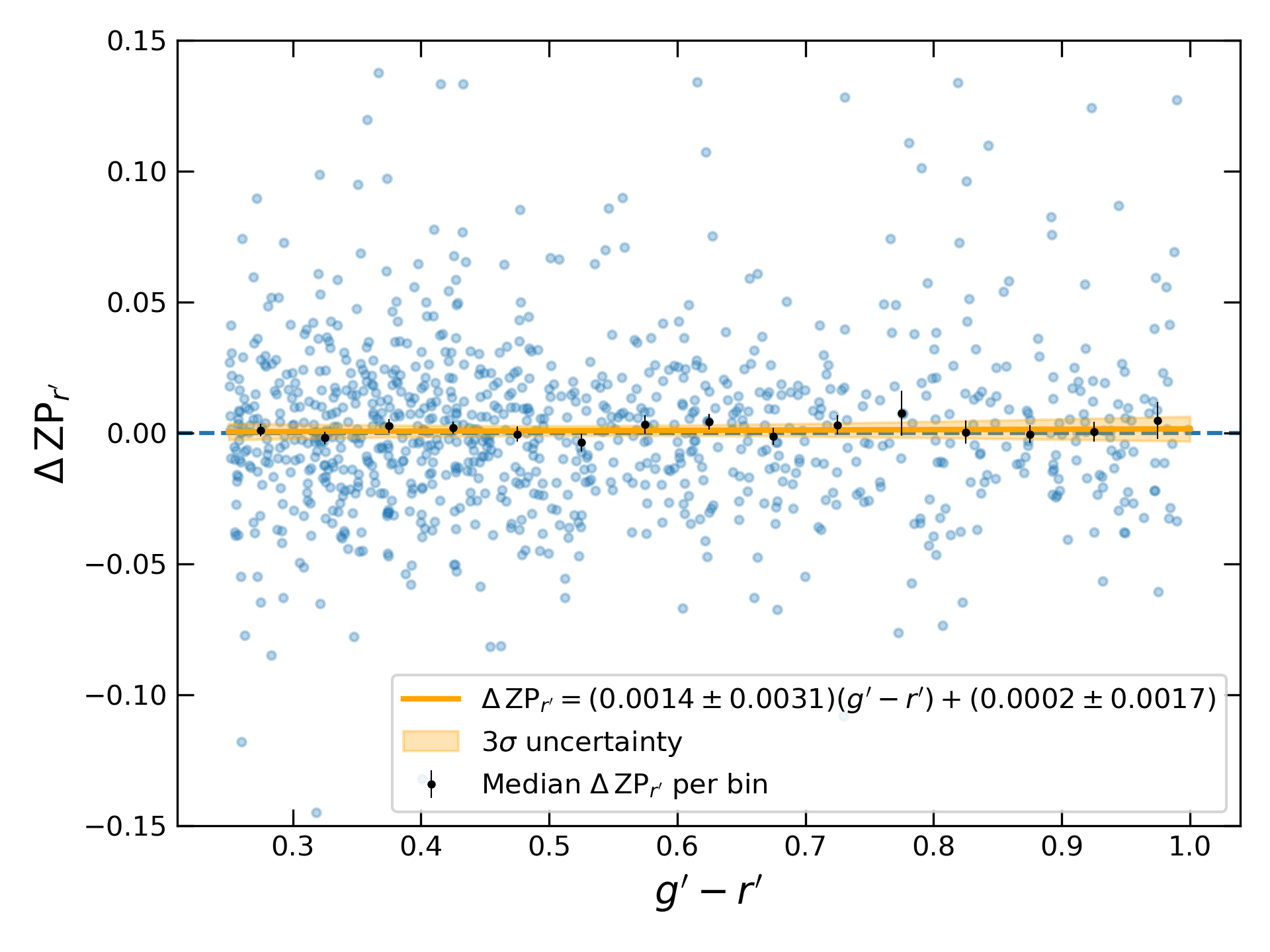}
    \caption{Zero-point residuals as a function of $g'-r'$ color for the $g'$ band (top) and $r'$ band (bottom). The orange lines show the weighted linear fits to the binned median residuals, and the shaded regions indicate the $3\sigma$ uncertainties.}
    \label{fig:colorterms}
\end{figure}

The data were binned in color with a bin width of $0.05$ mag. In each bin, we used the median $\Delta ZP$ and fitted a weighted linear relation, $\Delta ZP = a(g'-r') + b$. The resulting color-term fits for the $g'$ and $r'$ bands are shown in Figure \ref{fig:colorterms}, and the best-fit values and uncertainties are reported in Table \ref{tab:colorterms}.

\begin{deluxetable}{c|c|c}[ht]
    \tabletypesize{\small}
    \tablecaption{Color Correction Terms}
    \label{tab:colorterms}
    \tablehead{
        \colhead{}& \colhead{a} & \colhead{b}
    }
    \startdata
        $\Delta ZP_{g^\prime}$ & $-0.0495\pm0.0053$& $0.0246\pm0.0027$\\
        $\Delta ZP_{r^\prime}$ & $-0.0014\pm0.0031$ & $0.0002\pm0.0017$
    \enddata
\end{deluxetable}

Based on these results, we assume the color term in the $r'$ band to be negligible and adopt the empirical color-term correction only for the $g'$-band in the catalog processing using the measured $ g'-r'$ colors of our galaxy sample.

\section{Conversion of the Schechter Faint-End Slope}
\label{app:alpha_conversion}

At faint magnitudes, where the exponential cutoff of the Schechter function is
negligible, the luminosity function in the $R$ band scales as
\begin{equation}
    \Phi_R(M_R)\propto
    10^{-0.4(\alpha_R+1)M_R}.
\end{equation}
For a linear magnitude transformation
\begin{equation}
    M_{g'}=aM_R+b,
\end{equation}
the inverse relation is
\begin{equation}
    M_R=\frac{M_{g'}-b}{a}.
\end{equation}
The number of galaxies is conserved under this change of magnitude variable,
such that
\begin{equation}
    \Phi_{g'}(M_{g'})\,dM_{g'}
    =
    \Phi_R(M_R)\,dM_R.
\end{equation}
Since
\begin{equation}
    \frac{dM_R}{dM_{g'}}=\frac{1}{a},
\end{equation}
the transformed luminosity function is
\begin{equation}
    \Phi_{g'}(M_{g'})
    =
    \frac{1}{a}\,
    \Phi_R(M_R).
\end{equation}
Substituting $M_R=(M_{g'}-b)/a$ into the faint-end form gives
\begin{align}
    \Phi_{g'}(M_{g'})
    &\propto
    \frac{1}{a}
    10^{-0.4(\alpha_R+1)(M_{g'}-b)/a}
    \nonumber\\
    &=
    \frac{1}{a}
    10^{0.4(\alpha_R+1)b/a}
    10^{-0.4[(\alpha_R+1)/a]M_{g'}}.
\end{align}
The first two factors are independent of $M_{g'}$ and therefore only change the normalization of the luminosity function. The remaining magnitude-dependent term determines the faint-end slope. Comparing it with the corresponding $g'$-band form,
\begin{equation}
    \Phi_{g'}(M_{g'})
    \propto
    10^{-0.4(\alpha_{g'}+1)M_{g'}},
\end{equation}
yields
\begin{equation}
    \alpha_{g'}+1
    =
    \frac{\alpha_R+1}{a}.
\end{equation}
For the transformation adopted in this work, $a=0.974$, and hence
\begin{equation}
    \alpha_{g'}
    =
    \frac{\alpha_R+1}{0.974}-1.
\end{equation}

\section{Correlated Noise and Surface-brightness Depth}
\label{app:correlatednoise}

Because the resampling of the individual exposures can modify the single-pixel noise properties and introduce correlations between neighboring pixels, we test here how well the propagated error model represents the noise in the final coadded images and, consequently, the statistical surface-brightness depths derived from it. The uncertainties themselves are estimated in the individual exposures and propagated through the resampling and coaddition using the same weights and masks as for the science data, rather than being derived from the measured pixel-to-pixel scatter of the final coadd.

To test the effect of resampling-induced pixel correlations, we performed an $N\times N$ test following the procedure presented in \cite{KlugeERO}. We applied the test to the $g'$-band images of Coma, Ref1, and Ref2, and additionally to the Coma $u'$- and $r'$-band images.
All masks created for this work were applied, and the bright galaxy models had previously been subtracted. To reduce the contribution of large-scale background variations, we additionally subtracted a SExtractor background model using the same background parameters as for the dwarf catalogs (\texttt{BACK\_SIZE=32} and \texttt{BACK\_FILTERSIZE=3}). 
For each image, we restricted the analysis to unmasked regions with finite pixel values and a narrow range of propagated per-pixel uncertainties, thereby selecting regions with a relatively uniform noise level. For each box size $N$, we randomly placed 10,000 $N\times N$ boxes in these regions, requiring each box to be fully unmasked, and calculated the mean pixel value in each box.
We then determined the standard deviation of these box means, $\sigma_{N\times N}$.

For statistically independent pixels with approximately uniform variance, $\sigma_{N\times N}$ is expected to decrease as $1/N$. We therefore compared the measured $\sigma_{N\times N}$ with the prediction from the propagated error image, $\tilde{\sigma}_{\rm pix}/N$, where $\tilde{\sigma}_{\rm pix}$ is the median propagated per-pixel uncertainty over the same valid regions.

\begin{figure}[!ht]
    \centering
    \includegraphics[width=\linewidth]{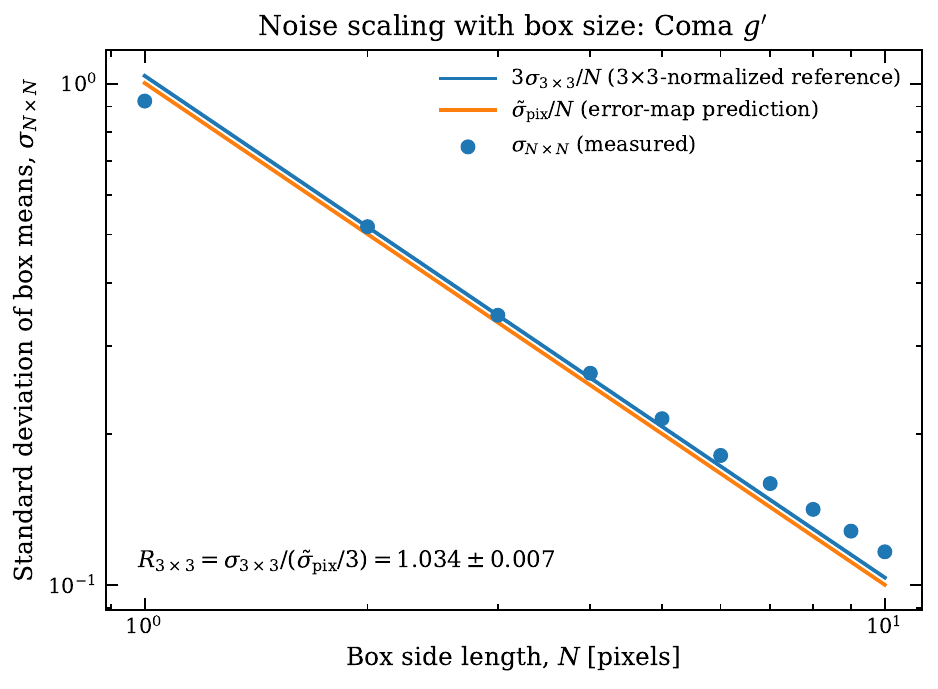}
    \includegraphics[width=\linewidth]{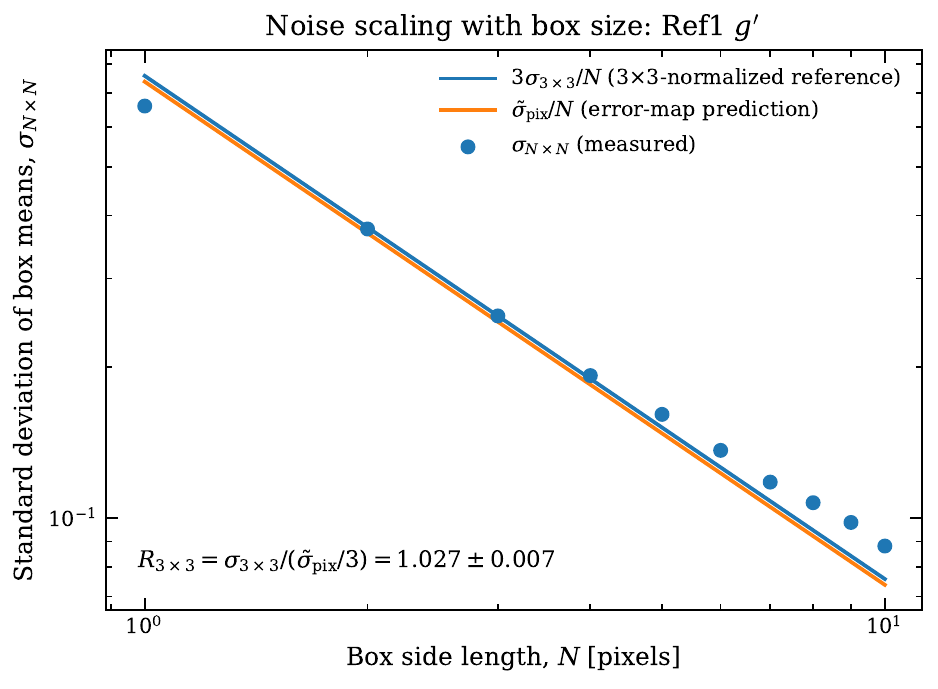}
    \includegraphics[width=\linewidth]{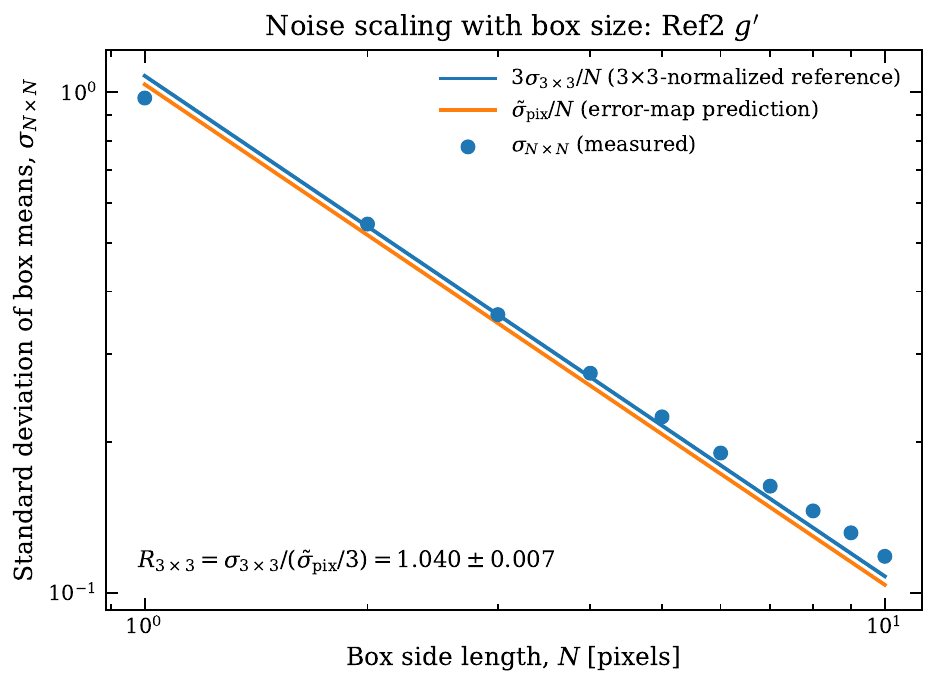}
    \caption{
    Standard deviation of the mean counts measured in randomly placed $N\times N$ boxes in the masked Coma, Ref1, and Ref2 $g'$-band images. The orange lines show the predictions $\tilde{\sigma}_{\rm pix}/N$ from the propagated error images, assuming statistically independent pixels. The blue lines show the corresponding $1/N$ scalings normalized to the measured $3\times3$ values.}
    \label{fig:correlatednoise_g}    
\end{figure}

\begin{figure}[!ht]
    \centering
    \includegraphics[width=\linewidth]{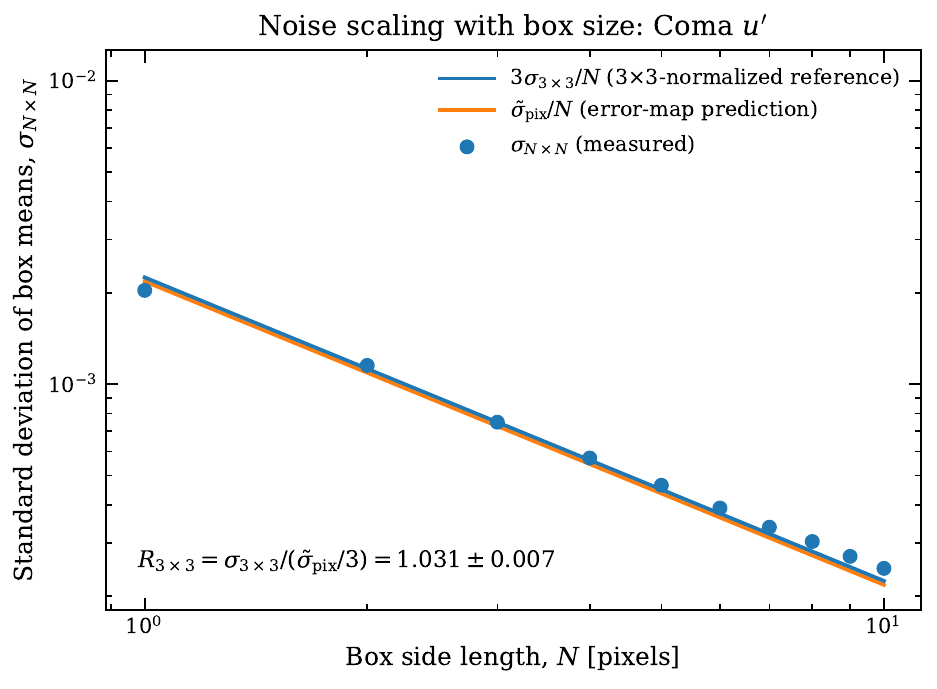}
    \includegraphics[width=\linewidth]{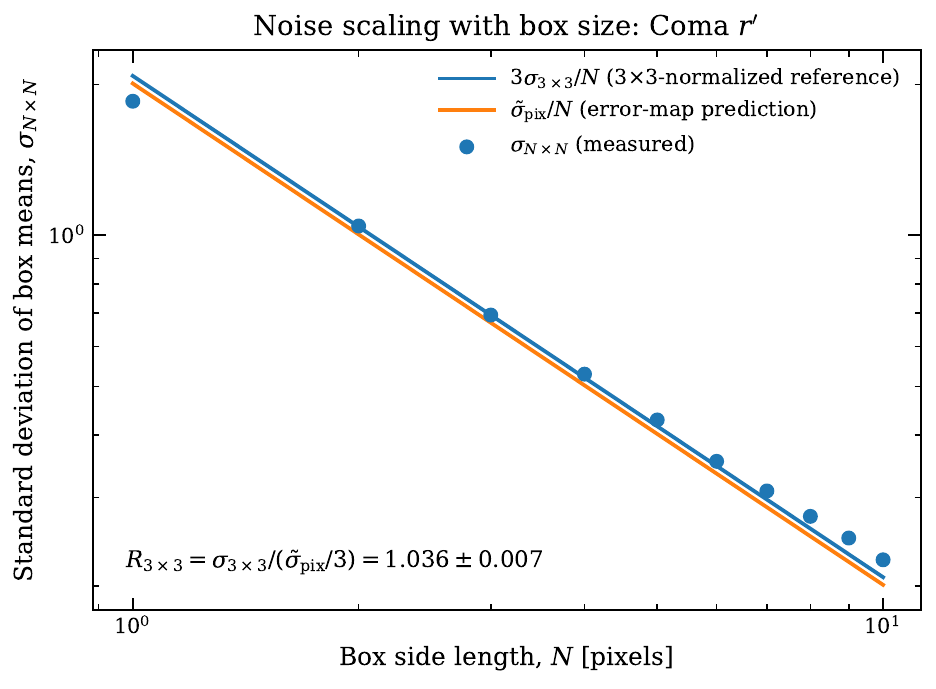}
    \caption{Standard deviation of the mean counts measured in randomly placed $N\times N$ boxes in the masked Coma $u'$- and $r'$-band images. The orange lines show the predictions $\tilde{\sigma}_{\rm pix}/N$ from the propagated error images, assuming statistically independent pixels. The blue lines show the corresponding $1/N$ scalings normalized to the measured $3\times3$ values.}
    \label{fig:correlatednoise_filters}
\end{figure}

For box sizes of $2\times2$, $3\times3$, and $4\times4$ pixels, the measured $\sigma_{N\times N}$ closely follows the expected $1/N$ scaling in all five images (Figures~\ref{fig:correlatednoise_g} and \ref{fig:correlatednoise_filters}). This indicates that pixel covariances introduced by the resampling do not significantly affect the noise scaling over these spatial scales. The deviations of the $1\times1$ measurements from this relation instead reflect the modification of the single-pixel variance by the resampling process. Importantly, our propagated error estimate does not rely on the measured single-pixel scatter in the coadded image and therefore does not inherit the corresponding bias seen for the $1\times1$ measurements.

The similar behavior in the three independent $g'$-band fields shows that this result is not specific to the Coma image, while the agreement between the Coma $u'$, $g'$, and $r'$ bands shows that it is also not specific to a particular filter. The latter is expected, since the photometric filter itself does not introduce resampling-induced pixel correlations.


The measured $3\times3$ scatter is only $(2.7\pm0.7)\%$ to $(4.0\pm0.7)\%$ higher than the prediction from the propagated error image across the five images, with $R_{3\times3}$ ranging from $1.027\pm0.007$ to $1.040\pm0.007$. This small difference could, at least in part, be caused by faint unmasked sources, which increase the measured standard deviation but are not included in the propagated error model. At larger box sizes, residual large-scale background variations become increasingly dominant and may therefore explain the growing deviation from the $1/N$ relation.

Overall, the close agreement between the measured and predicted $3\times3$ scatter across all five images indicates that our propagated error model provides a reasonable description of the statistical noise on these scales. The measured excess of $2.7$--$4.0\%$ likely represents an upper limit on any discrepancy arising from the error propagation itself, since faint unmasked sources can increase the measured scatter.
Even the largest measured difference corresponds to only $\simeq0.043$ mag in the inferred depth, which is smaller than the $0.1$ mag precision at which we report our surface-brightness limits.
Residual large-scale background variations constitute a separate systematic contribution and are not included in the propagated statistical error model. Accordingly, the quoted surface-brightness depths trace the propagated statistical uncertainties and do not include additional uncertainties arising from large-scale background structure.

\section{BCF and GLF Fit Posteriors}\label{sec:GLFposteriors}

\begin{figure*}[t!]
    \begin{center}
        \includegraphics[width=0.49\textwidth]{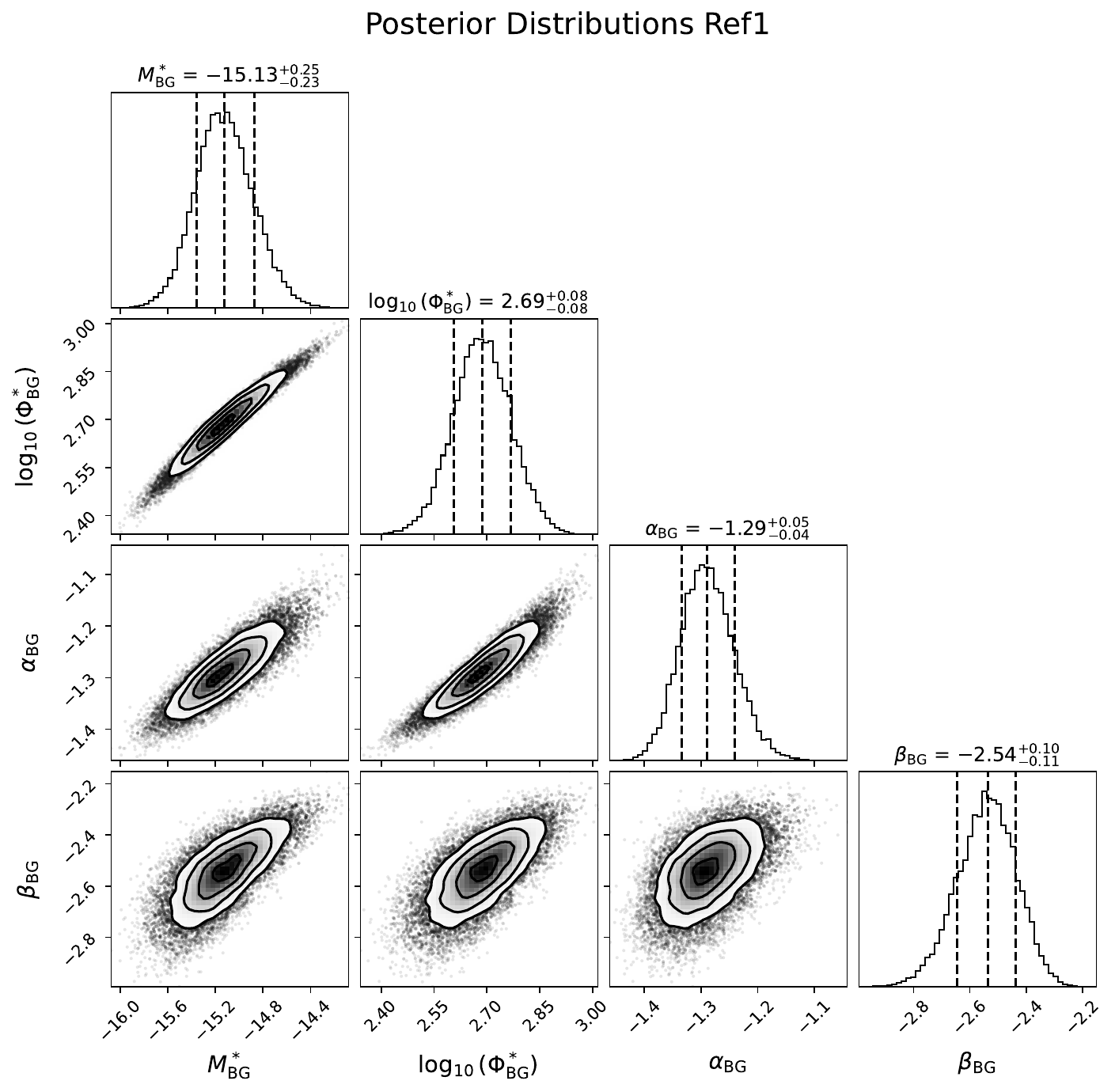}
        \includegraphics[width=0.49\textwidth]{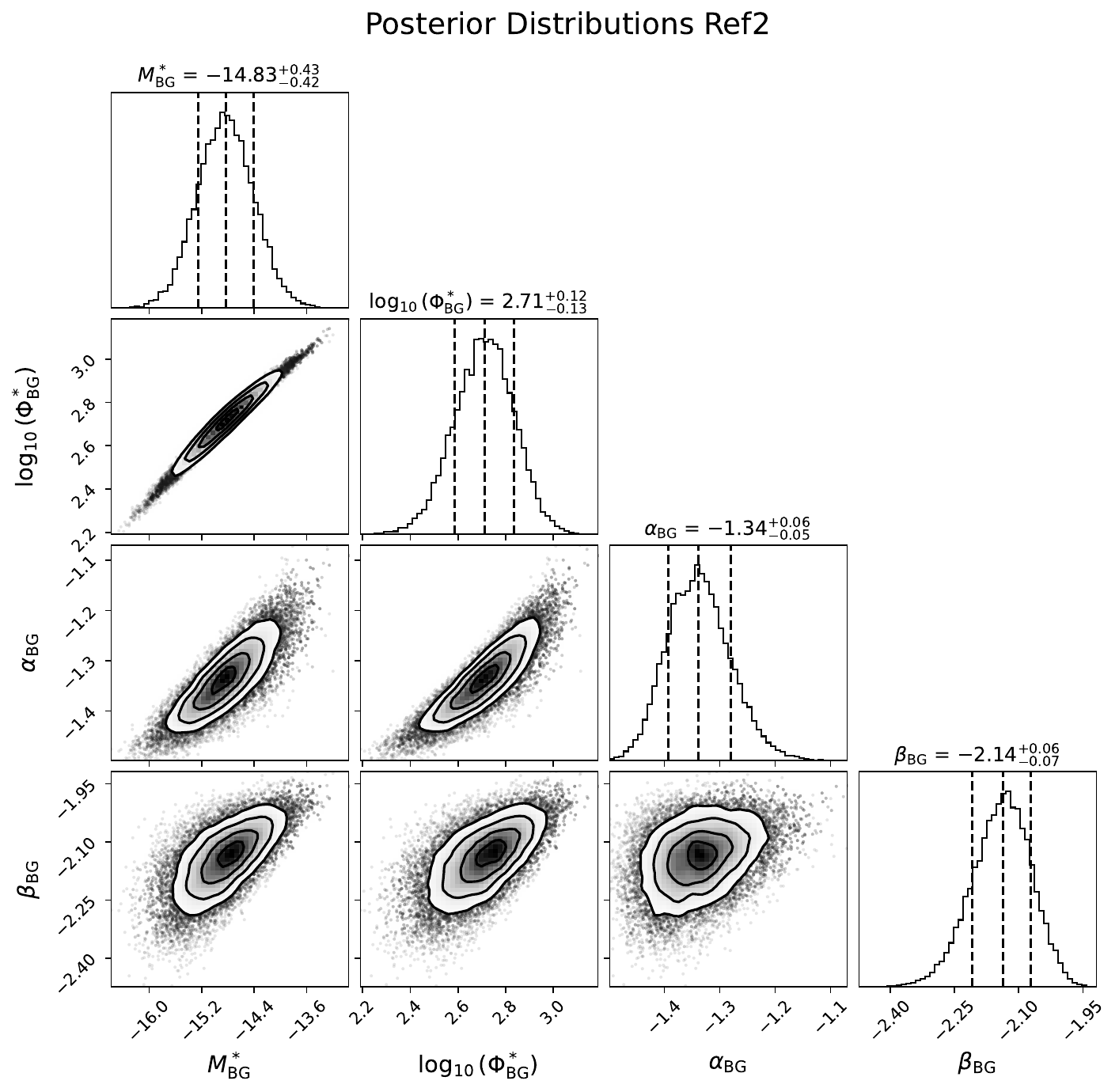}
        \caption{Posterior distributions of the nominal-magnitude fits of a BCF using a double power law for both reference fields individually (Ref1 on the left and Ref2 on the right). The contours correspond to the $0.5\sigma$, $1\sigma$, $2\sigma$, and $3\sigma$ levels.}
        \label{fig:PosteriorLFref_combined}
    \end{center}
\end{figure*}

\begin{figure*}[t!]
    \begin{center}
        \includegraphics[width=0.5\textwidth]{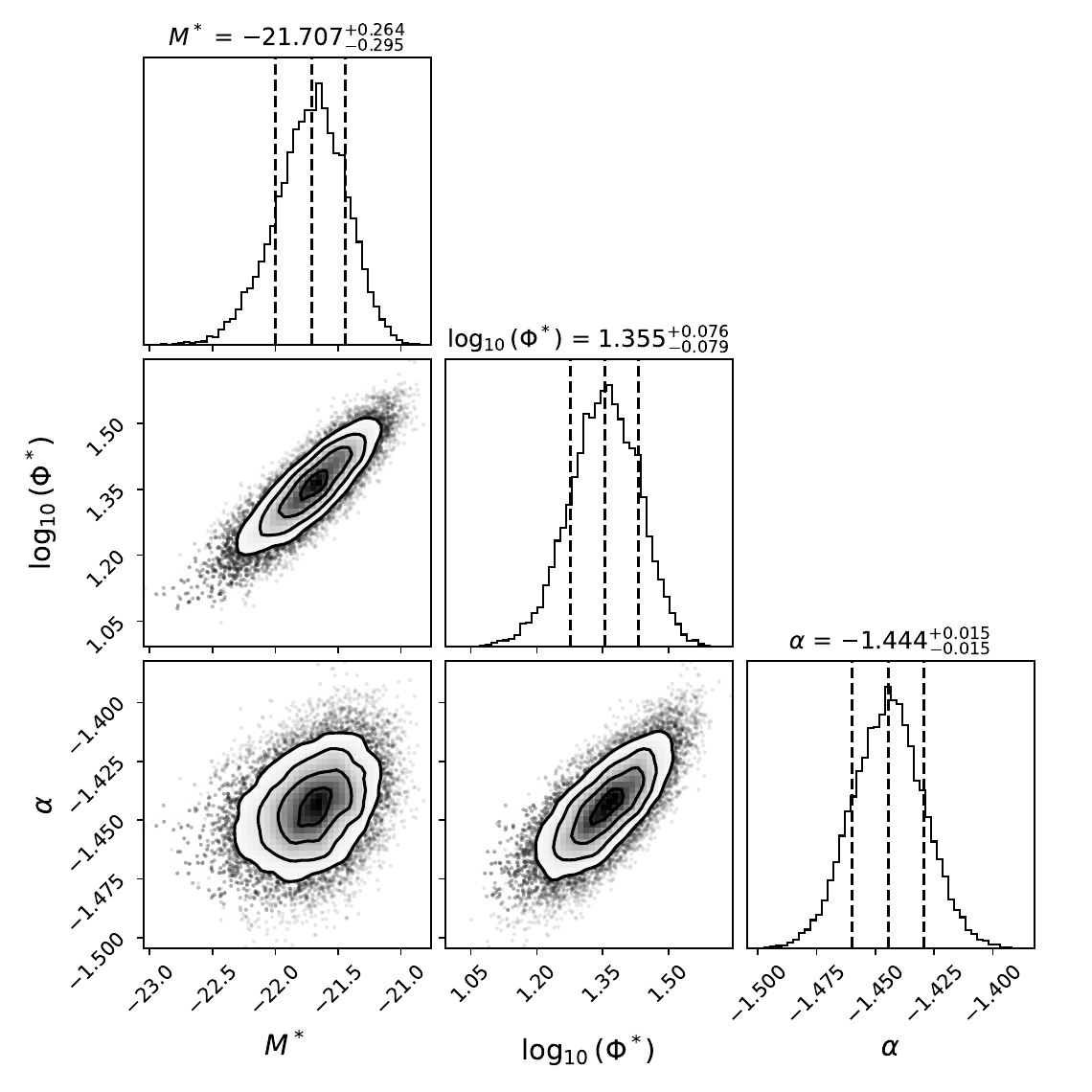}
        \caption{Posterior distribution of the nominal-magnitude fit of a GLF using a single Schechter function excluding NGC 4889, NGC 4874, and NGC 4839. The contours correspond to the $0.5\sigma$, $1\sigma$, $2\sigma$, and $3\sigma$ levels.}
        \label{fig:PosteriorLF}
    \end{center}
\end{figure*}

\begin{figure*}[ht]
    \begin{center}
        \includegraphics[width=\textwidth]{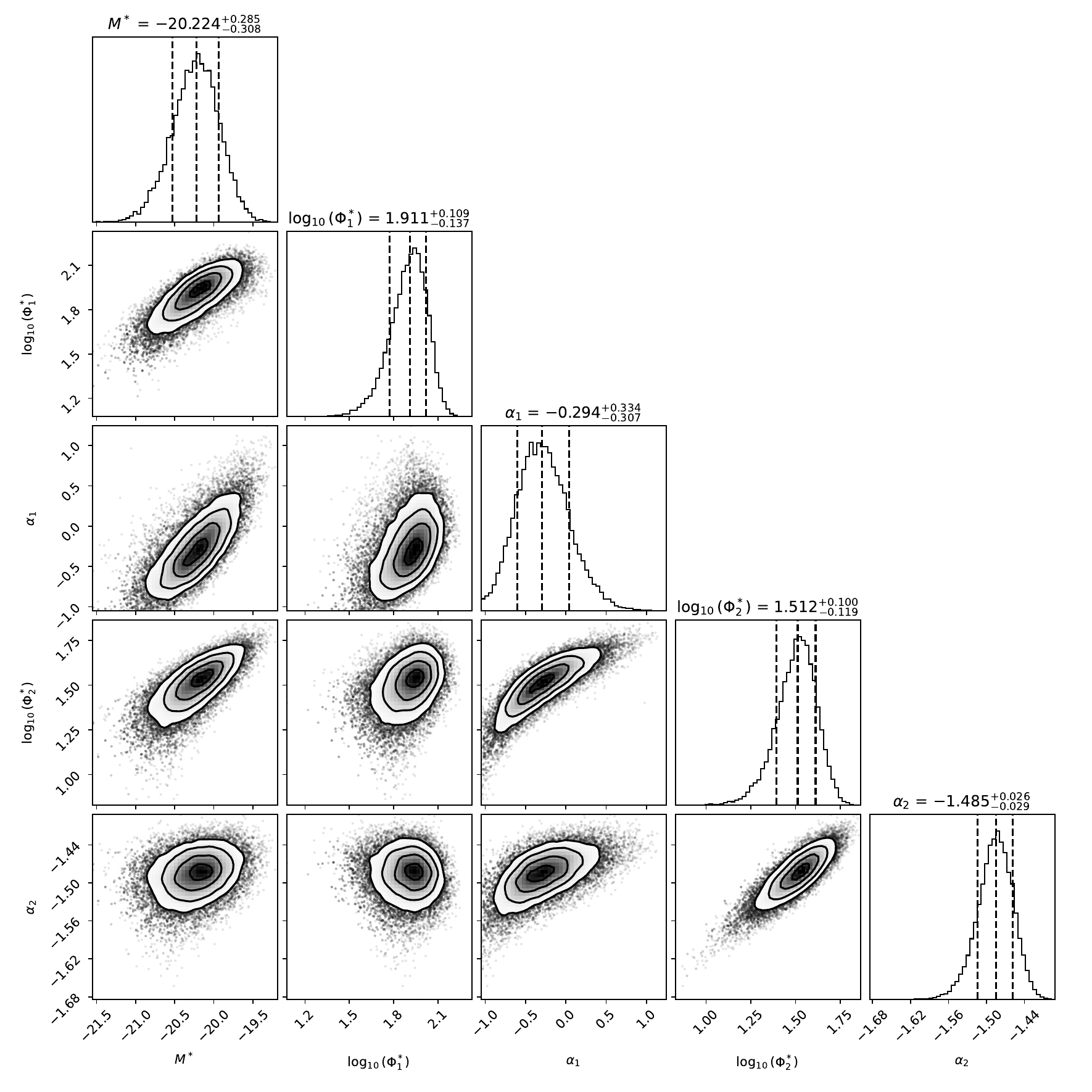}
        \caption{Posterior distribution of the nominal-magnitude fit of a GLF using a double Schechter function excluding NGC 4889, NGC 4874, and NGC 4839. The contours correspond to the $0.5\sigma$, $1\sigma$, $2\sigma$, and $3\sigma$ levels.}
        \label{fig:PosteriorLF_DS}
    \end{center}
\end{figure*}

\begin{figure*}[ht]
    \begin{center}
        
        \includegraphics[width=\textwidth]{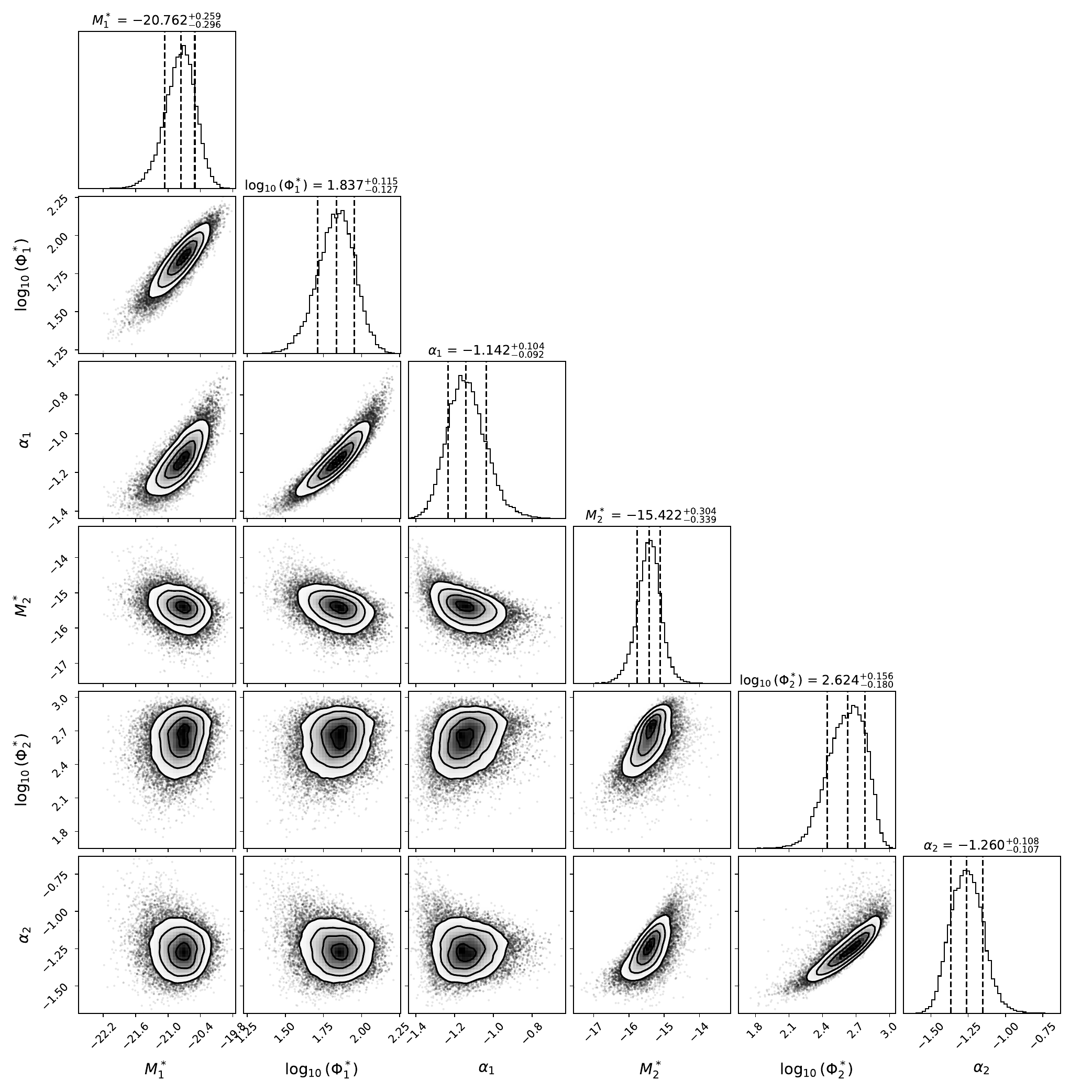}
        \caption{Posterior distribution of the nominal-magnitude fit of a GLF using a double Schechter function with two $M^*$ excluding NGC 4889, NGC 4874, and NGC 4839. The contours correspond to the $0.5\sigma$, $1\sigma$, $2\sigma$, and $3\sigma$ levels.}
        \label{fig:PosteriorLF_2MDS}
    \end{center}
\end{figure*}

\begin{figure*}[ht]
    \begin{center}
        \includegraphics[width=\textwidth]{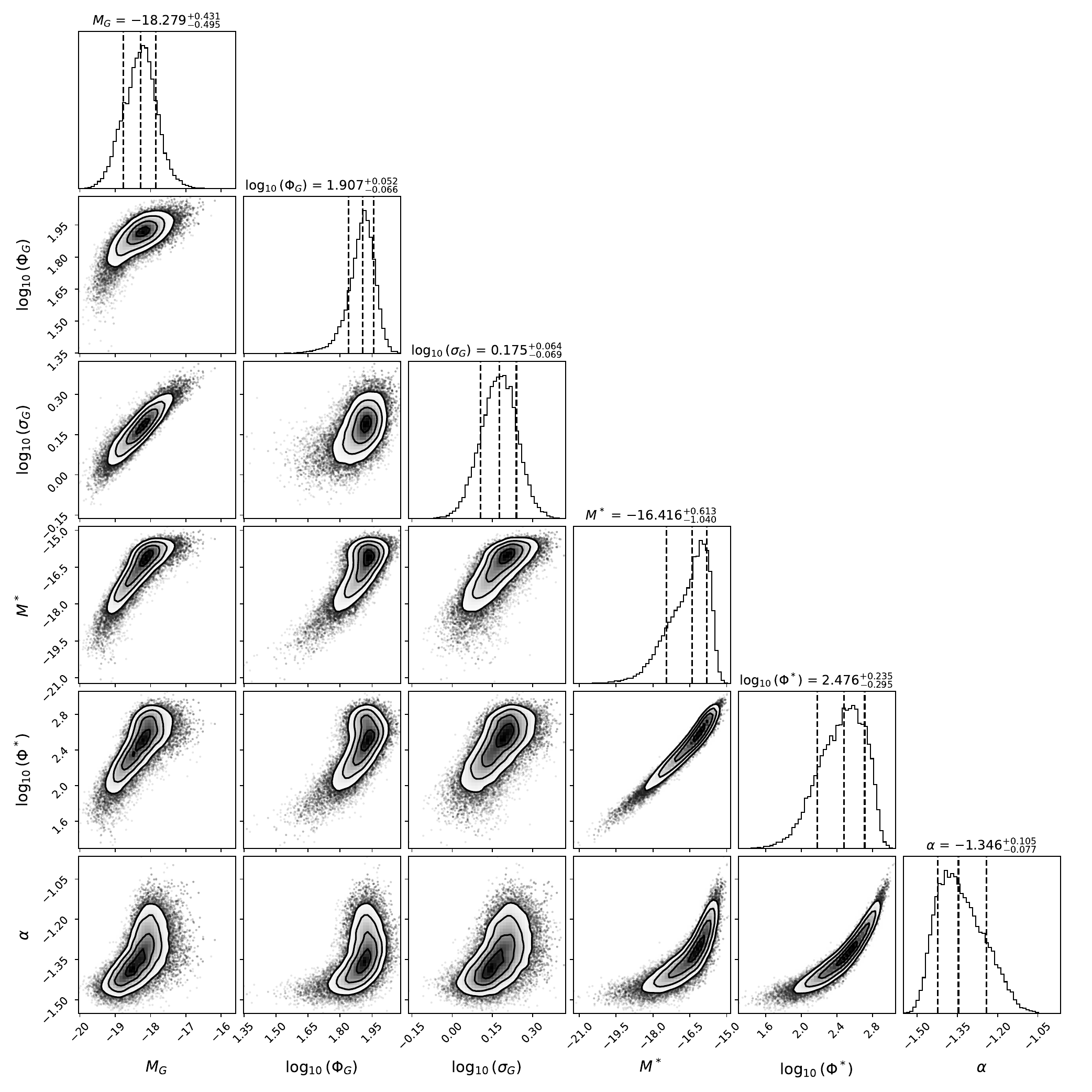}
        \caption{Posterior distribution of the nominal-magnitude fit of a GLF using a Gaussian + Schechter function excluding NGC 4889, NGC 4874, and NGC 4839. The contours correspond to the $0.5\sigma$, $1\sigma$, $2\sigma$, and $3\sigma$ levels.}
        \label{fig:PosteriorLF_GS}
    \end{center}
\end{figure*}

The posterior distributions of the nominal-magnitude BCF and GLF fits used for the model-selection analysis discussed in Sect.~\ref{sec:LFfunction} are shown in this Appendix. The BCF posteriors for the two reference fields are shown in Fig.~\ref{fig:PosteriorLFref_combined}. The GLF posteriors are shown for the single Schechter model in Fig.~\ref{fig:PosteriorLF}, the double Schechter model in Fig.~\ref{fig:PosteriorLF_DS}, the double Schechter model with two independent $M^*$ parameters in Fig.~\ref{fig:PosteriorLF_2MDS}, and the Gaussian + Schechter model in Fig.~\ref{fig:PosteriorLF_GS}. In all GLF fits, NGC~4889, NGC~4874, and NGC~4839 are excluded.





\end{document}

%% file: schechter_model_comparison_table.tex
\begin{deluxetable*}{lcc}
\tabletypesize{\scriptsize}
\tablewidth{\textwidth}
\tablecaption{
Comparison of nominal-magnitude GLF model fits used for model selection.
Quoted parameter values correspond to the posterior median with asymmetric
$16$th--$84$th percentile credible intervals.
For each model, the BIC and $\Delta{\rm BIC}$ are listed below the fit parameters.
The quantity $\Delta{\rm BIC}$ is defined as
$\Delta{\rm BIC}={\rm BIC}_{\rm model}-{\rm BIC}_{\rm single\,Schechter}$. 
The characteristic magnitudes are given in $g'\,\mathrm{mag}$, while the Gaussian width is reported as $\log_{10}(\sigma_{\rm G})$, with $\sigma_{\rm G}$ expressed in $g'\,\mathrm{mag}$, and
$\Phi^*$ and $\Phi_{\rm G}$ are reported as $\log_{10}$ values in units of $\mathrm{deg}^{-2}\,\mathrm{mag}^{-1}$; all slope parameters are dimensionless.
\label{tab:schechter-constraints}
}
\tablehead{
\colhead{Parameter} &
\colhead{Prior} &
\colhead{Posterior constraint / value}
}
\startdata
\multicolumn{3}{c}{\textbf{Single Schechter}} \\
\hline
$M^{*}$ & $\mathcal{N}(-21.469, 0.511^2)$ & $-21.71^{+0.26}_{-0.29}$ \\
$\log_{10}(\Phi^{*})$ & $\mathcal{N}(1.391, 0.179^2)$ & $1.355^{+0.076}_{-0.079}$ \\
$\alpha$ & $\mathcal{U}(-2.048, -0.848)$ & $-1.444^{+0.015}_{-0.015}$ \\
\hline
BIC & \nodata & $-53493.7$\\
$\Delta{\rm BIC}$ & \nodata & $0.0$ \\[0.8em]
\multicolumn{3}{c}{\textbf{Double Schechter}} \\
\hline
$M^{*}$ & $\mathcal{N}(-19.927, 0.793^2)$ & $-20.220^{+0.29}_{-0.31}$ \\
$\log_{10}(\Phi^{*}_{1})$ & $\mathcal{N}(1.922, 0.222^2)$ & $1.91^{+0.11}_{-0.14}$ \\
$\alpha_{1}$ & $\mathcal{U}(-1.051, 1.349)$ & $-0.30^{+0.34}_{-0.31}$ \\
$\log_{10}(\Phi^{*}_{2})$ & $\mathcal{U}(0.733, 2.533)$ & $1.51^{+0.10}_{-0.12}$ \\
$\alpha_{2}$ & $\mathcal{U}(-2.063, -0.863)$ & $-1.485^{+0.026}_{-0.029}$ \\
\hline
BIC & \nodata & $-53484.0$\\
$\Delta{\rm BIC}$ & \nodata & $9.7$ \\[0.8em]
\multicolumn{3}{c}{\textbf{Double Schechter with two $M^{*}$}} \\
\hline
$M^{*}_{1}$ & $\mathcal{N}(-20.812, 0.595^2)$ & $-20.76^{+0.26}_{-0.30}$ \\
$\log_{10}(\Phi^{*}_{1})$ & $\mathcal{N}(1.788, 0.298^2)$ & $1.84^{+0.11}_{-0.13}$ \\
$\alpha_{1}$ & $\mathcal{U}(-1.788, -0.588)$ & $-1.14^{+0.10}_{-0.09}$ \\
$M^{*}_{2}$ & $\mathcal{N}(-15.460, 1.129^2)$ & $-15.42^{+0.30}_{-0.34}$ \\
$\log_{10}(\Phi^{*}_{2})$ & $\mathcal{N}(2.624, 0.391^2)$ & $2.62^{+0.16}_{-0.18}$ \\
$\alpha_{2}$ & $\mathcal{U}(-1.831, -0.631)$ & $-1.26^{+0.11}_{-0.11}$ \\
\hline
BIC & \nodata & $-53487.4$\\
$\Delta{\rm BIC}$ & \nodata & $6.3$ \\[0.8em]
\multicolumn{3}{c}{\textbf{Gaussian + Schechter}} \\
\hline
$M_{\rm G}$ & $\mathcal{N}(-18.891, 1.082^2)$ & $-18.28^{+0.43}_{-0.49}$ \\
$\log_{10}(\Phi_{\rm G})$ & $\mathcal{N}(1.804, 0.181^2)$ & $1.907^{+0.052}_{-0.066}$ \\
$\log_{10}(\sigma_{\rm G})$ & $\mathcal{N}(0.097, 0.167^2)$ & $0.175^{+0.064}_{-0.069}$ \\
$M^{*}$ & $\mathcal{N}(-17.441, 1.631^2)$ & $-16.4^{+0.6}_{-1.0}$ \\
$\log_{10}(\Phi^{*})$ & $\mathcal{N}(2.261, 0.456^2)$ & $2.48^{+0.23}_{-0.29}$ \\
$\alpha$ & $\mathcal{U}(-1.977, -0.777)$ & $-1.35^{+0.10}_{-0.08}$ \\
\hline
BIC & \nodata & $-53486.8$\\
$\Delta{\rm BIC}$ & \nodata & $6.9$ \\
\enddata
\tablecomments{
Normal priors are written as $\mathcal{N}(\mu,\sigma^2)$,
where $\sigma$ is the standard deviation.
Uniform priors are written as $\mathcal{U}(a,b)$.
The BIC rows give the Bayesian information criterion for each model.
The $\Delta{\rm BIC}$ rows are computed relative to the single-Schechter model;
negative values indicate a lower BIC than the single-Schechter fit. These fits use the measured best-estimate galaxy magnitudes and are used for model selection only. The fiducial single-Schechter constraints, including the individual magnitude uncertainties as nuisance quantities, are presented in Section~\ref{sec:EddingtonBias}.
}
\end{deluxetable*}